\pdfoutput=1
\documentclass[a4paper, 11pt]{article}
\usepackage{jheppub, amssymb, amsmath, graphicx, feynmp-auto, hyperref, color,multirow, tabularx, arydshln, slashed, orcidlink} 
\usepackage[T1]{fontenc}
\definecolor{nicered}{rgb}{.7,.1,.1}
\definecolor{nicegreen}{rgb}{.1,.5,.1}
\definecolor{darkblue}{rgb}{0,0,.5}
\hypersetup{colorlinks, citecolor=nicegreen,linkcolor=nicered, urlcolor=darkblue}
\definecolor{lightgray}{gray}{0.6}

\renewcommand\P{{\mathcal P}}
\renewcommand\d{{\mathrm d}}
\newcommand\C{{\mathcal C}}

\title{\hspace*{-1.8pt}Beautiful Majorana Higgses at Colliders}

\author[a]{\hspace*{-1pt}Benjamin Fuks \orcidlink{0000-0002-0041-0566},}
\emailAdd{fuks@lpthe.jussieu.fr}
\affiliation[a]{\normalsize \it Laboratoire de Physique Th\'eorique et Hautes Energies (LPTHE),
UMR 7589, Sorbonne Universit\'e \& CNRS, 4 place Jussieu, 75252 Paris Cedex 05, France}

\author[b]{Jonathan Kriewald \orcidlink{0000-0002-3313-115X},}
\emailAdd{jonathan.kriewald@ijs.si}
\affiliation[b]{\normalsize \it  Jo\v zef Stefan Institute, Jamova 39, 1000 Ljubljana, Slovenia}

\author[b, c]{Miha Nemev\v{s}ek \orcidlink{0000-0003-1110-342X}}
\emailAdd{miha.nemevsek@ijs.si}
\affiliation[c]{\normalsize \it  Faculty of Mathematics and Physics, University of Ljubljana, 
Jadranska 19, 1000 Ljubljana, Slovenia}

\author[d, e]{and Fabrizio Nesti \orcidlink{0000-0003-4508-8141}}
\emailAdd{fabrizio.nesti@aquila.infn.it}
\affiliation[d]{\normalsize \it 
  Dipartimento di Scienze Fisiche e Chimiche, Universit\`a dell'Aquila, via Vetoio, I-67100, L'Aquila, Italy}
\affiliation[e]{\normalsize \it INFN, Laboratori Nazionali del Gran Sasso, I-67100 Assergi (AQ), Italy}

\date{\today}

\abstract{
We investigate a novel collider signature within the minimal Left-Right Symmetric Model, 
featuring a Higgs sector composed of a bi-doublet and two triplets.
Our study focuses on a region of the parameter space where the $SU(2)_R$ charged gauge 
boson $W_R$ lies in the multi-TeV regime (3–100~TeV) and the additional Higgs states play a 
significant role.
In this scenario, a heavy neutral Higgs boson $\Delta$ with a dominant $SU(2)_R$ triplet 
component can be produced in association with either a Standard Model Higgs boson or a 
massive weak boson.
The subsequent decay of the heavy Higgs into Majorana neutrinos $N$ results in displaced 
lepton signatures, providing a striking manifestation of lepton number violation. 
Additionally, we explore how the production of $b$-jets in these processes can enhance 
hadron-collider sensitivity to such signals. 
A particularly compelling channel, $pp \to b \bar{b} NN$, offers the exciting possibility 
of simultaneously probing the spontaneous mass origin of both Dirac fermions and Majorana states.
Based on an optimised event selection strategy and state-of-the-art Monte Carlo simulations, 
we outline the expected reach at the HL-LHC and future colliders. 
Our findings demonstrate that this channel probes a region of parameter space where the 
neutral Higgs triplet and heavy neutrino masses are relatively light ($m_\Delta \lesssim 250$~GeV,
$m_N \lesssim 80$~GeV), indirectly constraining the $W_R$ boson to the deep multi-TeV domain, 
with sensitivity extending up to 70-80~TeV, effectively turning the LHC into a precision machine.
}


\begin{document}

\maketitle
\flushbottom

%
%
\section{Introduction} \label{sec:Intro}
Explaining the origin of neutrino mass remains one of the most significant mysteries still 
left unresolved by the Standard Model (SM) of particle physics.
Perhaps the most elegant way of accounting for neutrino masses is the seesaw  
mechanism~\cite{Minkowski:1977sc, Gell-Mann:1979vob, Glashow:1979nm, Mohapatra:1979ia,Sawada:1979dis} 
that postulates the existence of new degrees of freedom, leading to light Majorana neutrinos.
Another aesthetic puzzle of the SM is the complete asymmetry of weak interactions that couple 
chirally to left-handed components only via the $SU(2)_L$ gauge group.
Within the minimal Left-Right~\cite{Pati:1974yy, Mohapatra:1974gc, Mohapatra:1979ia} symmetric model 
(LRSM), these two problems are resolved simultaneously by extending the gauge group to 
$SU(2)_L \times SU(2)_R \times U(1)_{B-L}$.
The $SU(2)_R$ gauge group gets spontaneously broken~\cite{Senjanovic:1978ev} at a scale $v_R$, 
which is necessarily above the electroweak scale and at least within a few TeVs.
The model enjoys an additional discrete Left-Right (LR) exchange symmetry, either in the form 
of $\cal P$ or $\cal C$ conjugation~\cite{Maiezza:2010ic}, that is broken together with the 
$SU(2)_L \times SU(2)_R \times U(1)_{B-L}$ symmetry. 
This gauge symmetry moreover necessarily requires three generations of right-handed neutrinos 
that may obtain a Majorana mass when the symmetry is broken by the vacuum expectation values 
(vevs) of a pair of scalar triplets $\Delta_{L,R}$ and a single scalar bi-doublet $\phi$.

Imposition of either LR discrete symmetry in the Yukawa sector severely restricts the flavour
structure of the LRSM, effectively requiring the right-handed quark mixing to be (nearly) equal 
to the CKM matrix for the case of ($\P$) $\C$ conjugation.
This leads to strong flavour constraints~\cite{Beall:1981ze}, mainly from kaon mixing, 
which push $v_R$ in the few TeV regime and the bi-doublets mass scale even
higher~\cite{Senjanovic:1979cta}.
These studies have been continuously updated~\cite{Zhang:2007da, Maiezza:2010ic, Bertolini:2014sua},
taking into account CP-violating constraints, electric dipole moments~\cite{Ecker:1985vv,
Kiers:2002cz, Maiezza:2014ala, Bertolini:2019out} and addressing the strong CP 
problem~\cite{Maiezza:2014ala, Kuchimanchi:2014ota,Senjanovic:2020int}.
The bottom-line is that even though the flavour constraints impose the typical LRSM mass scale 
to lie deep in the TeV region, collider searches are competitive and may constitute the first
evidence of new physics stemming from the LR symmetry.
One may also contemplate how to explain dark matter~\cite{Bezrukov:2009th, Nemevsek:2012cd}, 
which is an attractive possibility, however due to dilution constraints~\cite{Nemevsek:2022anh} 
the LR scale needs in this case to be far above the TeV scale~\cite{Nemevsek:2023yjl, Dev:2025fcv}.

In the LRSM there are two sources of neutrino mass~\cite{Mohapatra:1980yp}, a first one 
originating from Dirac mass terms and the right-handed Majorana mass terms (through a so-called
type~I seesaw), and another one directly connected to the vev of the $SU(2)_L$ Higgs 
triplet (type II).
Because of the LR symmetry, the two Dirac and Majorana mass sources are strongly 
related~\cite{Nemevsek:2012iq}, and the heavy-light neutrino mixing can thus be determined in 
a closed form using the Cayley-Hamilton formula~\cite{Kriewald:2024cgr}.
Ultimately, we would like to directly probe the full microscopic origin of neutrino masses by 
observing and separating the channels that depend on Majorana and Dirac couplings.
In the SM~\cite{Weinberg:1967tq}, this is straightforward: the Higgs boson $h$ decays to an 
$f\bar f$ fermion-antifermion pair with a rate $\Gamma(h \to f \overline f) \propto m_f^2$ 
(with $m_f$ referring to the mass of the fermion $f$).
This has been confirmed experimentally at least for the $b$ quark and the tau lepton, and 
indirectly for the top quark and through an upper bound for the
muon~\cite{ATLAS:2022vkf,CMS:2022dwd}.
In the LRSM, the right-handed neutrinos $N$ get their Majorana mass from the spontaneous 
breaking of $SU(2)_R$. 
This is reflected in the rate $\Gamma(\Delta \to NN) \propto m_N^2$ (where 
$\Delta \equiv \Delta_R^0$ stands for the neutral component of the $SU(2)_R$ Higgs 
triplet and $m_N$ for the heavy neutrino mass), a relation that should be tested directly 
in data. 
A handle on this challenge resides in the fact that the SM Higgs boson $h$ and LRSM 
$\Delta$ boson mix. 
This has two consequences: firstly it becomes possible to probe the spontaneous origin of 
the mass of the heavy neutrino $N$ via the $h\to NN$ decay~\cite{Maiezza:2015lza}, and 
secondly the $SU(2)_R$ Higgs triplet can be produced via gluon fusion, thus offering a 
direct access to a `Majorana Higgs' signature $gg \to \Delta \to NN$~\cite{Nemevsek:2016enw}.
In this work, we study further such opportunities offered within the LRSM to probe 
neutrino mass generation. 

The Higgs sector of the LRSM has been studied to various degrees of detail, from the 
original spontaneous LR symmetry breaking mechanism~\cite{Senjanovic:1975rk, 
Senjanovic:1978ev}, to the more complete treatment of the scalar mass 
matrices~\cite{Olness:1985bg, Gunion:1989in, Deshpande:1990ip, Kiers:2005gh, Maiezza:2016ybz} 
and determination of the constraints originating from electroweak observables and 
perturbativity~\cite{Maiezza:2016bzp, Chauhan:2018uuy, Mohapatra:2019qid}.
Recently~\cite{Kriewald:2024cgr}, we revisited this issue, diagonalised analytically the 
mass matrices and thus devised a physical input scheme with masses and mixing angles. 
This was then implemented through a \textsc{FeynRules}~\cite{Christensen:2008py, 
Alloul:2013bka}/UFO~\cite{Degrande:2011ua, Darme:2023jdn} model file enabling state-of-the-art 
Monte Carlo simulations at next-to-leading-order (NLO) in QCD and leading-order loop-induced 
processes, that we heavily use in the present work.

A number of searches at high energy colliders have investigated LRSM signatures in different
regions of the parameter space, and with qualitatively different final states. 
Perhaps the most model-independent limit comes from searches with di-jet~\cite{ATLAS:2019fgd,
CMS:2019gwf} and $tb$~\cite{CMS:2023gte, ATLAS:2023ibb} final states, since the left-handed
and right-handed CKM matrices are similar~\cite{Zhang:2007da, Maiezza:2010ic,
Senjanovic:2014pva, Senjanovic:2015yea}.
As soon as at least one right-handed neutrino is below the $W_R$ mass threshold, it can be
efficiently produced through the Keung-Senjanović (KS)~\cite{Keung:1983uu} channel that may
signal lepton-number violations if the final state particles are well separated in the 
transverse plane.  
This could exploit both the process $pp \to W_R^{\pm} \to \ell^{\pm} N \to \ell^{\pm}
\ell^{\pm} j j$ where light jets are produced, and the process process 
$pp \to W_R^{\pm} \to \ell^{\pm} N \to \ell^{\pm} \ell^{\pm} t b$ where a top-antibottom 
or an antitop-bottom pair is produced~\cite{Frank:2023epx}.
On the other hand, lowering the mass $m_N$ of the heavy neutrino~\cite{Nemevsek:2011hz}
results in merged non-isolated neutrino jets~\cite{Mitra:2016kov} that additionally 
become displaced before finally transitioning into a charged lepton and missing 
energy~\cite{Nemevsek:2018bbt}.
The latter signature can be efficiently constrained by recasting the bounds stemming 
from direct $W' \to \ell \nu$ searches~\cite{CMS:2022krd, CMS:2022ncp, ATLAS:2019lsy}, 
which yields the most significant direct bound on the mass of the $W_R$ boson $M_{W_R}$.
Obviously, the $W_R$ boson can be deeply off-shell and new physics signals could still 
be observed away from the resonance~\cite{Ruiz:2017nip}.
Currently, the most sensitive bounds hence range in the $5-6$~TeV region, depending on 
the flavour of the final state charged lepton and on the value of $m_N$. 
Finally, it has been shown that future colliders have the potential to push these 
limits beyond 30~TeV~\cite{Nemevsek:2023hwx}.

While the gauge sector of the model has been quite thoroughly investigated, including the
effects of gauge boson mixing~\cite{Dev:2015kca} and Dirac masses~\cite{Gluza:2016qqv,Das:2016akd,Das:2017hmg,Helo:2018rll}, 
the Higgs sector still offers opportunities.
In the minimal LRSM, flavour constraints push the bi-doublet in the $\mathcal{O}(20)$~TeV
range, beyond the reach of the LHC~\cite{Bertolini:2014sua, Bertolini:2019out} and the 
left-handed triplet also needs to be heavy if the $W_R$ boson is light~\cite{Maiezza:2016bzp}.
The $SU(2)_R$ scalar triplet on the other hand contains a singly-charged $\Delta_R^+$ state,
which is the mostly would-be-Goldstone eaten by the $W_R$ state, while its doubly-charged 
counterpart $\Delta_R^{++}$ and neutral component $\Delta$ are arbitrarily split and may be 
light, even lying around the TeV scale.
As a result, the mixing between the $h$ and $\Delta$ scalars can be significant, and 
this `Majorana Higgs' scenario can be probed both in the 
$gg \to h \to NN$~\cite{Maiezza:2015lza} and $gg \to \Delta \to NN$ 
channels~\cite{Nemevsek:2016enw}.
In this work we extend the analysis of~\cite{Nemevsek:2016enw}, and consider not only 
the gluon fusion production mode, but also the associated channels $pp \to \Delta X$ 
with $X = W, Z, h$.
The last process is particularly interesting from the conceptual point of view, because 
it offers a simultaneous handle on the Dirac mass origin of the bottom quarks originating 
from the SM Higgs decay, and the Majorana nature of the heavy neutrinos $N$ stemming 
from the $\Delta$ decay.
It thus manifestly signals lepton-number violation when $N \to \ell j j$ so that the 
final state comprises two same-sign or opposite-sign charged leptons and two $b$-jets, 
\textit{i.e.}\ an exciting novel smoking gun signature of the LRSM.
In this study, we consider the associated production process $pp \to \Delta h$, followed 
by the subsequent (and sometimes dominant) decays $h \to b \overline b$ and $\Delta \to NN$.
This hence leads to a `beautiful' Majorana final state with two $b$-jets and a pair of 
heavy neutrinos.
The heavy neutrinos then decay as $N \to \ell^\pm j j$ into charged leptons and jets, 
producing a manifestly lepton-number violating final state. 
In addition, the heavy neutrino $N$ can be significantly long-lived, decaying either 
inside the tracker, the muon chambers or even outside of a typical LHC detector.
With enough luminosity, these fairly soft final state objects can thus be used to probe 
very high ${W_R}$ boson mass values, extending well above the direct reach of the LHC.

We organise our discussion in the following way.
First we summarise the relevant features of the LRSM in Section~\ref{sec:Model}, pointing 
out the set of input parameters for our analysis, the relevant couplings and their role 
in generating the signals considered.
The phenomenological core of the paper lies in the following three sections, where we 
discuss $\Delta$ production in Section~\ref{sec:Production} and (displaced) decays and
branching ratios in Section~\ref{sec:Decays}, before providing details on our collider
analysis and the resulting sensitivities in Section~\ref{sec:Topos}.
In Section~\ref{sec:Outlook} we conclude and provide some future outlook.

%
%
\section{The minimal left-right symmetric model} \label{sec:Model}
The left-right symmetric model is a gauge extension of the SM based on the group 
\begin{equation}
  \mathcal G_{LR} = SU(3)_c \otimes SU(2)_L \otimes SU(2)_R \otimes U(1)_{B-L} \,.
\end{equation}
The whole model additionally enjoys a discrete LR-symmetry which exchanges the $SU(2)$ 
gauge group factors and acts on the fermionic and scalar fields as a generalised parity 
$\P$ or charge conjugation $\C$.
In addition to the usual left-handed fermion doublets $Q_L = (u_L, d_L)^T$ and 
$L_L = (\nu_L, \ell_L)^T$, the right-handed fermions are promoted to right-handed doublets
$Q_R = (u_R, d_R)^T$ and $L_R = (\nu_R, \ell_R)^T$, including thus three right-handed 
neutral leptons $\nu_R$.
Due to the introduction of right-handed neutrinos $\nu_R$, the gauge anomalies 
usually appearing due to the presence of the $U(1)_{B-L}$ symmetry are conveniently 
cancelled, and the electric charge is defined as $Q = T_L^3 + T_R^3 + \frac{B-L}{2}$.
The scalar sector of the LRSM comprises a bi-doublet field $\phi$ transforming as 
$(1, 2, 2, 0)$ under the gauge group, and two (complex) scalar triplets $\Delta_{L,R}$ 
respectively transforming as $(1, 3, 1, 2)$ and $(1, 1, 3, 2)$,
\begin{align}
  \phi &= \begin{pmatrix}\phi_1^{0*} & \phi_2^+ 
  \\
  \phi_1^- & \phi_2^0 \end{pmatrix}\,, 
  &
  \Delta_{L,R} &= 
  \begin{pmatrix}
    \frac{\Delta^+}{\sqrt{2}} & \Delta^{++}
    \\
    \Delta^0 & -\frac{\Delta^+}{\sqrt{2}}
  \end{pmatrix}_{L,R} \,.
\end{align}
After the simultaneous spontaneous symmetry breaking of $\mathcal G_{LR}\to SU(3)_c
\otimes U(1)_\mathrm{em}$ symmetry group and of the discrete LR-symmetry, the model 
predicts the vacuum structure,
\begin{align} \label{eq:vevs}
  \langle\phi\rangle &= 
  \begin{pmatrix}
  v_1 & 0
  \\
  0 & -e^{-i\alpha} v_2
  \end{pmatrix} \,, & 
  \langle\Delta_{L,R}\rangle &= 
  \begin{pmatrix}
    0 & 0
    \\
    v_{L,R} & 0
  \end{pmatrix} \, ,
\end{align}
with 
\begin{align}
  v^2 &= v_1^2 + v_2^2 \approx v_{\text{SM}}^2\, ,
  &
  0 \leq \tan\beta &= \frac{v_2}{v_1} \leq 1 \, .
\end{align}
In this case, $v_L \ll v \ll v_R$ and $v =174\,$GeV. 
The masses of the new heavy gauge bosons $W_R^{\pm}$ and $Z_R$ are then approximately given by
\begin{align}
  M_{W_R} &\simeq g v_R \, , & M_{Z_R} &\simeq \sqrt{3} \,M_{W_R} \, .
\end{align}
These two relations are valid up to $\mathcal O(v/v_R)$ corrections, and only when 
considering that parity is broken at low scales such that the $SU(2)_L$ and $SU(2)_R$ 
gauge couplings are equal $g_L = g_R \equiv g$.
The resulting mixing between the heavy and light charged gauge bosons, driven by 
$\tan \beta$, then plays only a marginal role, and will thus be omitted in our analysis.
For the full expressions of the gauge boson masses and mixings we refer 
to~\cite{Kriewald:2024cgr}.

After the breaking of the LRSM gauge symmetry, the scalar potential leads to mass 
matrices for all scalar fields depending on the vevs given in Eq.~\eqref{eq:vevs}. 
Their diagonalisation was explicitly solved in~\cite{Kriewald:2024cgr}, which showed that 
all scalar couplings could be expressed in terms of the scalar physical masses and mixings. 
The physical states then include, in addition to the SM-like Higgs boson $h$, several 
new neutral, singly-charged and doubly-charged scalar eigenstates.
As the $\Delta_L$ triplet fields are naturally decoupled in the vanishing $v_L$ limit, 
they will not be further discussed. 
We then focus on the four remaining neutral scalar fields, and in particular on a 
potentially light `Majorana Higgs' field $\Delta\equiv \Delta_R^0$. 
Among these four neutral scalar fields, the neutral bi-doublet components are required 
by flavour-changing neutral current constraints to be as heavy as 
$\sim 20\,$TeV~\cite{Bertolini:2019out}, and to have tiny mixings to the $\Delta$ and 
$h$ states~\cite{Nemevsek:2016enw}.
Our analysis is consequently restricted to the ($h$, $\Delta$) subsystem that is 
characterised by the masses $m_h$, $m_\Delta$ and the mixing angle $\theta$ among the 
$h$ and $\Delta$ states. 
Whereas this mixing is constrained by Higgs invisible decay and exotic 
searches~\cite{Dawson:2021jcl}, it can still be as large as approximately 20\% in the 
range of $m_\Delta$ values considered in this work.
This thus leaves open an interesting window for collider phenomenology.

Neglecting the mixings to the heavier scalar fields\footnote{The full expressions are (very)
lengthy. 
While they can be obtained from the \textsc{FeynRules} model developed 
in~\cite{Kriewald:2024cgr}, they do not offer any evident insights.}, the couplings of 
the $\Delta$ boson to a $WW$, $ZZ$ and SM $f\bar f$ pair are approximately given by
\begin{align} \label{eq:cvv}
  C_{\Delta WW} &\simeq \sin\theta \:g M_W \, ,
  &
  C_{\Delta ZZ} &\simeq \sin\theta \:\frac{g}{\cos\theta_w} M_Z \, ,
  &
  C_{\Delta f\bar f}^L &= C_{\Delta f\bar f}^R\simeq \sin \theta\: Y_{f\bar f} \, ,
\end{align}
where $\theta_w$ stands for the electroweak mixing angle, $M_W$ and $M_Z$ represent the 
masses of the $W$ and $Z$ boson respectively, and $Y_{f\bar f}$ is the Yukawa coupling of 
a pair of SM fermion-antifermion $f\bar f$. 
The $\Delta$-boson Yukawa couplings are thus proportional to the corresponding SM Higgs
couplings via the sine of the mixing angle $\theta$.
On the other hand, the triple scalar vertices are much more involved.
Again expanding the vertices in the limit of small mixing to the heavier Higgs states 
and small $\theta$ values, they can be written as~\cite{Nemevsek:2016enw}
\begin{equation} \label{eq:scalarcoups}
\begin{alignedat}{2}
  C_{hhh} &\simeq \frac{3 g (2 - 3 \theta^2)m_h^2}{4 M_W} \, ,
    \qquad & \
    C_{hh\Delta} &\simeq \frac{g \theta (\varepsilon\theta -1)(2 m_h^2 + m_\Delta^2)}{2 M_W}\,,
    \\
    C_{h\Delta \Delta} &\simeq \frac{g \theta(\theta + \varepsilon)(m_h^2 + 2 m_\Delta^2)}{2 M_W}\,,
    \qquad &\
    C_{\Delta\Delta\Delta} &\simeq \frac{3 g \varepsilon (2 - 3\theta^2)m_\Delta^2}{4 M_W} \, ,
\end{alignedat}
\end{equation}
in which $\varepsilon = v/v_R$. 
In particular, these expressions do not depend on $\tan \beta$, which appears only at 
higher orders. 

Passing on to fermions, the bi-doublet $\phi$ has Dirac Yukawa coupling to quarks and 
leptons, while the triplets $\Delta_{L,R}$ have Majorana Yukawa couplings to the 
left-handed and right-handed leptonic doublets, respectively,
\begin{align}
  \label{eq:LDirac}
  \mathcal L_Y^{\textrm{Dirac}} &= \bar Q_L \left( 
  Y_q \, \phi + \tilde Y_q \, \tilde \phi \right) Q_R +
  \bar L_L \left( Y_\ell \, \phi + \tilde Y_\ell \, \tilde \phi \right) L_R +
  \text{H.c.} \, ,
  \\[1ex]
  \label{eq:LMaj}
  \begin{split}
  \mathcal L_Y^{\text{Maj}} & = 
  \bar L_L^{c} i \sigma_2 \Delta_L Y_L L_L +
  \bar L_R^{c} i \sigma_2 \Delta_R Y_R L_R + \text{H.c.}
  \end{split}
\end{align}
Here, $Y_{q,\ell}$, $\tilde Y_{q,\ell}$ are the Dirac Yukawa matrices and $Y_{L,R}$ are
the Majorana ones. 
Spontaneous symmetry breaking induces masses for all the fermions upon insertion of 
the vevs given in Eq.~\eqref{eq:vevs}.
In the quark sector, the Dirac couplings lead to conventional quark masses and 
left-handed CKM mixing matrix, together with its right-handed analogue entering the 
right-handed charged current. 
As mentioned above, the ($\P$) $\C$ symmetry constrains the Yukawa matrices, with the
result that the quark mixing matrices are (almost) identical, up to possible new CP 
violating phases in the case of a $\C$ symmetry. 
For our study, we thus safely consider left-handed and right-handed quark mixings 
to be equal.
In the lepton sector, the vevs~\eqref{eq:vevs} induce standard Dirac masses for the 
charged leptons, and generate a type I+II seesaw mechanism for the neutrinos.   
Again, thanks to either a $\P$ or a $\C$ symmetry, the Dirac and Majorana masses are
connected~\cite{Nemevsek:2012iq, Senjanovic:2016vxw, Senjanovic:2018xtu}. 
In~\cite{Kriewald:2024cgr}, an explicit solution to the diagonalisation of the lepton 
sector was found by using the Cayley-Hamilton theorem, while in the case of a $\P$ symmetry
an algorithmic approach was developed in~\cite{Kiers:2022cyc}.
As a result, we adopt the heavy neutrino masses and mixings as physical inputs in addition 
to the standard light neutrino masses and mixings. 
These are directly accessible at collider studies through the charged 
current Lagrangian\footnote{The charged current includes also a flipped chirality current
induced by the Dirac (seesaw) mixing and the $SU(2)_L$ and $SU(2)_R$ gauge-boson mixing. 
The effect is however negligible in the range of $N$ masses considered in the 
present study.}
\begin{equation} \mathcal L_{cc}^\ell \simeq 
  \frac{g_L}{\sqrt{2}} \bar\ell_L \gamma_\mu U_\nu \nu\, W_L^\mu + 
  \frac{g_R}{\sqrt{2}} \bar\ell_R \gamma_\mu U_N N \,W_R^\mu \, ,
\end{equation}
where $U_\nu$ effectively coincides with the standard PMNS mixing matrix relevant for 
the light neutrino eigenstates $\nu$, and $U_N$ represents its analogue for the heavy
neutrinos mass eigenstates $N$. 
Choosing $m_{N_{1,2,3}}$ and $U_N$ as input parameters, we then consider a
benchmark case where the lightest heavy neutrino $N\equiv N_1$ can only decay to 
electrons and positrons, muons and antimuons, or democratically to all of them.  

The Majorana Yukawa coupling of the $\Delta$ states to leptons in Eq.~\eqref{eq:LMaj}
generates not only a mass matrix for the heavy neutrinos $M_N\simeq  Y_R v_R$, but 
also enables the decay modes $\Delta \to N_i N_j$.
Thus, the mass eigenstate $\Delta$ acts effectively as the Higgs field for the heavy 
Majorana neutrinos, hence its name `Majorana Higgs'.  
The connection is made explicit by writing the $\Delta NN$ Feynman rule in terms of 
the physical input parameters,
\begin{equation} \label{eq:CDNN}
  C^R_{\Delta NN} = \frac g{\sqrt{2}} \frac {M_N}{M_{W_R}} \cos\theta \, .
\end{equation}
A similar Feynman rule proportional to $\sin \theta$ couples an $NN$ pair to the SM Higgs
boson $h$, leading not only to the possibility of probing the $N$ mass matrix through 
the exotic $h$ decay~\cite{Maiezza:2015lza}, but allowing also for the production of a 
$\Delta h$ or a $\Delta\Delta$ pair via gluon fusion~\cite{Nemevsek:2016enw}. 
These two mechanisms lie at the heart of this work.

%
%
\section{Pair and associated production of \texorpdfstring{$\Delta$}{Delta} at 
\texorpdfstring{$pp$}{pp} colliders} \label{sec:Production}
The primary production mechanisms for the heavy Higgs boson $\Delta$ at proton-proton 
colliders considered in this work includes its associated production with a SM weak 
vector boson, $pp \to W^\pm \Delta$ and $pp \to Z \Delta$, as well as with the SM 
Higgs boson $h$ via gluon fusion, $gg \to h \Delta$. 
Additionally, we examine the pair production of heavy Higgs states, 
$gg \to \Delta \Delta$, which also proceeds through gluon fusion. 
As will be shown in the rest of this work, these channels provide complementary 
avenues for probing the properties of the heavy Higgs $\Delta$ and its connection with
heavy neutrinos in the LRSM.

\subsection{The \texorpdfstring{$\Delta$}{Delta}-strahlung process, \texorpdfstring{$pp\to V^*\to V\Delta$}{p p to V V Delta}} \label{sec:delta_strahlung}
\begin{figure}
  \centering
  \input{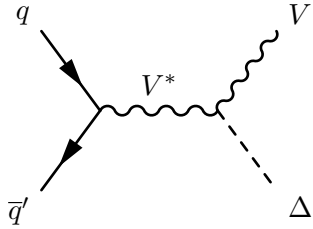}
  \vspace*{-1ex}
  \caption{Parton-level Feynman diagram for the `$\Delta$-strahlung' associated production
  of a heavy $\Delta$ boson with a massive SM weak boson $V$ that can be either a $W$ or 
  a $Z$ boson.} \label{fig:delta-strahlung}
\end{figure}

Just as for the SM Higgs boson $h$, the heavy Higgs $\Delta$ (of mass $m_\Delta$) can 
be produced abundantly at proton-proton colliders via the so-called `$\Delta$-strahlung
process’, $pp\to V^{\ast}\to V\Delta$, where $V = W, Z$ (see the Feynman diagram in 
Figure~\ref{fig:delta-strahlung}).
At leading order, the squared matrix element averaged over initial spins and colours
$\overline{|\mathcal{M}_{V\Delta}|}^2$ for the partonic process 
$q_1 \bar{q}_2\to V\Delta$, is given by
\begin{equation}
  \overline{|\mathcal M_{V\Delta}|}^2 = C_{\Delta VV}\, 
  \left(\left|g_L^{q_1 q_2}\right|^2 + \left|g_R^{q_1 q_2}\right|^2 \right)\, 
  \frac{M_V^2(2 \hat s + \hat t) - m_\Delta^2(m_V^2 - \hat t) - 
  \hat t(\hat s + \hat t)}{4 N_c M_V^2((M_V^2-\hat s)^2 + \Gamma_V^2 M_V^2)}\,,
\end{equation}
where $C_{\Delta VV}$ represents the effective coupling strength of the $\Delta VV$ 
vertex approximately given by Eq.~\eqref{eq:cvv}, while $g_{L,R}^{q_1 q_2}$ denote the 
left-handed and right-handed couplings of the weak boson $V$ to the initial quark pair. 
This expression additionally depends on the standard partonic Mandelstam variables
$\hat s = (p_1 + p_2)^2 = (p_3 + p_4)^2$, $ \hat t = (p_1 - p_3)^2 = (p_2 - p_4)^2$ and
$\hat u = (p_1 - p_4)^2 = (p_2 - p_3)^2$, where $p_{1,2}$ and $p_{3,4}$ denote the 
four-momenta of the initial-state and final-state partons, respectively. 
Furthermore, $M_V$ and $\Gamma_V$ refer to the mass and width of the produced 
weak boson, and $N_c = 3$ represents the number of colours.

Including the phase-space dependence, the parton-level differential cross section
$\d\hat\sigma$ is given by
\begin{equation}
  \frac{\d\hat\sigma(q_1\bar q_2\to V\Delta)}{\d\!\cos\theta} = 
  \frac{1}{32\pi\hat s}\, \lambda \left(1, \frac{m_\Delta^2}{\hat s}, 
  \frac{M_V^2}{\hat s} \right) \ \overline{|\mathcal M_{V\Delta}|}^2\,.
\end{equation}
where $\hat{\lambda}(x,y,z) = x^2 + y^2 + z^2 - 2(xy + yz + zx)$ is the usual
Källén function. 
The cosine of the scattering angle, $\cos \theta \in [-1,1]$, can be expressed 
in terms of the Mandelstam variables as
\begin{equation}
  \cos \theta = \frac{\hat t - \hat u}{\hat s \hat \lambda
  \left(1, m_{\Delta}^2/\hat s, M_{V}^2/\hat s\right)} \, .
\end{equation}
The total unpolarised production cross section at proton-proton colliders is then
obtained by convoluting the partonic cross section with the universal parton 
distribution functions (PDFs) $f_{q_1}$ and $f_{\bar q_2}$ and summing over all
possible initial quark pairs, in accordance with the QCD factorisation theorem. 
This yields
\begin{equation}
  \frac{\d^3\sigma(p p \to V \Delta)}{\d x_1 \d x_2 \,\d\!\cos\theta} = 
  \sum_{q_1 q_2 } \left(f_{q_1}(x_1, \mu_F)\, f_{\bar{q}_2}(x_2, \mu_F) + 
  (x_1\leftrightarrow x_2) \right) \, 
  \frac{\d\hat\sigma(q_1\bar q_2\to V\Delta)}{\d\!\cos\theta}\,,
\end{equation}
with
\begin{align}
  x_1 &\in \left[\frac{(m_{\Delta} + M_{V})^2}{s}, 1    \right] \, , & 
  x_2 &\in \left[\frac{(m_{\Delta} + M_{V})^2}{sx_1}, 1 \right] \, ,
\end{align}
where the hadronic centre-of-mass energy $s$ is related to the partonic one 
$\hat{s}$ through the Bjorken variables $x_1$ and $x_2$ as $\hat{s} = x_1 x_2 s$. 
In the following, the factorisation scale is set to $\mu_F = \sqrt{\hat{s}}$.

\subsection{Heavy Higgs production via gluon fusion}\label{sec:gg_fusion}
\begin{figure}
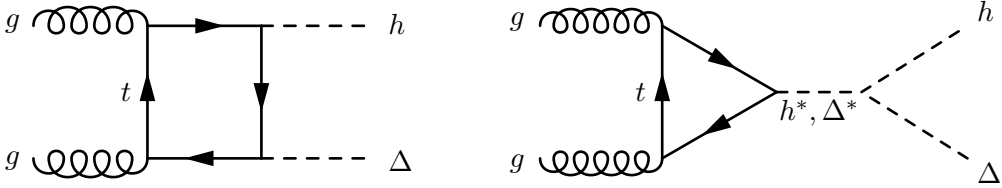

  \centering
  \input{figures/fig_gg_hD_box}
  \quad 
  \input{figures/fig_gg_hD_triangle}
  \caption{Representative box (left) and triangle (right) Feynman diagrams 
  contributing to the $gg\to h\Delta$ process via top-quark exchanges.}
  \label{fig:gg_hD}
\end{figure}

In addition to the $\Delta$-strahlung processes described in 
Section~\ref{sec:delta_strahlung}, the heavy $\Delta$ boson can also be produced in
pairs or in association with an SM Higgs boson via gluon fusion. 
In this section, we compute the corresponding loop-induced cross sections at leading
order in QCD, accounting for both triangle and box topologies. 
For the presented analytical results, we consider the limit where the Higgs bosons $h$
and $\Delta$ have purely scalar couplings to fermions. 
This simplification is motivated by the fact that including potential (small) mixings 
with heavy (pseudo-)scalars $A$ and $H$ states significantly complicates the amplitude
without significantly affecting the total cross section. 
However, our numerical analysis presented later in this paper relies on event 
generation with \textsc{MadGraph5\_aMC@NLO}~\cite{Alwall:2014hca} using the 
UFO~\cite{Degrande:2011ua, Darme:2023jdn} model file developed 
in~\cite{Kriewald:2024cgr}, where no such assumption is made. 
Consequently, any small effects of pseudo-scalar mixings on event shapes will be 
fully incorporated. 

The amplitude for $gg\to S_1 S_2$ (where $S_1 S_2 = h \Delta$ or $\Delta \Delta$) 
primarily receives contributions from triangle and box diagrams. 
Representative Feynman diagrams for $gg\to h\Delta$ are shown in Figure~\ref{fig:gg_hD}, 
the diagrams for $gg\to \Delta\Delta$ being similar. 
The amplitude decomposes as
\begin{equation} \label{eqn:ampfull}
  \mathcal M(gg\to S_1 S_2) = \frac{\alpha_s}{4 \pi} \frac{\delta_{ab}}{2}
  \left( \mathcal M_\triangle^{\mu\nu} + \mathcal M_\square^{\mu\nu}\right)
  \varepsilon_{1}^{\ast\mu}(p_1) \varepsilon_{2}^{\ast\nu}(p_2) \, ,
\end{equation}
where the individual triangle and box contributions can be expressed in terms of a 
few independent Lorentz structures involving the external four-momenta,
\begin{align}
  \mathcal M^{\mu\nu}_\triangle &= \mathcal M_\triangle^{00} g^{\mu\nu} + 
  \mathcal M_\triangle^{21} p_2^\mu p_1^\nu\,, 
  \\[0.7ex]
  \mathcal M^{\mu\nu}_\square &= \mathcal M_\square^{00} g^{\mu\nu} + 
  \mathcal M_\square^{21} p_2^\mu p_1^\nu + \mathcal M_\square^{31} p_3^\mu p_1^\nu 
  + \mathcal M_\square^{23} p_2^\mu p_3^\nu \, .
\end{align}
Here, $\varepsilon_{1,2}$ and $p_{1,2}$ represent the polarisation vectors and 
four-momenta of the initial-state gluons, while $p_3$ is the momentum of the $S_1$ state.
There is no dependence on the four-momentum $p_4$ of the $S_2$ scalar as we used 
energy-momentum conservation to remove its dependence. 
Moreover, the factor $\delta_{ab}/2$ accounts for the trace over the colour indices 
of the quark running into the loops, and $\alpha_s$ denotes the strong coupling constant.
Reducing the tensorial loop integrals to scalar Passarino-Veltman integrals using 
\textsc{Package-X}~\cite{Patel:2015tea, Patel:2016fam}, we obtain
\begin{align}
  \mathcal M_\triangle^{00} &= -4 \sum_{S_i, q} Y_{S_i}^q\, C_{S_i S_1 S_2} \, 
  \frac{2 B_0 - 8 C_{00} + \hat s C_0}{\hat s - m_{S_i}^2 + i \Gamma_{S_i}m_{S_i}} \,,
  \\
  \mathcal M_\triangle^{21} &= 8 \sum_{S_i, q} Y_{S_i}^q\, C_{S_i S_1 S_2} \, 
  \frac{C_0 - 4 C_{12}}{\hat s - m_{S_i}^2 + i \Gamma_{S_i}m_{S_i}} \,.
\end{align}
In these expressions, the Passarino-Veltman functions are abbreviated as $B_0 \equiv B_0(\hat{s}, m_q^2,
m_q^2)$ and $C_{ij} \equiv C_{ij}(0,\hat{s},0,m_q^2, m_q^2, m_q^2)$, following the 
conventions of \textsc{LoopTools}~\cite{Hahn:1998yk}, with $m_q$ standing for the mass 
of the quark running in the loop. 
The UV-divergent pieces in $B_0$ and $C_{00}$ cancel in $\mathcal{M}_\triangle^{00}$, 
ensuring its (UV-)finiteness. 
The trilinear scalar couplings $C_{S_iS_jS_k}$ can be approximated as in 
Eq.~\eqref{eq:scalarcoups} and the quark Yukawa couplings are normalised as in the 
Lagrangian of Eq.~\eqref{eq:LDirac}; the $m_{S_i}$ and $\Gamma_{S_i}$ refer to the mass 
and width of scalar $S_i$. 
Since the box amplitudes are significantly more complex, we provide their 
analytical expressions in Appendix~\ref{app:Box-amplitude} for completeness.

Squaring the amplitude in Eq.~\eqref{eqn:ampfull} and averaging over initial-state 
gluon polarisations yields
\begin{equation}
\begin{split}
  \overline{|\mathcal M|^2} &= \frac{\alpha_s^2(\mu_R)}{32 N_g^2 \pi^2}\, 
  \bigg\{4 |\mathcal M_{\triangle + \square}^{00}|^2 + \hat s \:\mathfrak{R}
  \left[\mathcal M_{\triangle + \square}^{00} 
  \mathcal M_{\triangle + \square}^{21 \ast} \right]
  \\[0.7ex]
  &{}
  + \left.(m_{S_1}^2 - \hat t)\, \mathfrak{R}\left[\mathcal M_{\triangle + \square}^{00}
  \mathcal M_{ \square}^{31 \ast} \right]
  + (m_{S_1}^2 - \hat u) \, 
  \mathfrak{R}\left[\mathcal M_{\triangle + \square}^{00} 
  \mathcal M_{\square}^{23 \ast}\right]\right.
  \\[0.7ex]
  & + \frac{1}{2}(m_{S_1}^2 - \hat t)(m_{S_1}^2 - \hat u) \, 
  \mathfrak{R}\left[\mathcal M_{\square }^{31} 
  \mathcal M_{\square}^{23 \ast} \right] \bigg\}  ,
\end{split}
\end{equation}
where $\mathcal{M}_{\triangle + \square}^{ij} = \mathcal{M}_{\triangle}^{ij} + 
\mathcal{M}_{\square}^{ij}$ and $N_g = 8$ represents the number of gluons. 
The differential partonic cross section follows as
\begin{equation}
  \frac{\d \hat \sigma}{\d\!\cos \theta} = \frac{1}{1 + \delta_{S_1 S_2}}\, 
  \frac{1}{32\pi \hat s}\,\hat \lambda \left(1, \frac{m_{S_1}^2}{\hat s}, 
  \frac{m_{S_2}^2}{\hat s} \right) \, \overline{|\mathcal M|^2}\,,
\end{equation}
where the scattering angle $\cos\theta$ is given by
\begin{equation}
  \cos \theta = \frac{\hat t - \hat u}{\hat s\, \hat\lambda\left(1, m_{S_1}^2/\hat s, 
  m_{S_2}^2/\hat s\right)} \,.
\end{equation}
The total hadronic cross section is finally obtained by convoluting the partonic 
cross section with the gluon PDFs, 
\begin{equation}
  \frac{\d^3\sigma}{\d x_1 \d x_2 \,\d\!\cos\theta} = f_g(x_1, \mu_F) f_g(x_2, \mu_F)\, 
  \frac{\d\hat \sigma}{\d \!\cos \theta}\,,
\end{equation}
with
\begin{align}
 x_1 &\in \left[\frac{(m_{S_1} + m_{S_2})^2}{s} , 1 \right] ,
 & 
 x_2 &\in \left[\frac{(m_{S_1} + m_{S_2})^2}{s \ x_1} , 1\right] .
\end{align}
The phase-space integration is detailed in Appendix~\ref{app:Phasespace} and can be 
non-trivial because of resonant effects, especially 
for $gg \to \Delta\Delta$ production. 
As in the SM process $g g \to h h$, the dominant contribution to the $g g \to h \Delta$
cross section arises from box diagrams, which interfere destructively with the 
triangle contributions.

%
\subsection{Cross sections for \texorpdfstring{$\Delta$}{Delta} production at 
proton-proton colliders}
\begin{figure}
  \centering 
  \includegraphics[width=0.49\linewidth]{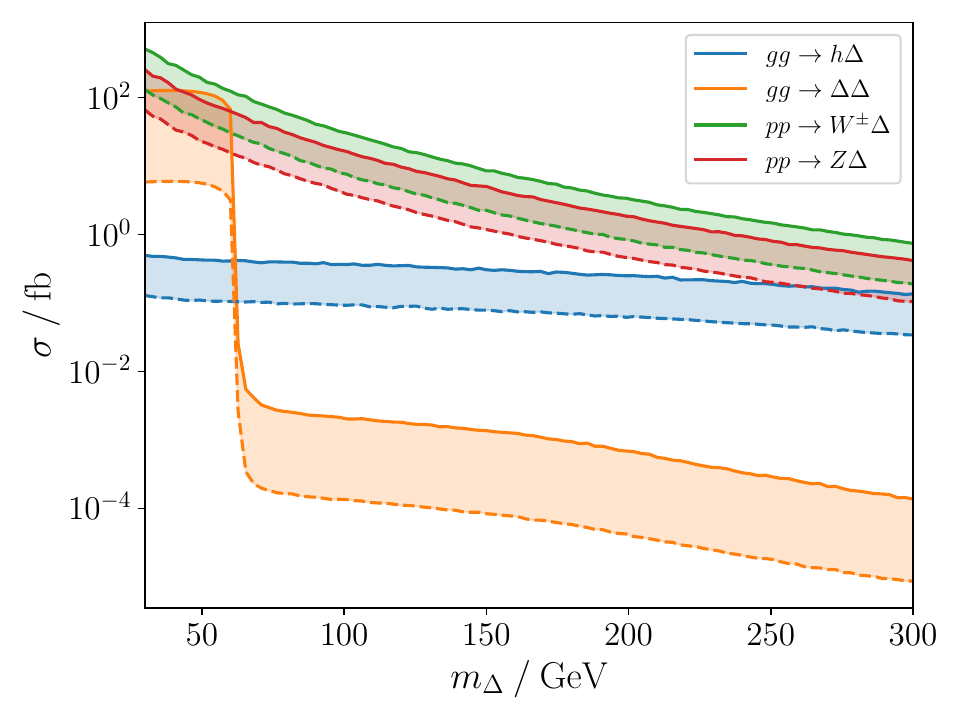}\hfill
  \includegraphics[width=0.49\linewidth]{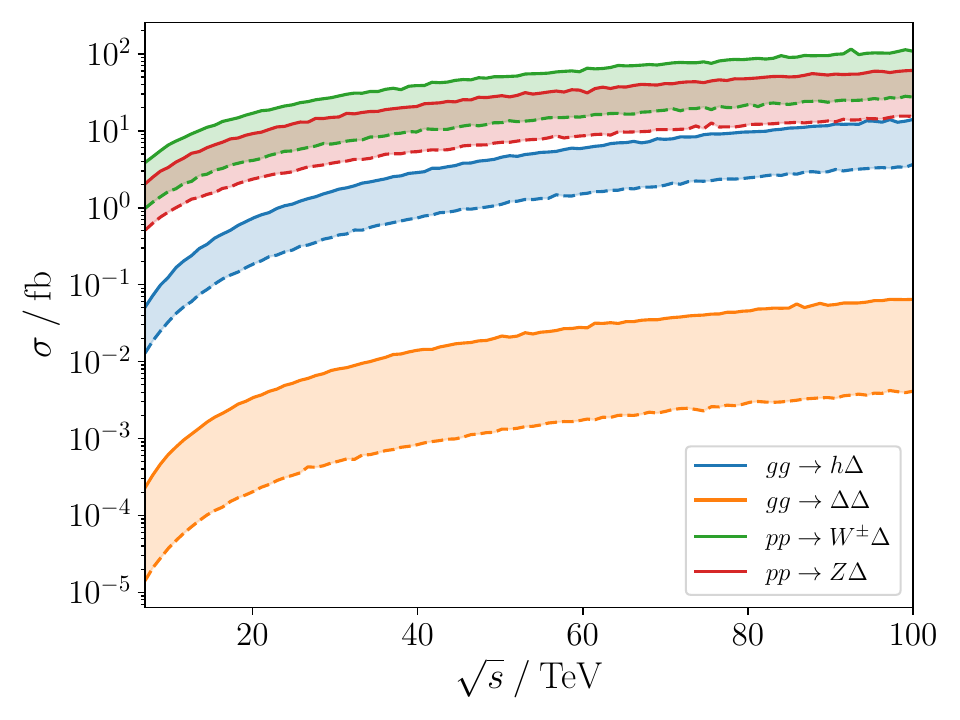}
  \vspace*{-2ex}
  \caption{Production cross sections for the different $\Delta$ production processes 
  studied in this work. 
  In the left panel, we present cross sections as a function of $m_\Delta$ at $\sqrt{s} = 14\:\mathrm{TeV}$ while in the right panel, we show cross sections as a function of $\sqrt{s}$ for a fixed $m_\Delta = 142\:\mathrm{GeV}$. In both panels, solid and dashed lines correspond to $\sin\theta = 0.1$ and $0.05$, respectively.
  }
  \vspace*{2ex}
  \label{fig:prod_xsec}
\end{figure}

In Figure~\ref{fig:prod_xsec}, we present the production cross sections for the 
different $\Delta$ production modes introduced in Sections~\ref{sec:delta_strahlung} 
and~\ref{sec:gg_fusion}. 
The left panel shows the variation of the cross sections with the heavy Higgs 
mass $m_\Delta$ for proton-proton collisions at $\sqrt{s}=14\:\mathrm{TeV}$, using 
the NLO set of NNPDF40 parton densities~\cite{NNPDF:2021njg} (\textit{i.e.}, 
\texttt{NNPDF40\_nlo\_as\_01180}, that corresponds to the identifier 331700 in LHAPDF
6.5.4~\cite{Buckley:2014ana}). 
In this figure, we explore two benchmark values for the scalar mixing angle, 
$\sin\theta = 0.1$ (solid lines) and $\sin\theta = 0.05$ (dashed lines). 
In the right panel, we instead examine the dependence of the rate on $\sqrt{s}$, 
fixing $m_\Delta = 142\:\mathrm{GeV}$ and varying $\sin\theta$ as above, between 
0.05 and 0.1. 

Our results have been derived analytically and cross-validated against \textsc{MG5aMC}, 
with the Passarino-Veltman functions being evaluated using a custom Python interface to
\textsc{LoopTools}~\cite{Hahn:1998yk}.
They indicate that associated production $pp \to V\Delta$ maintains sizeable cross 
sections above 1~fb across the entire considered parameter range. 
For $m_\Delta < m_h/2$, resonant pair production via gluon fusion, 
$gg \to h \to \Delta\Delta$, reaches comparable rates before sharply decreasing beyond 
the threshold. 
On the other hand, the associated Higgs production mode, $gg\to h\Delta$, is somewhat 
subdominant but remains significant enough to lead to a substantial event yield at the 
high-luminosity phase of the LHC.

%
%
\section{Heavy Higgs and neutrino decays and branching ratios} \label{sec:Decays}
In this section, we derive analytical expressions for all relevant decay modes of the 
heavy Higgs boson $\Delta$ and the heavy right-handed neutrino $N$. 
Additionally, we present numerical results to identify viable regions of parameter 
space where the decay $\Delta \to NN$ is either dominant or sufficiently sizeable for 
heavy Higgs production to yield a significant number of signal events at colliders. 
The study made in this section therefore provides insights into the interplay between 
the different decay channels of the considered states, and helps to delineate the 
region of the parameter space where left-right models could be probed to a new signatures.

%
\subsection{\texorpdfstring{$\Delta$}{Delta} decays}
Depending on its mass $m_\Delta$ and the mixing parameters of the scalar sector, the 
heavy Higgs boson $\Delta$ can undergo various two-body and three-body decays into
different final states. 
We begin our analysis by introducing the relevant analytical expressions, before 
examining their numerical behaviour for specific choices of input parameters. 
The partial decay widths related to two-body decay channels of $\Delta$ are given by
\begin{align}
  \Gamma(\Delta\to hh) &= \frac{|C_{\Delta h h}|^2}{32\pi m_\Delta}
  \sqrt{1 - \frac{4 m_h^2}{m_\Delta^2}} \, ,
  \\
  \Gamma(\Delta\to VV) &= \frac{|C_{\Delta V V}|^2}{16\pi m_\Delta ( 1 + \delta_{V})}
  \left(2 + \frac{m_\Delta^2}{4 M_V^2}\right)\sqrt{1 - \frac{4 M_V^2}{m_\Delta^2}} 
   \quad \text{with}\ V = W, \, Z \, ,
  \\
  \Gamma(\Delta\to f\bar f) &= \frac{m_\Delta}{16\pi}\sqrt{1 - \frac{4 m_f^2}{m_\Delta^2}}
  \left[\left(1 - \frac{2m_f^2}{m_\Delta^2}\right)\left(\left|C_{L}\right|^2 + 
  \left|C_R\right|^2\right) - 4\frac{m_f^2}{m_\Delta^2}\mathfrak{R}(C_L C_R^\ast)\right]  ,
\end{align}
where the symmetry factors are $\delta_Z = 1$ and $\delta_W = 0$. 
The coefficients $C_{\Delta hh}$ and $C_{\Delta VV}$ represent the trilinear scalar 
coupling and the scalar-vector interaction strengths, respectively, while the $C_{L,R}$ 
couplings encode the Yukawa interactions of the $\Delta$ state with fermions. 
In particular, for SM fermions, we have $C_L \approx C_R$ in the limit of negligible 
mixing between the scalar state $\Delta$ and the pseudo-scalar state $A$. 
On the contrary, in the case of $\Delta \to NN$, the right-handed coupling $C_R$ is
dominant, with an expression given by Eq.~\eqref{eq:CDNN}.
If $m_\Delta$ lies below one or more of the two-body kinematic thresholds, the $\Delta$
decays via three-body processes mediated by off-shell weak bosons $V$ or Higgs boson $h$,
that can contribute significantly. 
For an off-shell SM Higgs exchange, the differential decay width is given by
\begin{equation}
  \frac{\d\Gamma(\Delta\overset{h^*}{\to} h f\bar f)}{\d \hat s_1} =
  \frac{N_c\sqrt{\hat\lambda(1,\hat s_1, \hat m_h^2)\hat\lambda(1,\hat m_f^2, 
  \hat m_f^2)}|C_{\Delta h h }|^2 m_\Delta^3}{256\pi^3 
  ((\hat s_1 m_\Delta^2 - m_h^2)^2 + (m_h\Gamma_h)^2)} \hat s_1 
  \left(1 - 4 \frac{\hat m_f^2}{\hat s_1}\right)|Y_{ff}|^2\,,
\end{equation}
where $Y_{ff}$ denotes the Yukawa coupling of the SM Higgs to a fermion $f$ of mass $m_f$.
The `hatted' variables appearing in this expression are defined as 
$\hat s_1 = s_1/m_\Delta^2$ and $\hat m_i = m_i/m_\Delta$, while $m_h$ and $\Gamma_h$ 
refer to the SM Higgs boson mass and width. 
The total decay width is then obtained by integrating the differential partial width 
over $\hat{s}_1$ over the range
\begin{equation}
  4 \hat m_f^2 \leq \hat s_1 \leq (1 - \hat m_h)^2 \, .
\end{equation}
For three-body decays via an off-shell vector boson, the expression of the 
corresponding partial decay width reads
\begin{equation}\begin{split}
  & \frac{\d\Gamma(\Delta\overset{V^*}{\to} Vf_i\bar f_j)}{\d\hat s_1} = \frac{N_c\sqrt{\lambda_V\lambda_{ij}}|C_{\Delta VV}|^2 m_\Delta^3}{256\pi^3((\hat s_1 m_\Delta^2 - M_V^2)^2 + (m_V\Gamma_V)^2)} \left(\hat s_1 + \frac{\lambda_V}{12 \hat M_V^2 }\right)
  \\
    &\qquad \times \left[\left(2 - \Sigma_{ij}\left(1 + \frac{3\beta_V^2\lambda_V}{12\hat M_V^2\hat s_1 + \lambda_V}\right) - \Delta_{ij}^2\left(1 - \frac{3\beta_V^2\lambda_V}{12\hat M_V^2\hat s_1 + \lambda_V}\right)\right)\right.\\
    &\qquad  \times \left.\left(\left|g_L\right|^2 + \left|g_R\right|^2\right) + 12 \frac{\hat m_i\hat m_j}{\hat s_1}\left(1 - \frac{\beta_V^2\lambda_V}{12\hat M_V^2\hat s_1 + \lambda_V}\right)\mathfrak{R}(g_L g_R^\ast)\right]\,,
\end{split}\end{equation}
where we defined
\begin{equation}\begin{split}
  \beta_V = 1 - \frac{\hat s_1}{\hat M_V^2} \, , \qquad
  \lambda_V = \hat\lambda(1, \hat s_1, \hat M_V^2) \,, \qquad
  \lambda_{ij} = \hat\lambda(1, \hat m_i^2, \hat m_j^2)\\
  \Sigma_{ij} = \hat m_i^2 + \hat m_j^2\,, \qquad
  \Delta_{ij} = \hat m_i^2 - \hat m_j^2 \,.
\end{split}\end{equation}
Here, $g_{L,R}$ denote the left- and right-handed gauge couplings of the involved 
vector boson to fermions.
The full partial decay width is then obtained after integration over $\hat{s}_1$ 
within the boundaries
\begin{equation}
  (\hat m_i + \hat m_j)^2 \leq \hat s_1 \leq (1 - \hat M_V)^2 \,.
\end{equation}
Finally, we note that the loop-induced $\Delta$ decays to $\gamma\gamma$, $\gamma Z$, 
and $gg$ final states are also possible. 
However, these channels are subdominant~\cite{Nemevsek:2016enw} and will not be 
considered further.

\begin{figure}
    \centering
    \includegraphics[width=0.49\linewidth]{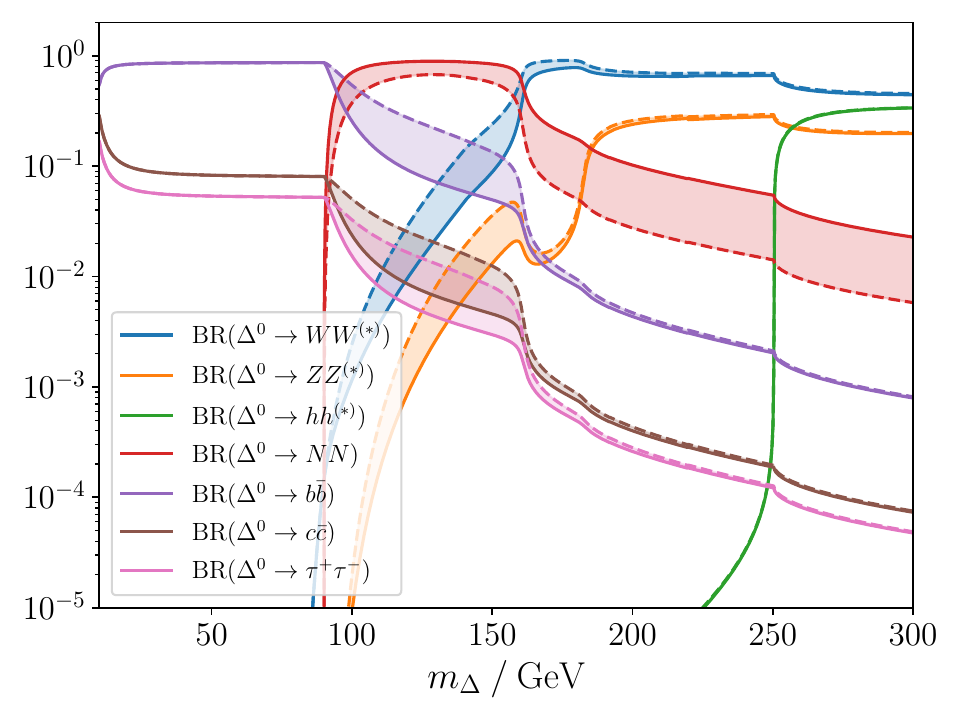}\hfill \includegraphics[width=0.49\linewidth]{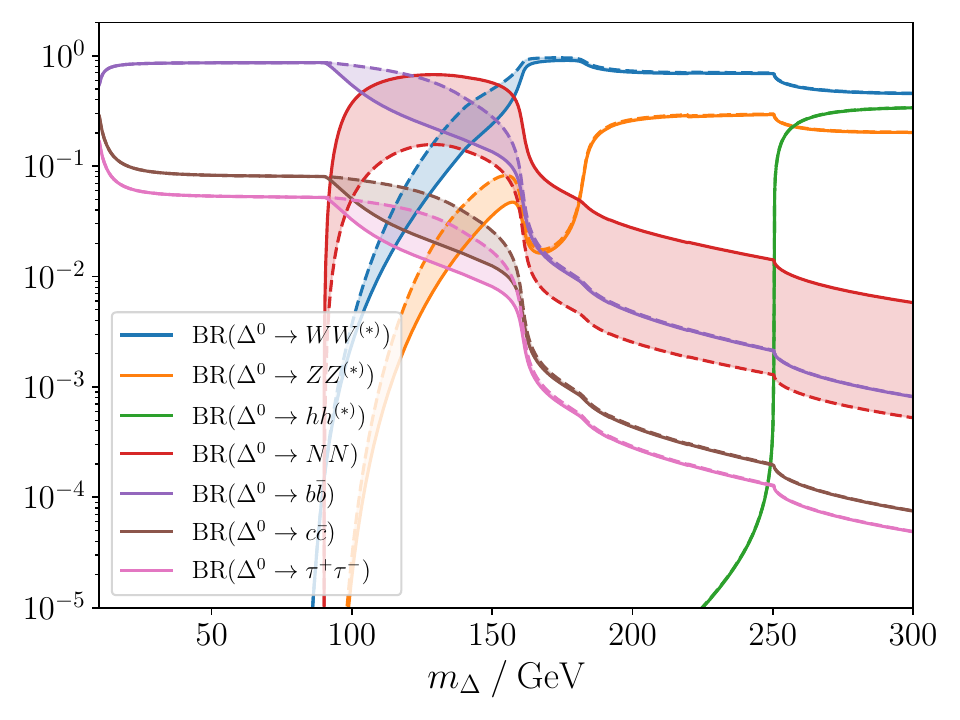}
    \vspace*{-2ex}
    \caption{Branching ratios of $\Delta$ for $m_{N_i} = 45\:\mathrm{GeV}$. In the left panel, we consider a setup where we have fixed $M_{W_R}$ to $6\:\mathrm{TeV}$, with the solid and dashed lines showing predictions for $\sin\theta = 0.05$ and $0.1$ respectively. In the right panel, we fix $\sin\theta = 0.1$ and vary $M_{W_R}$ from $6\:\mathrm{TeV}$ (solid) to $20\:\mathrm{TeV}$ (dashed). \label{fig:spaghetti_N45}}\vspace{.7cm}
    \includegraphics[width=0.49\linewidth]{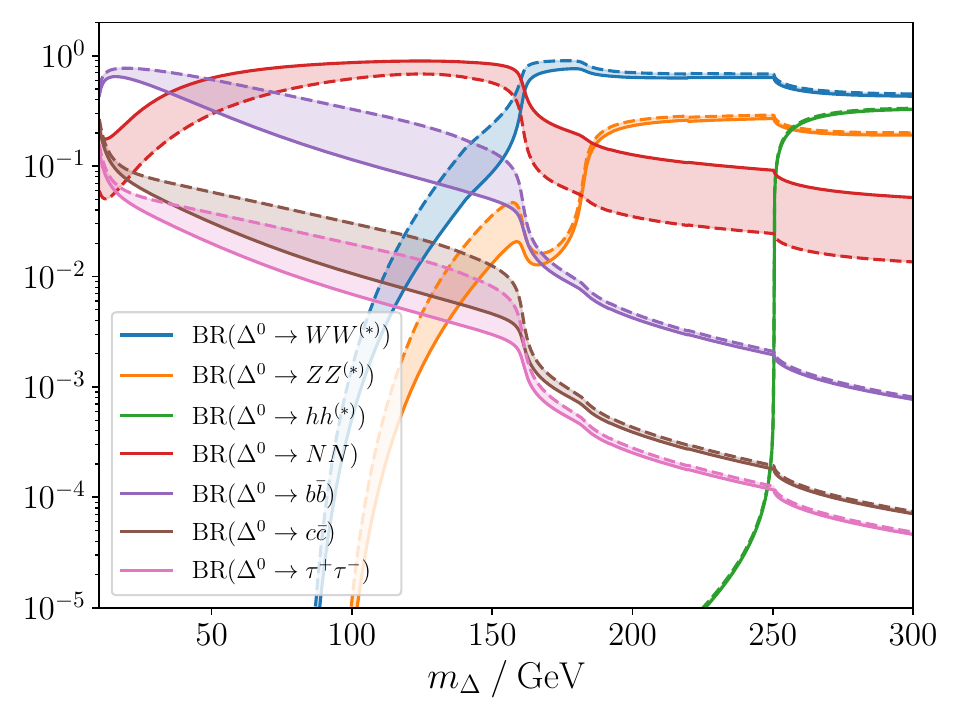}\hfill \includegraphics[width=0.49\linewidth]{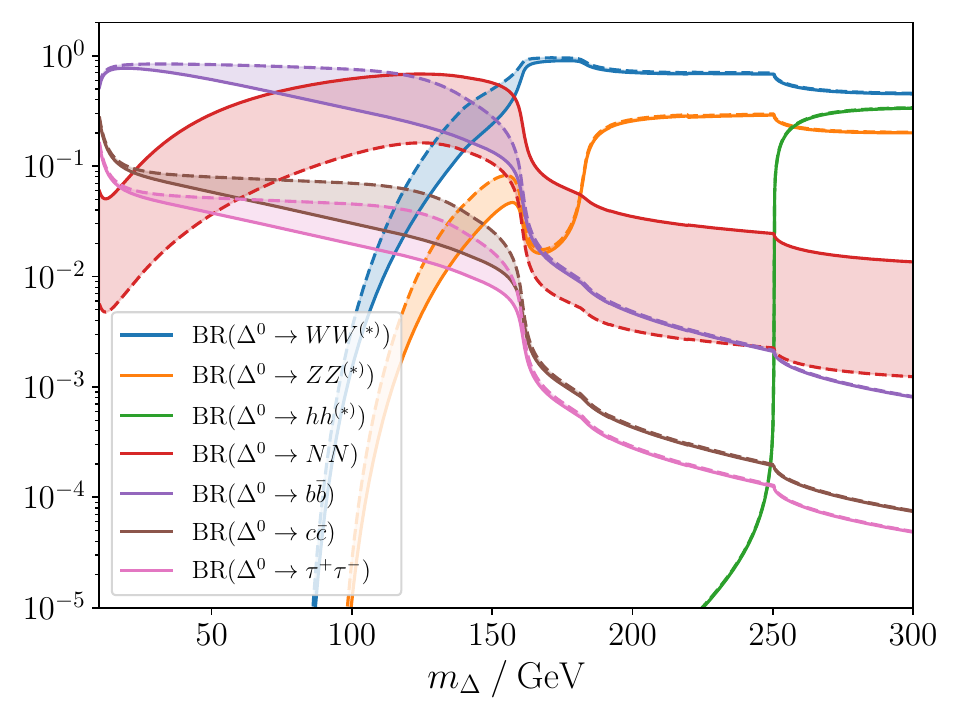}%
    \vspace*{-2ex}
    \caption{Same as Figure~\ref{fig:spaghetti_N45} for $m_{N_i} = m_\Delta/3$. \label{fig:spaghetti_Nv}}
\end{figure}

In Figure~\ref{fig:spaghetti_N45}, we present the dominant branching ratios of the heavy Higgs $\Delta$ as a function of its mass $m_\Delta$. The left panel illustrates the impact of varying the sine of the mixing angle, $\sin\theta$, between $5\%$ (solid lines) and $10\%$ (dashed lines), while keeping the mass of the right-handed $W$ boson fixed at $M_{W_R} = 6\:\mathrm{TeV}$. The thickness of the band embeds variations of $\sin\theta$ between these two values. In contrast, the right panel explores the dependence on $M_{W_R}$ after setting $\sin\theta = 10\%$ and varying $M_{W_R}$ between $6\:\mathrm{TeV}$ (solid) and $20\:\mathrm{TeV}$ (dashed), the variation being again embedded in the band thickness. In both cases, we assume a fixed heavy neutrino mass of $m_{N_i} = 45\:\mathrm{GeV}$. To investigate the impact of the heavy neutrino mass, we display in Figure~\ref{fig:spaghetti_Nv} the same branching ratios, but now setting $m_{N_i} = m_\Delta/3$. 

\begin{figure}
    \centering
    \includegraphics[width=0.49\linewidth]{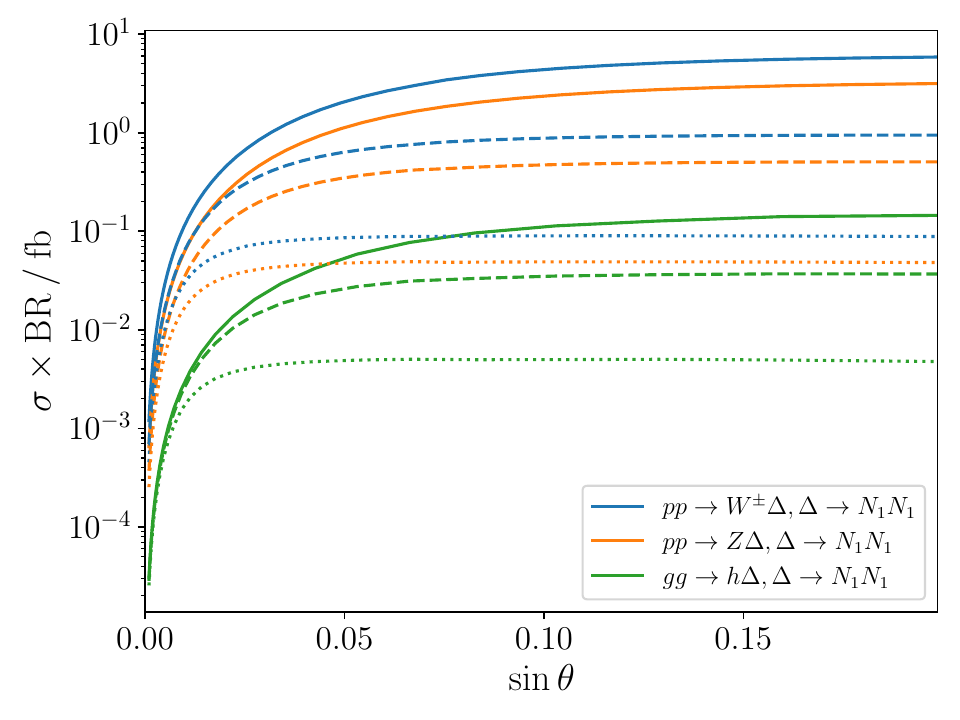}\hfill
    \includegraphics[width=0.49\textwidth]{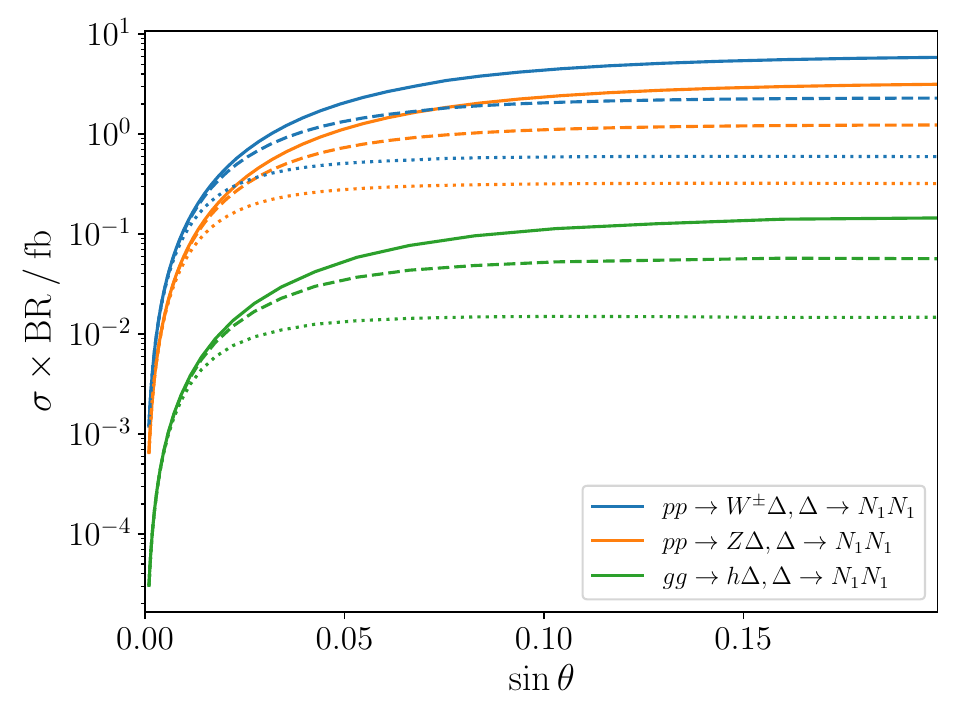}%
    \vspace*{-2ex}    
    \caption{Product of the dominant cross sections for $\Delta$ production with the branching ratio $\mathrm{BR}(\Delta\to N_1N_1)$, shown as a function the sine of the scalar mixing angle $\theta$. We consider $W\Delta$ production (blue), $Z\Delta$ production (orange) and $h\Delta$ production (green), and different mass spectra. In the left panel, we vary $m_\Delta$ and fix it to 135~GeV (solid), 160~GeV (dashed) and 180~GeV (dotted), with $m_{N_1} = m_\Delta/3$. In the right panel, we instead vary $M_{W_R}$ and fix it to 6~TeV (solid), 10~TeV (dashed) and 20~TeV (dotted) with $m_\Delta = 135\:\mathrm{GeV}$ and $m_{N_1} = 45\:\mathrm{GeV}$.}\label{fig:xsec_BR_plots}
\end{figure}

As can be observed, the branching ratio of the decay $\Delta\to NN$ (red curve) dominates over a broad range of $m_\Delta$ values. However, it exhibits a slight decrease with increasing $\sin\theta$, which is attributable to a relative enhancement of the competing two-body decay $\Delta\to b\bar b$ as well as the three-body decays $\Delta\to X f\bar f$ (where $X = W, Z, h$). The branching ratio $\mathrm{BR}(\Delta\to NN)$ further decreases for increasing $M_{W_R}$ values, a consequence of Eq.\eqref{eq:CDNN}. Since the heavy neutrino Yukawa couplings scale as $y_N \sim m_N / M_{W_R}$, a larger $W_R$-boson mass leads to a smaller Yukawa coupling and thus a suppressed partial decay width for $\Delta\to NN$, assuming a fixed $m_N$. This interplay between $M_{W_R}$ and $\mathrm{BR}(\Delta\to NN)$ has important phenomenological implications when considering the production of heavy neutrinos via $\Delta$ decays at colliders. As previously discussed in Section~\ref{sec:Production}, the production cross section of $\Delta$ increases with $\sin\theta$. This means that although $\mathrm{BR}(\Delta\to NN)$ decreases with $\sin\theta$, the overall number of heavy neutrino events produced via intermediate $\Delta$ production can remain significant due to the enhanced production rate. This effect is illustrated in Figure~\ref{fig:xsec_BR_plots}, where we display the product of the production cross section $\sigma(pp\to\Delta)$ (distinguishing the different production channels) with the branching ratio $\mathrm{BR}(\Delta\to NN)$ for different values of $m_\Delta$ and $M_{W_R}$ (see caption for details). Notably, for $\sin\theta \gtrsim 1\%$, the variation in $\sigma\times \mathrm{BR}$ is relatively mild, justifying our choice of $\sin\theta = 10\%$ for the remainder of our analysis.

\subsection{Right-handed neutrino decay and lifetime}\label{sec:Displacement}
In the minimal LRSM, the heavy right-handed neutrinos $N$ typically decay via off-shell $W$- and $W_R$-boson exchanges, leading to three-body final states. However, if their masses are sufficiently large, two-body decays into a charged lepton and an on-shell SM-like $W$ boson can become relevant. The decay width for a massive fermion transitioning into another fermion and a massive vector boson is given by
\begin{equation}
    \Gamma(f_1\!\to\! f_2 V) =\frac{m_1 \lambda_{f_2V}}{16\pi}\bigg[\frac{|g_L|^2 + |g_R|^2}{2} \left(3\left(1 + \frac{m_2^2 - M_V^2}{m_1^2}\right) + \lambda_{f_2V}\frac{m_1^2}{M_V^2}\right)
    - \frac{m_2}{m_1}\mathfrak R (g_Lg_R^{\ast}) \bigg]\,,
\end{equation}
where $\lambda_{f_2V} = \hat\lambda(1, m_2^2/m_1^2, M_V^2/m_1^2)$, with $m_1$ and $m_2$ denoting the masses of the fermions $f_1$ and $f_2$, respectively, and $g_{L,R}$ representing the gauge couplings associated with the boson~$V$.

For the full three-body decay width of a heavy neutrino into a charged lepton and a quark-antiquark pair, both $W_L$ and $W_R$ interactions contribute. The relevant couplings arise either via left-right mixing, which links the SM fields to the heavy neutrino, or through direct Dirac neutrino mixing. In the parameter space of interest where $m_N \lesssim m_W$, the dominant contribution to the three-body decay $N\to \ell^\pm q\bar q'$ comes from $W_R$-boson exchange. The corresponding partial decay width can be approximated as~\cite{Nemevsek:2023hwx}
\begin{equation}
    \Gamma(N_k\to \ell_\alpha^\pm q_i \bar q_j) \simeq 2 \frac{\alpha_w^2 m_{N_k}^5}{128 \pi M_{W_R}^4}\left|V_{R,\, ij}^\text{CKM}\right|^2\left|U_{R,\, \alpha k}^\text{PMNS}\right|^2(1-8x + 8x^2 - x^4 - 12x^2 \log x)\,,
\end{equation}
where $x = m_q^2/m_{N_k}^2$ with $m_q$ being the heavier of the two final-state quarks. Moreover, the matrices $V_R^\text{CKM}$ and $U_R^\text{PMNS}$ represent the right-handed CKM and PMNS matrices, respectively, and $\alpha_w$ is the weak coupling constant. The full expression for the partial width $\Gamma(N_k\to \ell^\pm_\alpha q_i \bar q_j)$, including interference effects between $W_L$ and $W_R$ contributions, as well as the impact of the masses of the charged lepton and quarks, is too lengthy to be displayed here. However, in our numerical analysis, we fully account for these effects, which have been shown to be small for $m_N \gtrsim 10\:\mathrm{GeV}$~\cite{Nemevsek:2023hwx}. Furthermore, we neglect hadronisation effects, which become significant for $m_N \lesssim 10\:\mathrm{GeV}$. Consequently, our collider analysis is limited to scenarios with heavier neutrinos.

%
Although heavy neutrinos can decay through multiple channels, their total decay width can be quite small. As a result, their collider signatures may include displaced vertices and leptons. To compute the number of events where a heavy neutrino decays at a given transverse displacement $d_{xy}$ (which corresponds to the location of the secondary vertex, rather than the transverse impact parameter), we must convolute the production cross sections $\sigma(p p \to h\Delta, V\Delta)$ with the exponential probability distribution governing particle decays,
\begin{equation}
    P_N(d_{xy}) = \frac{1}{\langle d_{xy}\rangle}\exp\left(-\frac{d_{xy}}{\langle d_{xy}\rangle}\right)\,.
\end{equation}
Here, the average transverse displacement is given by
\begin{equation}
    \langle d_{xy}\rangle = \frac{p_T^\text{lab}(N)}{m_N} \tau_N\,,
\end{equation}
where $\tau_N$ is the proper lifetime of the decaying heavy neutrino. Since this depends on the transverse momentum of the heavy neutrino in the laboratory frame $p_T^\text{lab}(N)$, the average transverse displacement must be computed on an event-by-event basis. To determine $p_T^\text{lab}(N)$, we must boost the heavy neutrino's four-momentum from the rest frame of the produced $\Delta$ state into the laboratory frame. This transformation is performed using the boost vector $\vec\beta_\Delta$ and corresponding Lorentz factor $\gamma_\Delta$, 
\begin{equation}
    \vec p_N^\text{lab} = \vec p_N^\Delta + \gamma_\Delta \vec\beta_\Delta\left(\frac{\gamma_\Delta}{1 + \gamma_\Delta}\vec \beta_\Delta \cdot \vec p_N^\Delta - E_N^\Delta\right)\,,
    \label{eqn:boost_pn}
\end{equation}
where, in this expression, the superscript $\Delta$ denotes the rest frame of $\Delta$. Consequently, the quantities $E_N^\Delta$ and $\vec p_N^\Delta$ represent the heavy neutrino energy and momentum in the $\Delta$ rest frame. Moreover, the $\Delta$ boost and Lorentz factor are obtained from its energy and momentum in the laboratory frame $E_\Delta^\text{\,lab}$ and $\vec p_\Delta^\text{\,lab}$, 
as $ \vec\beta_\Delta = \vec p_\Delta^\text{\,lab}/E^\text{\,lab}_\Delta$, $ \gamma_\Delta = (1 - |\vec\beta_\Delta|^2)^{-1/2}$.
In order to determine the fully differential cross section for heavy neutrino $N$ production and decay via intermediate $\Delta$ production, we need to calculate the cross sections for the three-body processes $pp \to h NN$ and $pp \to V NN$. To achieve this, we begin by partitioning the three-body phase space $\Phi_3(\hat{s})$ into two components: a two-body phase-space component $\Phi_2(\hat{s}; m_{h,V}^2, \hat{s}_\Delta)$ relevant for the $2 \to 2$ production process ($pp \to h\Delta$ or $pp\to V\Delta$), and another two-body phase-space component $\Phi_2(\hat{s}_\Delta; m_N^2, m_N^2)$ which corresponds to the $1 \to 2$ decay process $\Delta\to NN$. The full phase space is then factorised as
\begin{equation}
    \Phi_3(\hat s) = \int \d \hat s_\Delta\, \Phi_2(\hat s; m_{V,h}^2, \hat s_\Delta) \, \Phi_2(\hat s_\Delta; m_N^2, m_N^2)\,.
\end{equation}
To simplify this expression further, we rely on the narrow-width approximation that can be enforced on the intermediate $\Delta$ propagator,
\begin{equation}
    \frac{1}{(\hat s_\Delta - m_\Delta^2)^2 + (\Gamma_\Delta m_\Delta)^2} \simeq \frac{\pi}{\Gamma_\Delta m_\Delta}\delta(\hat s_\Delta - m_\Delta^2)\,,
\end{equation}
which leads to a factorised form for the fully differential cross section,
\begin{equation}
    \frac{\d\sigma(pp\to h NN)}{\d x_1 \d x_2 \d\!\cos\theta_\Delta \d\phi_N\, \d\!\cos\theta_N} \simeq \frac{\d \sigma(pp\to h\Delta)}{\d x_1 \d x_2\, \d\!\cos\theta_\Delta} \, \frac{1}{\Gamma_\Delta}\, \frac{\d\Gamma(\Delta\to NN)}{\d\phi_N\, \d\!\cos\theta_N}\,.
\label{eq:factorisedsigma}\end{equation}
In this expression, $\phi_N$ and $\theta_N$ are the azimuthal and polar angles of the heavy neutrino $N$ in the rest frame of the $\Delta$ scalar boson, while $\theta_\Delta$ is the polar angle of the $\Delta$ state in the laboratory frame. Additionally, the dependence on the azimuthal angle $\phi_\Delta$ has been trivially omitted due to the rotational symmetry along the collision axis. For illustration, this expression and the following ones considers a process featuring a final-state SM Higgs boson $h$. Similar expressions can be naturally derived for the other two production processes.

In order to determine the event yields as a function of the heavy neutrino displacement, we note that the differential partial decay width $\d \Gamma(\Delta \to NN)$ is independent of $\phi_N$ and $\theta_N$, so the angular integration of the expression in Eq.~\eqref{eq:factorisedsigma} can be trivially performed. However, to estimate the transverse displacement of the heavy neutrino $N$ in the laboratory frame, we must calculate its transverse momentum $p_T^\text{lab}(N)$, which retains a dependence on $\phi_N$, $\theta_N$ and $\theta_\Delta$. The differential event distribution $\d \mathcal{N}$ with respect to the transverse displacement $d_{xy}$ can then be determined and written as
\begin{equation}
    \frac{\d\mathcal N}{\d(d_{xy})} = \mathcal L\int\!\frac{\d\sigma(pp\to h\Delta)}{\d x_1 \d x_2\, \d\!\cos\theta_\Delta} \, \frac{\mathrm{BR}(\Delta\to NN)}{4\pi} \, \frac{{\rm e}^{-d_{xy}/\langle d_{xy} \rangle}}{\langle d_{xy}\rangle} \, \d x_1\, \d x_2\, \d\!\cos\theta_\Delta\, \d\phi_N\, \d\!\cos\theta_N\,,
\label{eqn:event_dist}\end{equation}
where $\mathcal{L}$ represents the integrated luminosity. To estimate the number of events $\mathcal{N}$ decaying with a certain transverse displacement $[d_{xy}^\text{min}, d_{xy}^\text{max}]$, it is finally sufficient to integrate this result over the displacement range. This gives, up to detector efficiencies,
\begin{equation}
    \mathcal N = \int_{d_{xy}^\text{min}}^{d_{xy}^\text{max}}\frac{\d\mathcal N}{\d(d_{xy})} = \int \d\Phi_3 \frac{\d\sigma}{\d\Phi_3} \left(e^{-\frac{d_{xy}^\text{min}}{\langle d_{xy}\rangle}}-e^{-\frac{d_{xy}^\text{max}}{\langle d_{xy}\rangle}}\right).\label{eqn:dtminmax}
\end{equation}
This expression can be refined further to account for the displaced vertex related to the decay of the second produced heavy neutrino, which leads to an additional exponential distribution. Since $\vec{p}_{N_1}^\Delta = -\vec{p}_{N_2}^\Delta$ (where the notation $N_1$ and $N_2$ stand for the two final-state neutrinos), the transverse momentum $p_T^\text{lab}(N_2)$ can be immediately derived, leading to an equivalent expression for $\langle d_{xy}(N_2)\rangle$.

\begin{figure}
    \centering
    \includegraphics[width=0.49\linewidth]{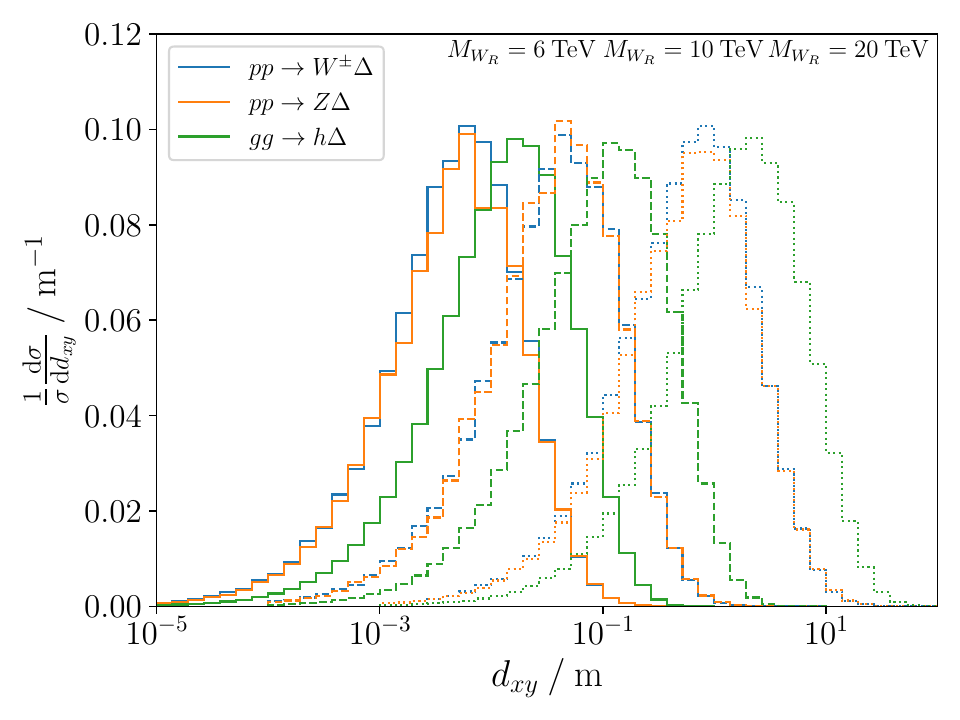} \hfill
    \includegraphics[width=0.49\linewidth]{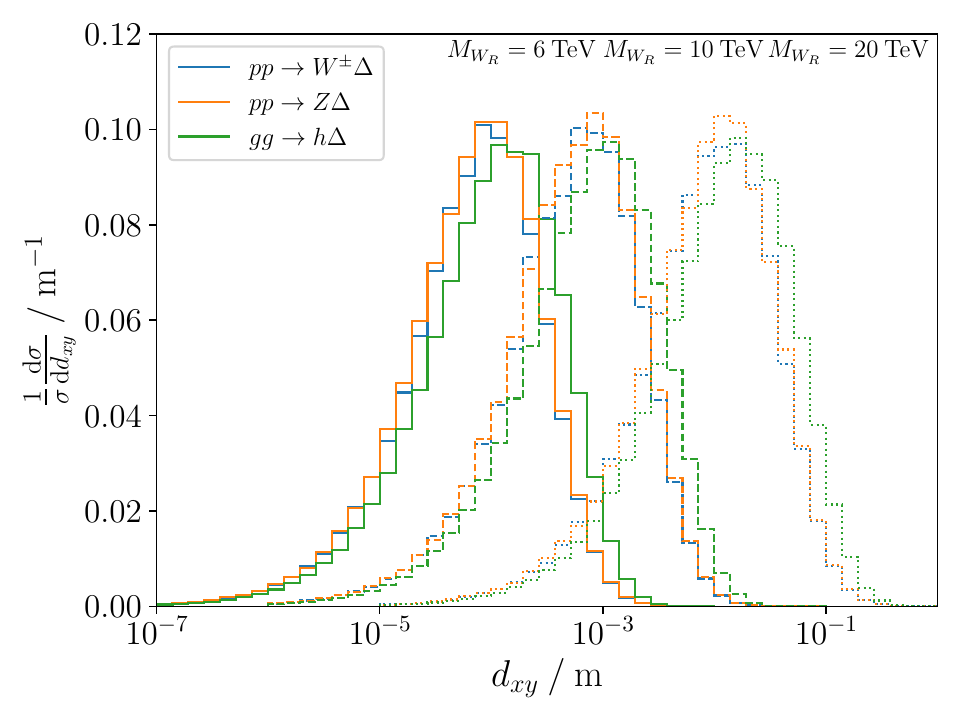}
    \vspace{-2ex} 
    \caption{Normalised distributions of the transverse displacement of heavy neutrinos originating from the decay of a $\Delta$ scalar produced in association with a $W$ boson (blue), $Z$ boson (orange), or Higgs boson $h$ (green). We consider scenarios where $m_\Delta = 60\:\mathrm{GeV}$ and $m_{N_1} = 20\:\mathrm{GeV}$ (left), as well as $m_\Delta = 135\:\mathrm{GeV}$ and $m_{N_1} = 45\:\mathrm{GeV}$ (right). Solid, dashed and dotted lines indicate $M_{W_R} = 6\,,10\,,20\:\mathrm{TeV}$ respectively. \label{fig:dTdist}}
\end{figure}

In Figure~\ref{fig:dTdist}, we show the (semi-)analytically obtained distributions of events with respect to their transverse displacement $d_{xy}$ for various example mass spectra. The presented results rely on the analytical expressions derived in Section~\ref{sec:Production} and this section, and have been further validated against Monte Carlo simulations. For this purpose, hard-scattering events were generated using \textsc{MG5aMC}, with heavy particle decays handled by \textsc{MadSpin}~\cite{Artoisenet:2012st} and \textsc{MadWidth}~\cite{Alwall:2014bza}, while parton showering and hadronisation were simulated with \textsc{Pythia~8}~\cite{Bierlich:2022pfr}. Event analysis was performed using the \textsc{MadAnalysis 5} framework~\cite{Conte:2012fm, Conte:2014zja, Conte:2018vmg, Araz:2021akd}. 

The colour code in Figure~\ref{fig:dTdist} represents displaced $N$ decays from $pp \to W^\pm \Delta$ (blue), $pp \to Z \Delta$ (orange), and $gg \to h \Delta$ (green) production. In the left panel, we fix $m_\Delta = 60 \, \mathrm{GeV}$ and $m_{N_1} = 20 \, \mathrm{GeV}$, while in the right panel, we choose instead $m_\Delta = 135 \, \mathrm{GeV}$ and $m_{N_1} = 45 \, \mathrm{GeV}$. Finally, solid, dashed, and dotted lines correspond to $M_{W_R} = [6, 10, 20] \, \mathrm{TeV}$. The displacement relevant for heavy neutrinos originating from $gg \to h \Delta$ production is slightly larger compared to that stemming from $\Delta$-strahlung processes ($pp \to V \Delta$), due to the higher production threshold and therefore larger boost of the $\Delta$ state. Additionally, the displacement distribution is shifted to larger values with growing $M_{W_R}$ masses. Consequently, for very heavy $W_R$ bosons, a significant displacement ($d_{xy} > 0.1 \, \mathrm{mm}$) can be expected for many events. For light neutrinos $N$ and very heavy $W_R$ boson, the expected displacement can even exceed the typical size of the inner tracker of an LHC detector ($R_\text{Tracker} \lesssim 30 \, \mathrm{cm}$), and the decay could occur in the muon system ($8 \, \mathrm{m} \lesssim R_\text{MS} \lesssim 13 \, \mathrm{m}$).

%
%
\section{Sensitivities at run-3 and HL-LHC} \label{sec:Topos}
Up to the present day, a plethora of experimental searches for a heavy $W_R$ boson (often called a $W'$ boson) have been performed across a variety of channels. Among these, di-jet or $t\bar b$ resonance searches probe the resonant production and decay of the $W_R$ boson into two light jets~\cite{ATLAS:2019fgd, CMS:2019gwf} or a $t\bar b$ pair~\cite{CMS:2023gte, ATLAS:2023ibb}, and place a lower bound on the mass of the $W_R$ boson of $M_{W_R} \gtrsim 4.5\:\mathrm{TeV}$. Furthermore, searches for a single high-$p_T$ lepton accompanied by large missing transverse energy have been conducted by both the ATLAS and CMS collaborations~\cite{CMS:2022krd, CMS:2022ncp, ATLAS:2019lsy}. These provide competitive constraints on the $pp\to W_R\to \ell N$ channel, particularly if the heavy neutrino $N$ is long-lived enough to escape detection.

However, one of the most promising LRSM signals stems from the so-called Keung-Senjanović (KS) mechanism~\cite{Keung:1983uu}, where the process $pp\to \ell^+ N$ is followed by the heavy neutrino decay $N\to \ell^+ j j$. This signature hence features a high-$p_T$ lepton alongside a same-sign secondary lepton and a pair of light jets originating from $N$ decay. In addition, a variant of the KS mechanism has also been recently proposed in the context of third-generation quarks, thereby relying on the process $pp\to N\to \ell^+ N$ followed by the decay $N\to \ell^+ t \overline{b}$~\cite{Frank:2023epx}. Searches for the KS process have been conducted by both the ATLAS and CMS collaborations~\cite{CMS:2021dzb, ATLAS:2023cjo}. They also account for possible displacements due to a displaced heavy neutrino decay, and consider scenarios where several final-state particles merge into a single detector-level object due to the large boost of the heavy neutrino $N$. Depending on the heavy neutrino mass, these searches typically constrain $M_{W_R} \gtrsim 6-7\:\mathrm{TeV}$ for $m_N \gtrsim 50\:\mathrm{GeV}$.

In contrast, the scalar sector of the LRSM has received significantly less experimental attention. In the following subsections, we propose searches for $\Delta$ production and decay, and assess numerically their potential via a dedicated and extensive sensitivity analysis.

\subsection{Displaced Majorana Higgses}\label{sec:majoranahiggses}
The dominant production channels for the $\Delta$ Higgs boson across most of the parameter space involve the so-called $\Delta$-strahlung processes, where it is produced in association with an SM vector boson, \textit{i.e.}\ $pp\to V\Delta$ with $V=W^\pm, Z$ (see Figure~\ref{fig:prod_xsec}). The subsequent decay of the $\Delta$ boson via the $\Delta\to N N$ channel and the heavy neutrino decay $N\to \ell^\pm j j$ result in a final state comprising multiple leptons and jets. Moreover, these decays often exhibit sizeable displacements, as described in Section~\ref{sec:Displacement}, thus occurring within the inner tracker or even in the muon system of a typical LHC detector. We recall that this is particularly true if the $W_R$ boson is significantly heavy and the heavy neutrino $N$ remains relatively light, \textit{ie.}\ the mass configuration to which our study is dedicated.

To analyse this signal, we perform detailed simulations using the \textsc{Feynrules}/UFO model developed in~\cite{Kriewald:2024cgr}. Hard-scattering event generation, including the decays of unstable heavy particles, is carried out using \textsc{MG5aMC}, \textsc{MadSpin}, and \textsc{MadWidth}. The generated events are then processed with \textsc{Pythia 8} for parton showering and hadronisation, followed by a fast detector simulation using \textsc{Delphes}~\cite{deFavereau:2013fsa} that relies on the anti-$k_T$ algorithm~\cite{Cacciari:2008gp} as implemented in \textsc{FastJet}~\cite{Cacciari:2011ma} for event reconstruction. Event generation is achieved by assuming the following input model parameters,
\begin{align}
    \tan\beta &= 0.1\,, & \sin\theta &= 0.1 \,, & \eta &= \phi=0 \, ,
    & m_{N_1} = m_\Delta/3\,,
\end{align}
and we assume that all other LRSM heavy states to be sufficiently decoupled so that their contributions can be safely neglected. Additionally, we set the right-handed analogue of the PMNS matrix to be diagonal, and we focus on final states only containing electrons and muons.

We preselect events containing at least two isolated lepton tracks, each satisfying the transverse momentum requirement $p_T(\ell)>10\:\mathrm{GeV}$ and the pseudo-rapidity constraint $|\eta(\ell)|<2.4$. Furthermore, each lepton must be separated, in the transverse plane, from the nearest jet $j_c$ by $\Delta R(\ell,j_c)>0.4$. To improve the reconstruction of soft leptons which are characteristic of the mass scales probed, we modify several \textsc{Delphes} parameters. Specifically, we adopt the standard ATLAS detector card while setting $\Delta R_\text{max} = 0.3$ (corresponding to the lepton cone size) and $(p_T)_\text{ratio}^{max} = 0.12$ for lepton isolation, following the loose lepton criteria outlined in~\cite{ATLAS:2023dxj}. We validate these modifications by comparing with a soft-lepton implementation within the SFS framework~\cite{Araz:2020lnp, Araz:2021akd, Araz:2023axv} as integrated in \textsc{MadAnalysis~5}. The analysis is then divided into two distinct signal regions, each targeting specific decay displacements. We first consider a `Tracker Region’ where we require the presence of two displaced vertices each associated with a lepton track, with their transverse displacement satisfying $0.1\:\mathrm{mm} < d_{xy}(\ell_1,\ell_2) < 30\:\mathrm{cm}$. This ensures that heavy neutrino decays occur within the inner tracker volume of a typical LHC detector (see for instance~\cite{ATLAS:2023nze}). Secondly, we define a `Muon System Region’  corresponding to the selection of events where the heavy neutrino decays occur within the muon system, imposing a transverse displacement criterion of $8\:\mathrm{m} < d_{xy} < 13\:\mathrm{m}$.

The goal of this analysis is to pioneer an exploration of the LHC sensitivity to the signal proposed. Therefore, we only consider the simple cuts described above. We nevertheless emphasise that any more precise estimate would require a full detector simulation including in particular a more accurate description of muon system clustering, a task that lies well beyond our scope.

To control backgrounds, we consider hadronic decays of the hard-scattering $W$ or $Z$ bosons, and require two same-sign leptons. We therefore simulate SM backgrounds from $t\bar t + X$ production with $X = h, Z, W$, as well as for the multiboson processes $pp\to VV$ and $pp\to VVV$ with $V = W^\pm, Z$. After applying the selection criteria, we impose an additional requirement on the invariant mass of the reconstructed vertices, $m_{\mathrm{vert}}>10\:\mathrm{GeV}$, which eliminates practically all backgrounds. The remaining background events then primarily originate from cosmic rays, which can be effectively vetoed following methods such as those described in~\cite{ATLAS:2019fwx}. As a result, we consider our analysis to be essentially background-free and define the signal significance as
\begin{equation}
    Z = \sqrt{S}\,.
\end{equation}
The signal hypothesis is then rejected at 95\% confidence level if the number of signal events $S \gtrsim 3$.

\begin{figure}
    \centering
    \includegraphics[width=0.5\linewidth]{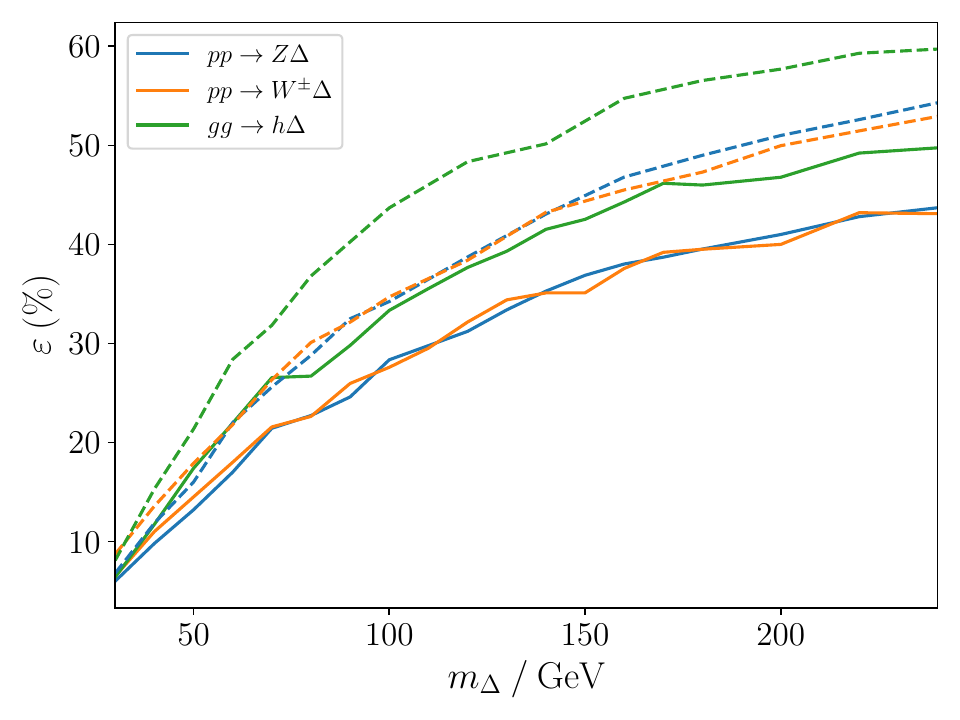}%
    \vspace*{-2ex}
    \caption{Signal efficiencies for $\Delta$ production (and decay into displaced leptons and jets) in association with a $W$ boson (blue), $Z$ boson (orange), or Higgs boson $h$ (green). Solid and dashed lines correspond to di-electron and di-muon final states, respectively.}
    \label{fig:eff}
\end{figure}

The signal selection outlined above leads to efficiencies of approximately 40--50\% across most of the parameter space and for all $\Delta$ production channels. Tables~\ref{tab:eff_6}--\ref{tab:eff_30} in Appendix~\ref{app:efftables} present signal efficiencies obtained with \textsc{MadAnalysis} and our SFS implementation for a variety of LRSM mass spectra, assuming flavour-democratic decays of heavy neutrinos, \textit{i.e.}\ $\mathrm{BR}(N\to e^+ j j) = \mathrm{BR}(N\to e^- j j) = \mathrm{BR}(N\to \mu^+ j j) = \mathrm{BR}(N\to \mu^- j j) = 25\%$. Figure~\ref{fig:eff} displays the corresponding efficiencies obtained with \textsc{Delphes} for events in which two same-flavour leptons are reconstructed, showing their dependence on the $\Delta$ boson mass. Signal efficiencies decrease significantly at lower $m_\Delta$, as fewer soft leptons satisfy isolation requirements. Also, as expected, final states containing muons exhibit slightly higher efficiencies due to the better muon reconstruction performance in the ATLAS detector parametrisation.

\begin{figure}
    \centering \includegraphics[width=0.49\linewidth]{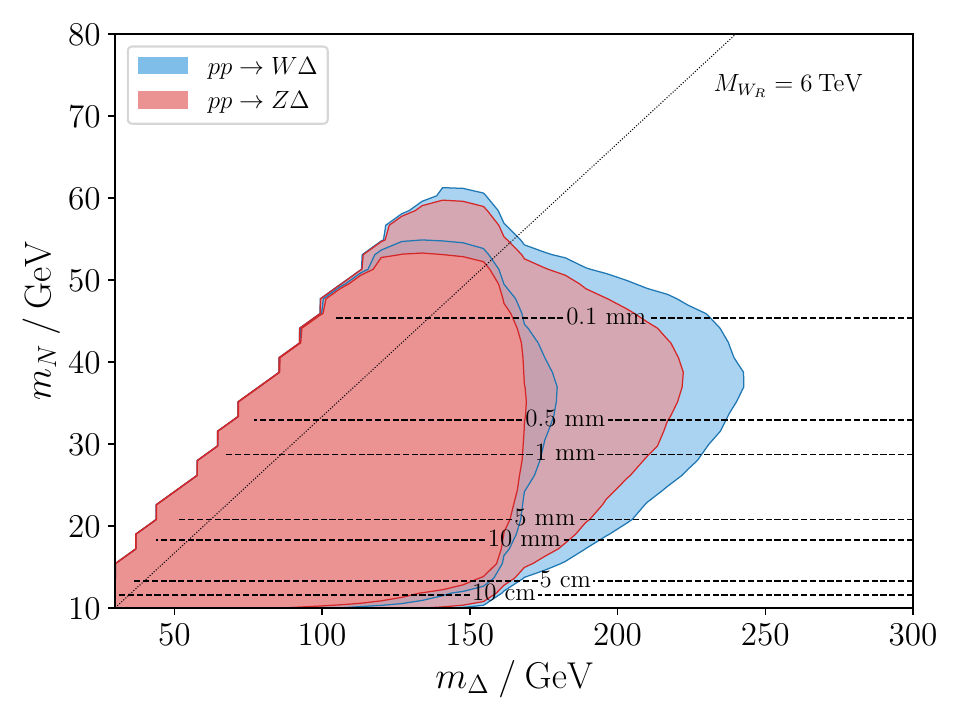}\hfill
    \includegraphics[width=0.49\linewidth]{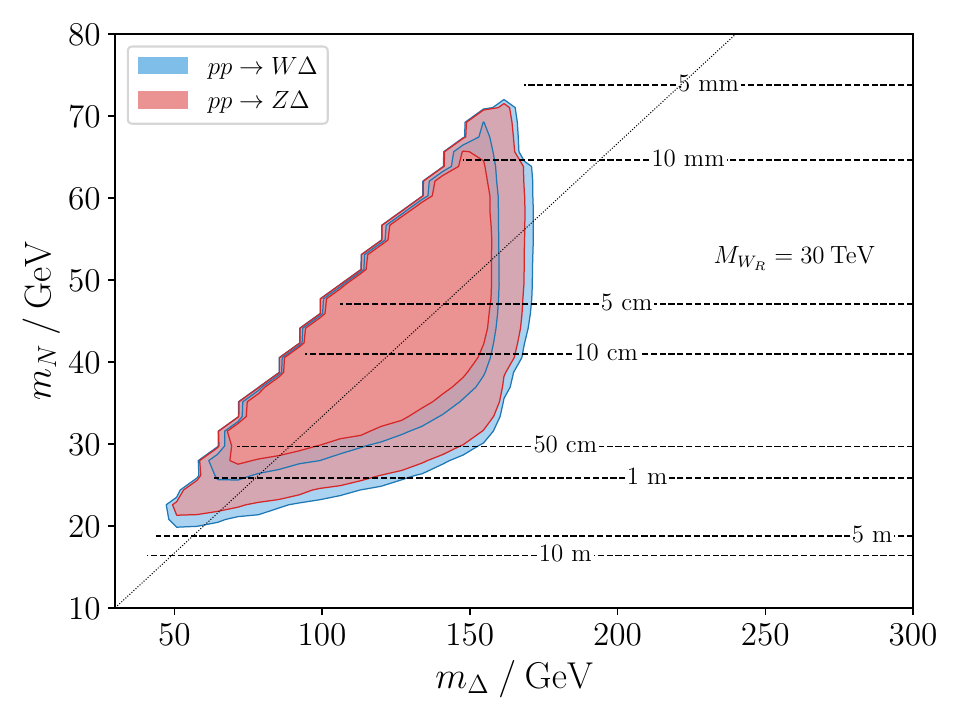}
    \vspace*{-2ex}
    \caption{Sensitivity contours in the $(m_\Delta, m_N)$ plane for the $pp\to W^\pm\Delta$ (blue) and $pp\to Z\Delta$ $\Delta$ (red) production processes, with the inner (outer) contours corresponding to an integrated luminosity of $300\:\mathrm{fb}^{-1}$ ($3000\:\mathrm{fb}^{-1}$). The horizontal dashed lines indicate the proper lifetime of the heavy neutrino $N$, while the diagonal dotted line marks benchmark scenarios with $m_N = m_\Delta/3$. The left and right panels show results for $M_{W_R} = 6$ and $30\:\mathrm{TeV}$, respectively.\label{fig:sens_Delta_N}}
    \end{figure}
 
Figure~\ref{fig:sens_Delta_N} presents the projected sensitivities in the $(m_\Delta, m_N)$ plane for two benchmark values of the $W_R$ boson mass: $M_{W_R} = 6\:\mathrm{TeV}$ (left) and $30\:\mathrm{TeV}$ (right). The dominant production processes, $pp\to W^\pm \Delta$ (blue) and $pp\to Z\Delta$ (red), are considered for integrated luminosities of $300\:\mathrm{fb}^{-1}$ (solid) and $3000\:\mathrm{fb}^{-1}$ (shaded), corresponding to the expected luminosity of the Run~3 and high-luminosity phase of the LHC, respectively. The figures also display the heavy neutrino lifetime isolines. The sensitivity reach depends significantly on $M_{W_R}$. For $M_{W_R} = 6\:\mathrm{TeV}$, the displacement is sizeable but remains within the detector size, allowing broad coverage in $m_N$ and $m_\Delta$ up to approximately $m_N \simeq 70\:\mathrm{GeV}$ and $m_\Delta \simeq 220\:\mathrm{GeV}$. Beyond this mass range, the $\Delta\to NN$ decay becomes subdominant (see also Figures~\ref{fig:spaghetti_N45} and \ref{fig:spaghetti_Nv}) and sensitivity is lost. On the other hand, for a heavier $W_R$ boson, the displacement of  the heavy neutrino increases beyond the tracker volume, significantly reducing the accessible region of the parameter space.  To further characterise the sensitivity, we consider the benchmark relation $m_\Delta = 3 m_N$, which is represented by the diagonal line in the figure. This choice optimises the coverage in the $(m_\Delta, m_N)$ plane and enables us to quantitatively determine the indirect sensitivity to the $W_R$ boson, and compare with conventional searches.
   \begin{figure}    \includegraphics[width=0.51\linewidth]{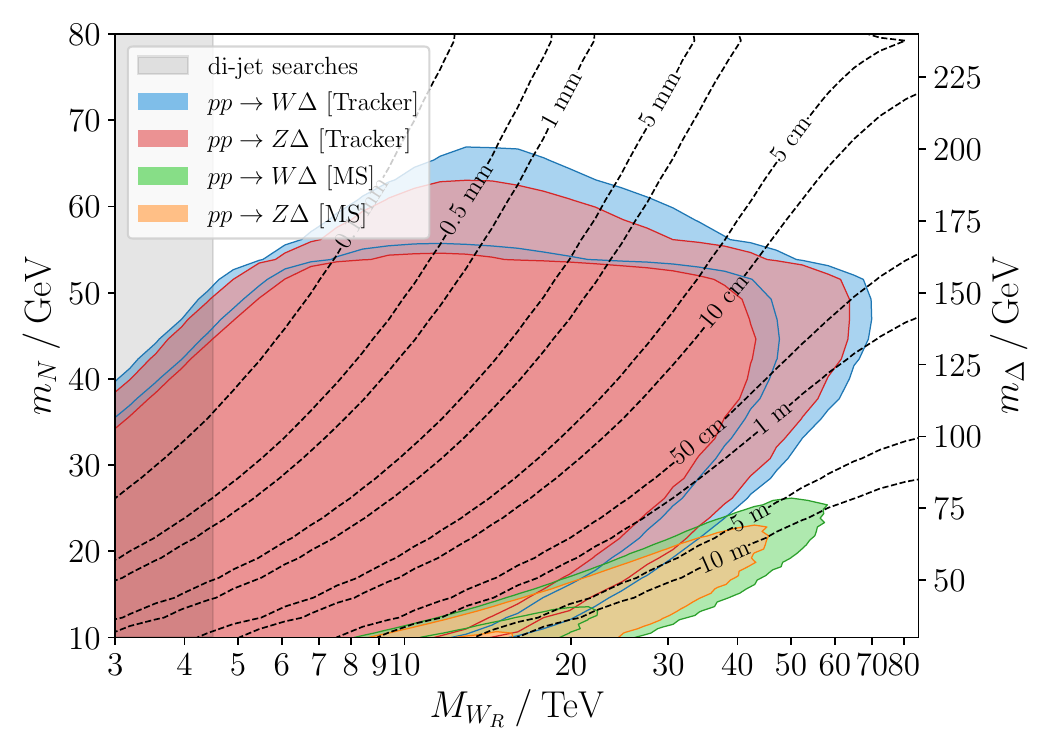}\hfill  
    \includegraphics[width=0.47\linewidth]{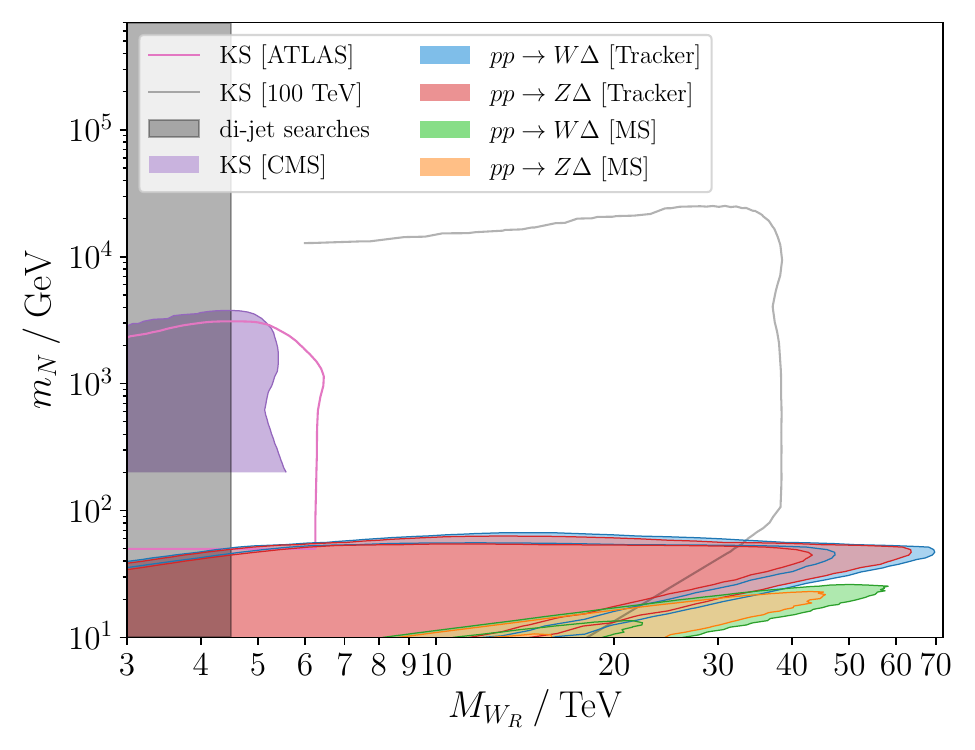}%
    \vspace*{-2ex}
    \caption{Sensitivity contours in the $(m_{W_R}, m_N = m_\Delta/3)$ plane as determined from Tracker Region analysis and Muon System (MS) Region analysis, and for the $pp\to W^\pm\Delta$ (blue and green) and $pp\to Z\Delta$ (red and orange) production processes. The dashed lines mark again the proper lifetime of the heavy neutrino $N$. Our results are compared, in the right panel, with the current reach of di-jet searches searches (grey region), existing ATLAS and CMS searches for the KS process (pink and purple), and the projected sensitivity at a future 100~TeV proton-proton collider. \label{fig:sens_main}}
\end{figure}

The resulting sensitivities, still based on $pp\to W^\pm \Delta$ and $pp\to Z\Delta$, are shown in Figure~\ref{fig:sens_main}. The left panel highlights the independent contributions from the Tracker Region (blue and red) and the Muon System Region (green and orange). Sensitivity extends up to $m_\Delta \simeq 180\:\mathrm{GeV}$, beyond which the $\Delta \to ZZ$ decay becomes dominant, followed by the $\Delta \to hh$ mode (see Figures~\ref{fig:spaghetti_N45} and \ref{fig:spaghetti_Nv}), so that the signal cross section time branching ratio becomes negligible. The indirect sensitivity to the $W_R$ boson mass reaches up to $M_{W_R} \simeq 70-80\:\mathrm{TeV}$, corresponding to an $SU(2)_R$-breaking scale of approximately $v_R\simeq\mathcal{O}(100\:\mathrm{TeV})$ for $N$ masses around $m_N \simeq 40-50\:\mathrm{GeV}$. For lower $m_N$, the displacement increases beyond the tracker volume, and the muon system signal region becomes the dominant detection channel. This enables sensitivity for $m_N \simeq 10-20\:\mathrm{GeV}$. Our results hence significantly surpass the current $W_R$ bounds from direct di-jet searches (grey region), which are primarily constrained to lower $W_R$ masses. This is emphasised in the right panel of Figure~\ref{fig:sens_main}, where we compare our projected reach with existing $W_R$ limits not only from di-jet searches, but also from CMS and ATLAS searches for the KS process (pink and purple). Additionally, we show the projected reach for the KS process at a future 100~TeV circular collider (grey contour), as taken from~\cite{Nemevsek:2023hwx}. This comparison highlights the importance of our proposed search strategy, which provides indirect access to $W_R$ bosons much deeper in the multi-TeV range than any other search by means of soft and displaced objects.

%
%
\subsection{The \texorpdfstring{$b\bar b NN$}{b-bbar NN}  signature}
\begin{figure}
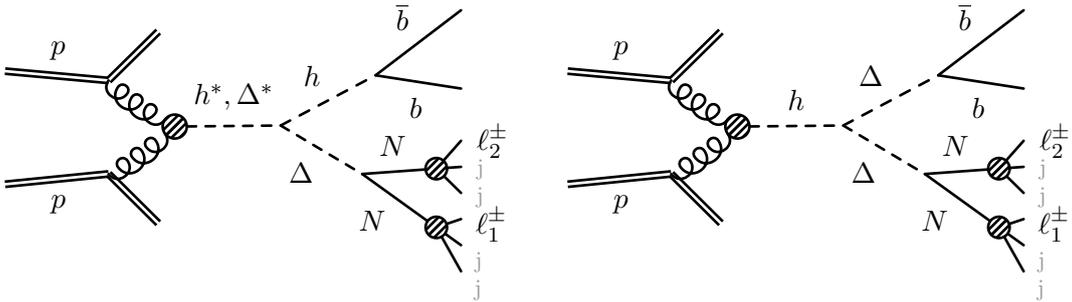

   \noindent 
   \hspace*{0.2em} \input{figures/fig_pp_hD_bbNN}\qquad
  \input{figures/fig_pp_DD_bbNN}\hfill\hfill
  \vspace*{-1ex}
  \caption{Representative Feynman diagrams for the production of beauty and Majorana neutrino pairs in the associated $pp\to h \Delta$ channel (left) and pair production $pp\to \Delta\Delta$ mode via mostly on-shell $h$-boson exchange (right). \label{fig:pp_bbNN}}
\end{figure}

In addition to the primary production channels of $\Delta$ discussed so far, we also considered in Section~\ref{sec:gg_fusion} its pair production via a potentially resonant Higgs boson exchange, $gg\to h^{(*)} \to \Delta\Delta$, as well as its associated production with a Higgs boson, $gg\to h\Delta$. In this section, we focus on these two processes, with again a subsequent decay of the $\Delta$ state into heavy neutrinos, $\Delta\to NN$, followed by the decay $N\to \ell^\pm j j$. To maximise signal rates, we restrict our analysis of the associated production channel to cases where the Higgs boson decays into a pair of bottom quarks, $h\to b\bar{b}$. Similarly, for the pair production mode, we consider scenarios in which one $\Delta$ boson decays into heavy neutrinos, while the other decays into a $b\bar{b}$ pair, such a decay mode being mediated via the mixing of the $\Delta$ scalar with the SM Higgs boson. These production mechanisms, illustrated schematically by the diagrams of Figure~\ref{fig:pp_bbNN}, present an intriguing opportunity to simultaneously probe the spontaneous mass generation of \textit{Dirac particles} (through the final-state $b$-jets) and the spontaneous mass generation of \textit{Majorana states} $N$. Moreover, if both production channels are observed, their relative signal strengths could provide a direct handle on the $h\Delta$ mixing angle.

For our analysis, we apply the same selection criteria as in Section~\ref{sec:majoranahiggses}. Since the Higgs boson is significantly heavier than the weak bosons ($m_h > M_Z, M_W$), the production threshold and, consequently, the boost of the $\Delta$ state are larger than in the $\Delta$-strahlung case previously studied. This subsequently results in greater lepton displacements, as discussed in Section~\ref{sec:Displacement}. The corresponding efficiencies, shown in Figure~\ref{fig:eff} and Table~\ref{tab:eff_dds}, confirm that these modes exhibit slightly higher efficiencies compared to the $\Delta$-strahlung production channels. However, the $\Delta$ pair production via $s$-channel Higgs exchange is only sizeable if the intermediate Higgs boson can be resonantly produced ($m_\Delta < m_h/2$), limiting its coverage of the parameter space. Moreover, we emphasise that in this low-mass regime, the lepton isolation requirement rejects a significant fraction of events, leading to signal efficiencies of only $1\%-5\%$ for $m_\Delta\in [30,60]\:\mathrm{GeV}$.

\begin{figure}
    \centering
    \includegraphics[width=0.49\linewidth]{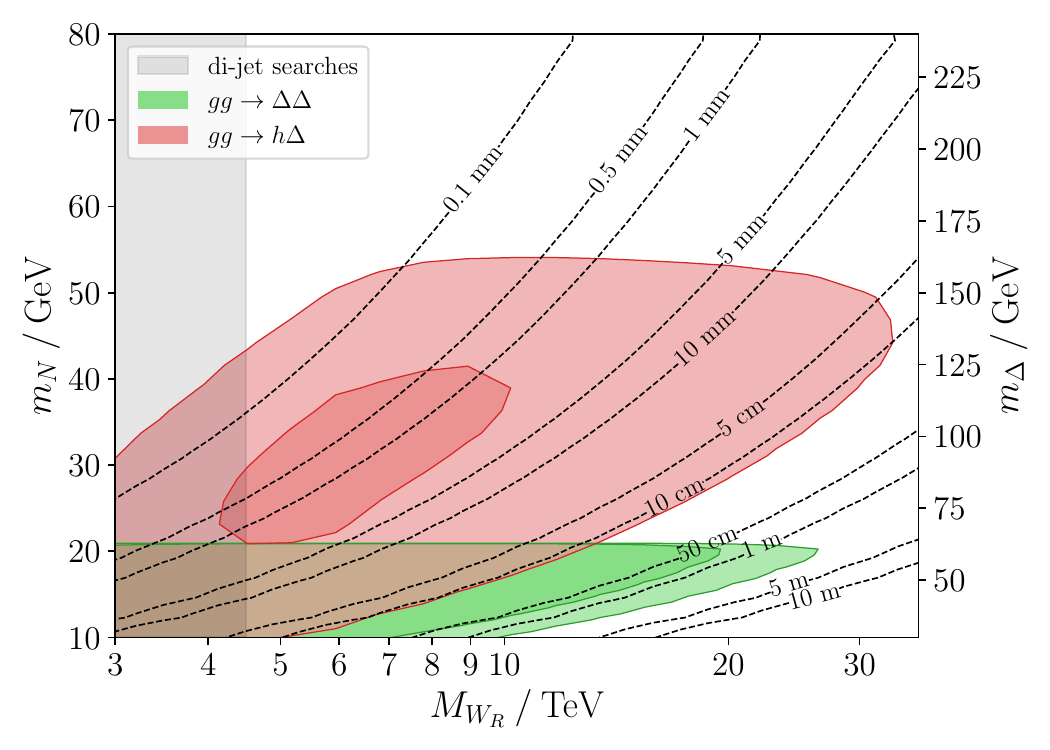}\hfill
    \includegraphics[width=0.49\linewidth]{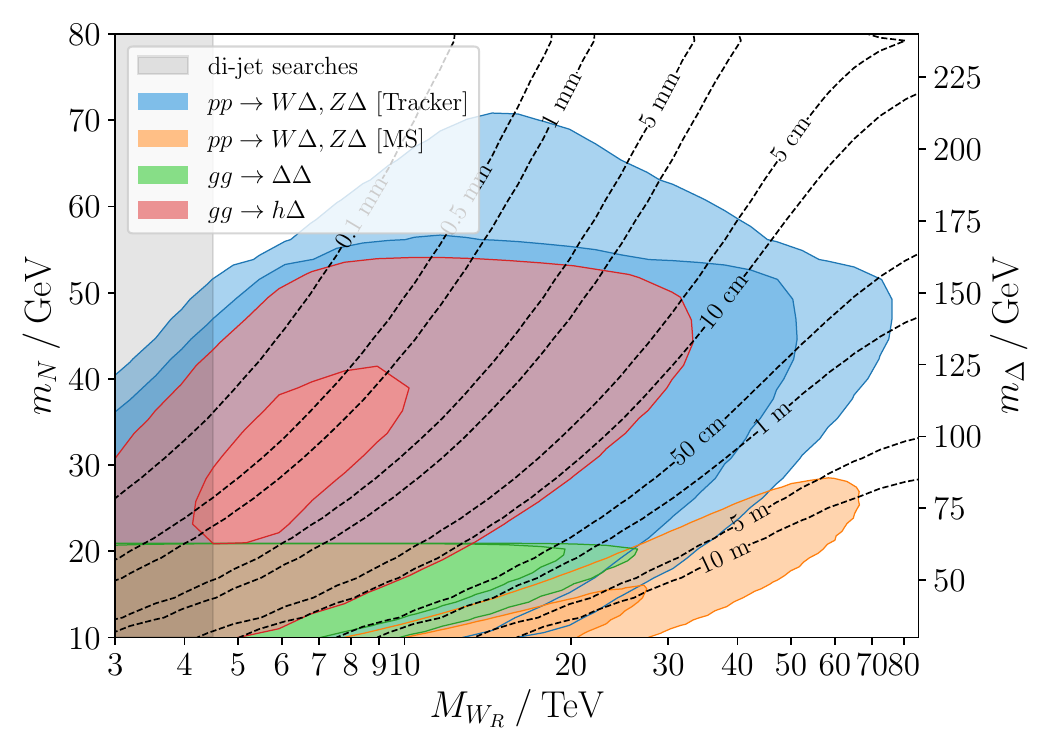}%
    \vspace*{-2ex}
    \caption{Sensitivity contours in the $(m_{W_R}, m_N = m_\Delta/3)$ plane for the $b\bar{b} NN$ signature (left panel), distinguishing the $\Delta h$ associated production mode (red) and the $\Delta\Delta$ pair production mode (green). The inner(outer) contours correspond to an integrated luminosity of $300(3000\:\mathrm{fb}^{-1}$). The dashed lines mark the  lifetime of the heavy neutrino $N$, and the shaded grey region represents the exclusion from di-jet searches. In the right panel, we compare this reach to that originating from the study of $\Delta$-strahlung processes in the Tracker (blue) and Muon System  (orange).\label{fig:sens_bbNN_all}}
    \vspace*{-2ex}
\end{figure}

Due to its comparatively lower production cross section (see Figure~\ref{fig:prod_xsec}), the sensitivity reach for the associated production channel $gg\to h\Delta$ is reduced relative to the two $\Delta$-strahlung channels. In addition, the pair production mode is further suppressed by the two $\Delta$ branching ratios. We indeed recall that $\mathrm{BR}(\Delta\to b\bar{b}) > \mathrm{BR}(\Delta\to NN)$ in most of the relevant parameter space  for $m_\Delta \leq m_h/2$, and that $\mathrm{BR}(\Delta\to NN)$ is small (see Figures~\ref{fig:spaghetti_N45} and \ref{fig:spaghetti_Nv}). Nevertheless, as will be shown, the $b\bar{b} NN$ final state remains a promising signature within the reach of the LHC, offering a complementary discovery channel alongside the $\Delta$-strahlung processes studied in the previous section. This is quantified in Figure~\ref{fig:sens_bbNN_all}, where we present projected sensitivities assuming integrated luminosities of $300\:\mathrm{fb}^{-1}$ (inner contours) and $3000\:\mathrm{fb}^{-1}$ (outer contours). The left panel displays both $b\bar{b} NN$ channels along with the proper lifetime of the heavy neutrino $N$. The reach for associated production mode (red contours) is found to extend up to $M_{W_R}\simeq 30\:\mathrm{TeV}$ and $m_{\Delta}\simeq 160\:\mathrm{GeV}$. In contrast, the sensitivity for the pair production mode (green contours) decreases sharply as $m_\Delta$ approaches $m_h/2$, since the resonance enhancement is lost. However, in the resonant regime, the larger production cross section allows the reach to extend up to $M_{W_R}\simeq 25\:\mathrm{TeV}$. In the right panel of Figure~\ref{fig:sens_bbNN_all}, we directly compare the sensitivity of the $b\bar{b} NN$ channels with the $\Delta$-strahlung modes, combining the $pp\to W\Delta$ and $pp\to Z\Delta$ signals for displaced vertex searches in both the tracker (blue contour) and the muon system (orange contour). This figure highlights the complementarity of the different search strategies explored in this study, demonstrating that a comprehensive experimental programme can substantially extend the reach for minimal LRSM bosons deep in the multi-TeV regime.

\renewcommand{\arraystretch}{1.3} 
\setlength{\tabcolsep}{8pt} 

\section{Conclusion and Outlook} \label{sec:Outlook}
Apart from low-energy probes such as neutrinoless double-beta decay or lepton-number violating meson decays, heavy Majorana neutrinos could, in principle, be produced at the LHC and lead to distinct lepton-number violating signatures with same-sign dileptons. 
The observation of such processes would then provide conclusive evidence for the Majorana nature of neutrinos. 
Furthermore, if the mass of heavy Majorana neutrinos $N$ arises from spontaneous symmetry breaking, then a corresponding `Majorana Higgs' boson $\Delta$ should exist as the source of their mass. 
Building on previous studies, we explored several production mechanisms for this $\Delta$ scalar, focusing on both its pair production via gluon fusion and its associated production with a SM $W$, $Z$, or Higgs boson. 
Through a detailed analytical analysis, we have demonstrated that a significant number of events could be expected in the datasets of the upcoming LHC runs, highlighting promising avenues for discovery.

We frame our study within the minimal left-right symmetric model as an illustrative UV-complete framework to explore the potential signatures arising from $\Delta$ production and decay at colliders. We have considered a $\Delta\to NN$ final state, where the neutrino $N$ is long-lived so that its decay leads to displaced vertex signatures in typical LHC detectors. As in the left-right framework, the $N$ lifetime is primarily governed by the heavy neutrino mass and the $SU(2)_R$ breaking scale, and therefore the mass of the heavy $W_R$ gauge boson, exploiting $N$ reconstruction either in the tracker or in the muon system of a detector grants access to regimes of very large $W_R$ masses. We obtain an indirect sensitivity reaching $M_{W_R} \simeq 70-80\:\mathrm{TeV}$, effectively turning the LHC into a precision probe of left-right symmetry breaking. This reach surpasses the expected sensitivity of direct heavy neutrino searches via the KS mechanism at a future $pp$ collider operating at $\sqrt{s}=100\:\mathrm{TeV}$, which is limited to $M_{W_R} \simeq (20)\, 40\:\mathrm{TeV}$ for $m_N \simeq (10)\, 1000\,\mathrm{GeV}$. The processes that we explored thus probe a complementary portion of the LRSM parameter space, and could serve as a powerful indirect discovery tool potentially guiding future searches for heavy resonances via the KS mechanism. In addition, we have also examined associated $\Delta$ production with a SM Higgs boson decaying into a $b\bar b$ pair. The resulting signature comprising two $b$-jets and two long-lived heavy neutrinos then offers the exciting possibility to simultaneously establish the spontaneous mass origin of {\it Dirac fermions} and {\it Majorana states}, although the sensitivity expressed in terms of the $W_R$ boson mass is comparatively lower. 

Our analyses could be further improved by including the $pp\to t\bar t\Delta$ production 
channel with cross sections lying typically one order of magnitude below the one expected
for $\Delta$-strahlung.
Another foreseeable improvement would be to also consider semi-visible final states 
where one heavy neutrino $N$ decays inside the inner detector while the other one escapes 
detection. 
Additionally, $\Delta$ pair production yields a $\Delta L=4$ signature via 
$gg\to\Delta\Delta\to 4N$ decays, whose observation and low rate could be related to 
$0\nu4\beta$ decay processes~\cite{Heeck:2013rpa, Barabash:2019enn}. 
If the $\Delta$ mass is large enough to allow for decays into pairs of 
$W$, $Z$, or Higgs bosons, the lepton-number violating signal yield considered in this 
study drastically decreases, correspondingly opening the door to searches for heavier 
$\Delta$ states through (partially) resonant multi-boson final states. 
Moreover, if the mixing between the SM Higgs boson and the $\Delta$ scalar is below
approximately $1\%$, the $\Delta$-strahlung production modes become highly suppressed. 
In this regime, $\Delta$ decays into the $NN$ final state however still largely dominate 
so that the resulting signals could be probed at a future (very) high-energy hadron
collider through the process $pp\to W_R\Delta$. 
Depending on the decay mode of the $W_R$ boson, this channel could lead to an intriguing
$\Delta L=4$ final state, providing a rare and striking signature of lepton number
violation at unprecedented energy scales.

\section*{Acknowledgments}
JK and MN are supported by the Slovenian Research Agency under the research core funding 
No. P1-0035 and in part by the research grants J1-3013 and N1-0253, while the work of BF has been partly supported by Grant ANR-21-CE31-0013 (project DMwithLLPatLHC) from the French \emph{Agence Nationale de la Recherche}. 
The work of BF, JK and MN has received further support by the bilateral project 
Proteus PR-12696/Projet 50194VC.


\appendix

%
%
\section{Box amplitudes for \texorpdfstring{$gg\to S_{1} S_{2}$}{gg -> S1 S2}
} \label{app:Box-amplitude}
The amplitude in Eq.~\eqref{eqn:ampfull} receives contributions from both triangle and 
box diagrams with an internal quark loop.
We present below expressions for the box contributions, split in several parts according 
to the permutation of external momenta for clarity.
The contributions to the box amplitudes can be written as
\begin{eqnarray}
    \mathcal M_\square^{00} &=& i \sum_q Y_{S_1}^q Y_{S_2}^q\Big(\mathcal F_\square^1( \hat s, \hat t, \hat u) + \mathcal F_\square^1( \hat s, \hat u, \hat t) + \mathcal G_\square^1( \hat s, \hat t, \hat u)\Big)\,,\\
    \mathcal M_\square^{21} &=& i \sum_q Y_{S_1}^q Y_{S_2}^q\Big(\mathcal F_\square^2( \hat s, \hat t, \hat u) + \mathcal F_\square^2( \hat s, \hat u, \hat t) + \mathcal G_\square^2( \hat s, \hat t, \hat u)\Big)\,,\\
    \mathcal M_\square^{31} &=& i \sum_q Y_{S_1}^q Y_{S_2}^q\Big(\mathcal F_\square^3( \hat s, \hat t, \hat u) + \mathcal F_\square^4( \hat s, \hat u, \hat t) + \mathcal G_\square^3( \hat s, \hat t, \hat u)\Big)\,,\\
    \mathcal M_\square^{23} &=& i \sum_q Y_{S_1}^q Y_{S_2}^q\Big(\mathcal F_\square^4( \hat s, \hat t, \hat u) + \mathcal F_\square^3( \hat s, \hat u, \hat t) + \mathcal G_\square^4( \hat s, \hat t, \hat u)\Big)\,,
\end{eqnarray}
where the Yukawa couplings are 
$\frac{1}{\cos\theta}Y_h^q \approx \frac{1}{\sin\theta}Y_\Delta^q \approx \frac{m_q}{v}$
and the loop-functions are 
\begin{eqnarray}
    \mathcal F_\square^1(x,y,z) &=& -4 (B_0(x) + B_0(y)) - 2\Big(x C_0(0,0,x) + (m_{S_1}^2 - x - z)C_0(0,m_{S_2}^2,y)\nonumber\\
    &\phantom{=}&{} - (m_{S_1}^2 - y)C_0(0,m_{S_1}^2, y) + (8m_q^2 - y - z)C_0(m_{S_1}^2, m_{S_2}^2,x) - 8 C_{00}(0,x,0)\nonumber\\
    &\phantom{=}&{}+ x(4 m_q^2 - y)D_0- 4(8 m_q^2 - y - z)D_{00}\Big)\,,\\[0ex]
    \mathcal F_\square^2(x,y,z) &=& -4 \Big( C_0(0, m_{S_2}^2, y)  +   2C_2(0,m_{S_2}^2,y) + 4 C_{12}(0,x,0)-  C_0(0,x,0)\nonumber\\
    &\phantom{=}& {}+ (y - 4 m_q^2)D_0 - (m_{S_1}^2 - y)(D_1 + D_2)+  (m_{S_1}^2 - z)D_3\nonumber\\
    &\phantom{=}& {}+2(8 m_q^2 - y- z)(D_{13} + D_{23})\Big)\,,\\[0ex]
    \mathcal F_\square^3(x,y,z) &=& -4 \Big(C_0(m_{S_1}^2, m_{S_2}^2, x) + C_0(y, m_{S_1}^2, 0) - 2 C_1(y, m_{S_2}^2,0) + x(D_1 + D_2) \nonumber\\
    &\phantom{=}&{}+ 2(8m_q^2 - y - z)(D_{12} + D_{22})  + (m_{S_1}^2 - z)D_2\Big)\,,\\[0ex]
    \mathcal F_\square^4(x,y,z) &=& 4 \Big(C_0(m_{S_1}^2, m_{S_2}^2, x) + C_0(0, m_{S_2}^2, y)+ 2 C_1(y, m_{S_2}^2,0) \nonumber\\
    &\phantom{=}&{}+ 2(8 m_q^2 - y - z)D_{23} + x D_3 + (y - m_{S_1}^2)D_2\Big)\,.
\end{eqnarray}
The quantities $D_{ij}\equiv D_{ij}(0,m_{S_1}^2, m_{S_2}^2, 0, y,x)$, and we further 
abbreviate the Passarino-Veltman scalar integrals as
\begin{eqnarray}
    B_0(x_1) &\equiv& B_0(x_1, m_q^2, m_q^2)\label{eqn:PV1}\\[0ex]
    C_{ij}(x_1, x_2, x_3) &\equiv& C_{ij}(x_1, x_2, x_3, m_q^2, m_q^2, m_q^2)\label{eqn:PV2}\\[0ex]
    D_{ij}(x_1, x_2, x_3, x_4, x_5, x_6) &\equiv& D_{ij}(x_1, x_2, x_3, x_4, x_5, x_6, m_q^2, m_q^2, m_q^2, m_q^2)\label{eqn:PV3}\,,
\end{eqnarray}
consistent with the convention and notation of \textsc{LoopTools}~\cite{Hahn:1998yk}. We further have
\begin{eqnarray}
    \mathcal G_\square^1(x,y,z) &=& 4 (B_0(y) + B_0(z)) - 2\Big((m_{S_1}^2 - z)(C_0(0, z, m_{S_1}^2) - C_0(0, m_{S_2}^2, y)) \nonumber\\
    &\phantom{=}&{} + (m_{S_1}^2 - y)(C_0(y, 0, m_{S_1}^2) - C_0(m_{S_2}^2, 0, z)) + x (C_0(0, m_{S_2}^2, y) + C_0(m_{S_2}^2, 0, z))\nonumber\\
    &\phantom{=}& {}+ x(4m_q^2 - m_{S_1}^2)D_0 + (m_{S_1}^2 - y)(m_{S_1}^2 - z)D_0 - 4(8m_q^2 - y - z)D_{00}\Big)
    \\[0ex]
    \mathcal G_\square^2(x,y,z) &=& 4\Big(C_0(0,m_{S_2}^2, y) - C_0(m_{S_2}^2, 0, z) + 2C_2(0,m_{S_2}^2, y) \nonumber
    \\
    &\phantom{=}&{}- 2(C_1(m_{S_2}^2, 0, z) + C_2(m_{S_2}^2, 0, z))  + (z - m_{S_1}^2)D_3 + (y - m_{S_1}^2) D_2\nonumber
    \\
    &\phantom{=}&{}+ (4 m_q^2 - m_{S_1}^2)D_0 + 2 (y + z - 8 m_q^2)D_{23}\Big)\,,
    \\[0ex]
    \mathcal G_\square^3(x,y,z) &=& 4\Big(C_0(y, 0, m_{S_1}^2) + C_0(m_{S_2}^2,0,z) + 2 C_1(m_{S_2}^2,0,z) + 2 C_2(m_{S_2}^2,0,z) \nonumber
    \\
    &\phantom{=}&{} - 2 C_2(0, m_{S_2}^2, y)+ (x +2y + 3z - m_{S_1}^2 - 16 m_q^2)D_2 \nonumber
    \\
    &\phantom{=}&{}+ (z - m_{S_1}^2)D_1 + 2(y + z - 8 m_q^2)(D_{12} + D_{22})\Big)\,,
    \\
    \mathcal G_\square^4(x,y,z) &=& 4\Big(C_0(0,z,m_{S_1}^2) - C_0(0,m_{S_2}^2,y)  + 2 C_0(m_{S_2}^2, 0, z) + 2C_1(m_{S_2}^2, 0, z) \nonumber
    \\
    &\phantom{=}&{} +  2C_2(m_{S_2}^2, 0, z)- 2C_2(0,m_{S_2}^2,y)  + (m_{S_1}^2 - y)(D_0 + D_1 + D_2) \nonumber
    \\
    &\phantom{=}&{}+ x D_3+ 2(8m_q^2 - y - z)(D_{13} + D_{23}) \Big) \, ,
\end{eqnarray}
with this time $D_{ij}\equiv D_{ij}(m_{S_1}^2, 0, m_{S_2}^2, 0, y, z)$.
In order to bring the amplitudes to this compact form we have used the following 
symmetry relations of the Passarino-Veltman functions
\begin{eqnarray}
    C_0(x,y,z) &=& C_0(x,z,y) = C_0(z,y,x) + \text{ cyclic perm.}\,,
    \\[0ex]
    C_1(x,y,z) &=& C_2(z, y, x)\,,
    \\[0ex]
    D_{0,2,00,13,22}(a,b,c,d, x,y) &=& D_{0,2,00,13,22}(d,c,b,a,x,y)\,,
    \\[0ex]
    D_{1}(a,b,c,d, x,y) &=& D_{3}(d,c,b,a,x,y)\,,
    \\[0ex]
    D_{12}(a,b,c,d, x,y) &=& D_{23}(d,c,b,a,x,y)\,,
    \\[0ex]
    D_{11}(a,b,c,d, x,y) &=& D_{33}(d,c,b,a,x,y)\,,
\end{eqnarray}
which hold for coinciding internal masses (or propagator poles), as in Eqs.~(\ref{eqn:PV1}-\ref{eqn:PV3}).

%
%
\section{Phase space integration}
\label{app:Phasespace}
The differential cross sections derived in this work are numerically 
integrated over the phase space with a Monte Carlo method such as the \textsc{Vegas} 
algorithm~\cite{Lepage:1977sw, Lepage:2020tgj}. Moreover, PDF values are obtained with the Python interface of \textsc{LHAPDF 6.5.4}~\cite{Buckley:2014ana}.
For better numerical stability and in order to derive distributions, it is convenient to change the integration variables from the Bjorken variables $x_{1,2}$ to the reduced invariant mass $\tau = m_\text{inv}/s$ and rapidity $y$
\begin{eqnarray}
    x_{1,2} &=& \sqrt{\tau}\exp(\pm y)\,,\qquad \tau = x_1 x_2\,,\qquad y = \frac12 \log\left(x_1/x_2\right)\,,\qquad \d x_1 \d x_2 = \d\tau \d y\,,\nonumber\\[0ex]
    \tau &\in&\left[\frac{(m_{1} + m_{2})^2}{s},1\right]\,,\qquad y\in\left[-\log\left(\frac{1}{\sqrt{\tau}}\right),\log\left(\frac{1}{\sqrt{\tau}}\right)\right]\,.
\end{eqnarray}
Depending on the mass configuration of the final state scalars of masses $m_1$ and $m_2$, their production can be enhanced via massive $s$-channel resonant exchanges (\textit{e.g.}\ the triangle diagrams in $gg\to h\to\Delta\Delta$), leading to a sharply peaked integrand in the variable $\tau$.
This poses a problem for Monte Carlo integrators, and it is thus convenient to change the integration variables to flatten the pole in the $\tau$ direction and precondition the integrator to ensure smooth sampling around the pole.
Assuming a single resonance, the pole structure admits a Breit-Wigner shape
\begin{equation}
  \mathrm{BW}(\tau) \simeq \frac{1}{((\tau s - M_R^2)^2 + \Gamma_R^2 M_R^2} \, ,
\end{equation}
where $M_R$ and $\Gamma_R$ are the real and imaginary parts of the propagator pole of 
the resonance $R$, that is in this case the mass and the total width of the 
$s$-channel mediator.
The integration variable $\tau$ can then be transformed symmetrically around the 
resonance as
\begin{eqnarray}
    \tau(u) &=& \frac{M_R^2}{s} + \frac{\Gamma_R M_R}{s}\tan\left[\pi\left(u -\frac{1}{2}\right)\right]\,,
    \qquad
    \frac{\d\tau}{\d u} = \pi \frac{\Gamma_R M_R}{s \cos^2\left(\pi(u - 1/2)\right)}\,,\nonumber\\
    u_\text{min/max} &=& \frac{1}{2} + \frac{1}{\pi}\arctan\left(\frac{s \,\tau_\text{min/max} - M_R^2}{\Gamma_R M_R}\right)\,.
\end{eqnarray}

\begin{figure}
    \centering
    \includegraphics[width=0.45\linewidth]{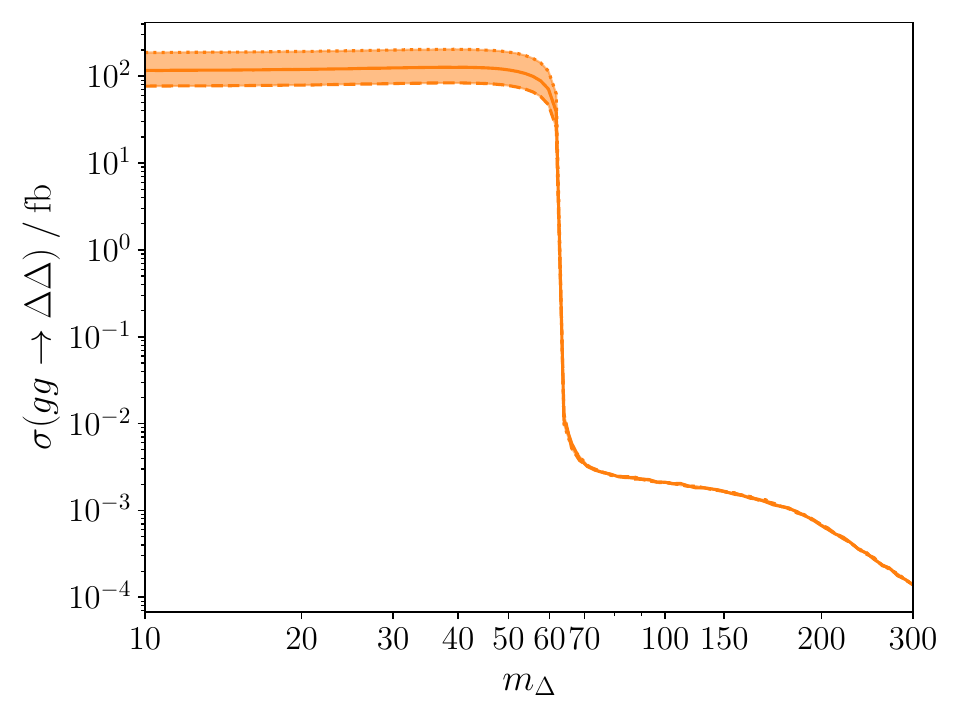}
    \vspace*{-2ex}    
    \caption{Variation of the $gg\to \Delta\Delta$ production cross section with respect 
    to $m_\Delta$ and the Higgs-width $\Gamma_\mathrm{tot}(h)$.
    The solid line denotes the central value $\Gamma(h) = 3.7\:\mathrm{MeV}$, while 
    the envelope spanned by dashed (dotted) lines denotes $\Gamma(h) = 5.6\:\mathrm{MeV}$ 
    ($\Gamma(h) = 2.3\:\mathrm{MeV}$).}
    \label{fig:gg_dd}
\end{figure}

For numerical computations, the Higgs boson width has been fixed to the central value returned by current measurements, $\Gamma_\mathrm{tot}(h) = 3.7_{-1.4}^{+1.9}\:\mathrm{MeV}$~\cite{ParticleDataGroup:2024cfk}.
A smaller (larger) value 
would lead to a significant increase 
(decrease) of the $gg\to h\to\Delta\Delta$ cross section, as shown in 
Figure~\ref{fig:gg_dd} where we vary the Higgs width within the current 
$1\sigma$ band.
The resulting variation  is found to be significant only in the resonant regime, as expected.

\section{Efficiency tables}\label{app:efftables}
In this appendix we present signal efficiencies resulting from the isolation 
requirements and reconstruction performance outlined in Section~\ref{sec:Topos}.
Selected events include exactly two reconstructed leptons with 
$p_T(\ell) > 10 \:\mathrm{GeV}$, $|\eta(\ell)|<2.4$ and are well isolated from the closest 
jet $j_c$ with $\Delta R(\ell,j_c) > 0.4$.

\begin{table}\small
\renewcommand{\arraystretch}{1.15}
  \centering
  \begin{tabularx}{.9\textwidth}{c c|*{6}{c}}
   \multicolumn{2}{c|}{$pp\to h \Delta \to b\bar b \, N N$}& $e^+ e^-$ & $\mu^+ \mu^-$ & $e^\pm \mu^\mp$ & $e^\pm e^\pm$ & $\mu^\pm \mu^\pm$ & $e^\pm \mu^\pm$\\ \hline
  \multirow{2}{*}{$m_\Delta = 50$~GeV}  & Prompt & 0.2\% & 0.2\% & 0.5\% & 0.1\% & $0.2\%$ & 0.4\%\\
  & Long-lived & 4.6\% & 4.7\% & 9.7\% & 4.2\%& 4.6\% & 9.4\% \\ \hdashline
  \multirow{2}{*}{$m_\Delta = 100$~GeV} & Prompt & 2.0\% & 1.9\% & 3.8\% & 2.0\% & 1.8\% & 3.2\%\\
  & Long-lived & 2.6\% & 2.4\% & 4.8\% & 2.3\% & 2.3\% & 5.0\% \\ \hdashline
  \multirow{2}{*}{$m_\Delta = 150$~GeV} & Prompt & 5.9\% & 5.8\% & 12.0\% & 5.7\% & 6.1\% & 10.8\%\\
  & Long-lived & 0.6\% & 0.4\% & 0.8\% & 0.4\% & 0.3\% & 0.8\% \\ \hdashline
  \multirow{2}{*}{$m_\Delta = 200$~GeV} & Prompt & 6.6\% & 7.0\% & 13.9\% & 6.5\% & 6.8\% & 14.2\%\\
  & Long-lived & 0.4\% & 0.3\% & 0.5\% &0.1\% & 0.1\% & 0.1\%\\ \hdashline
  \multirow{2}{*}{$m_\Delta = 250$~GeV} & Prompt & 6.8\% & 6.6\% & 12.6\% & 6.2\% & 6.6\% & 13.2\%\\
  & Long-lived & 0.3\% & 0.2\% & 0.6\% &0.2\% & 0.1\% &0.2\% \\
\end{tabularx}\vspace{.3cm}
  \begin{tabularx}{.9\textwidth}{c c|*{6}{c}}
   \multicolumn{2}{c|}{$pp\to Z \Delta \to b\bar b \, N N$} & $e^+ e^-$ & $\mu^+ \mu^-$ & $e^\pm \mu^\mp$ & $e^\pm e^\pm$ & $\mu^\pm \mu^\pm$ & $e^\pm \mu^\pm$\\ \hline
  \multirow{2}{*}{$m_\Delta = 50$~GeV}  & Prompt & $0.1\%$ & $0.1\%$ & $0.1\%$ & $0.1\%$ & $-$ & $0.1\%$ \\
  & Long-lived & $1.1\%$ & $1.1\%$ & $2.5\%$ & 1.3\% & 1.2\%  & 2.3\%\\ \hdashline
  \multirow{2}{*}{$m_\Delta = 100$~GeV} & Prompt & 1.2\%& $1.1\%$ & $2.3\%$ & $1.1\%$ & $1.1\%$ & $2.4\%$ \\
  & Long-lived & 1.2\% & 1.3\%& 2.1\% & 1.1\% & 1.1\% & 2.4\%\\ \hdashline
  \multirow{2}{*}{$m_\Delta = 150$~GeV} & Prompt & 4.5\% & 5.1\%& 9.3\% & 4.4\%& 4.5\% & 8.9\% \\
  & Long-lived & 0.3\% & 0.2\%& 0.5\% & 0.2\% & 0.2\%& 0.3\% \\ \hdashline
  \multirow{2}{*}{$m_\Delta = 200$~GeV} & Prompt & 6.3\% & 6.7\%& 12.6\%& 5.8\%& 6.2\% & 12.5\%\\
  & Long-lived & 0.2\% & 0.1\%& 0.3\% & 0.1\% & $-$& 0.2\%\\ \hdashline
  \multirow{2}{*}{$m_\Delta = 250$~GeV} & Prompt & 5.7\% & 6.0\% &11.5\% & 6.2\% &5.4\% & 11.9\% \\
  & Long-lived & 0.3\% & 0.2\% &0.4\% & $-$ & 0.1\% & 0.2\%\\
\end{tabularx}\vspace{.3cm}
\begin{tabularx}{.9\textwidth}{c c|*{6}{c}}
\multicolumn{2}{c|}{$pp\to W \Delta \to jj \, N N$} & $e^+ e^-$ & $\mu^+ \mu^-$ & $e^\pm \mu^\mp$ & $e^\pm e^\pm$ & $\mu^\pm \mu^\pm$ & $e^\pm \mu^\pm$\\ \hline
  \multirow{2}{*}{$m_\Delta = 50$~GeV}  & Prompt & $-$ & $-$ & $-$ & $-$ & $-$ & $-$ \\
  & Long-lived & $1.1\%$ & $0.9\%$ & $1.8\%$ & $1.0\%$ & $0.8\%$ & $1.9\%$  \\ \hdashline
  \multirow{2}{*}{$m_\Delta = 100$~GeV}  & Prompt & $0.9\%$ & $1.0\%$ & $2.0\%$ & $0.9\%$ & $0.9\%$ & $1.9\%$ \\
  & Long-lived & $0.8\%$ & $0.9\%$ & $1.8\%$ & $0.7\%$ & $0.7\%$ & $1.6\%$  \\ \hdashline
  \multirow{2}{*}{$m_\Delta = 150$~GeV}  & Prompt & $4.6\%$ & $4.5\%$ & $9.6\%$ & $4.4\%$ & $4.8\%$ & $8.8\%$ \\
  & Long-lived & $0.1\%$ & $0.1\%$ & $0.1\%$ & $-$ & $-$ & $0.1\%$  \\ \hdashline
  \multirow{2}{*}{$m_\Delta = 200$~GeV}  & Prompt & $6.7\%$ & $6.5\%$ & $12.7\%$ & $5.8\%$ & $7.0\%$ & $12.7\%$ \\
  & Long-lived & $-$ & $-$ & $-$ & $-$ & $-$ & $-$  \\ \hdashline
  \multirow{2}{*}{$m_\Delta = 250$~GeV}  & Prompt & $6.0\%$ & $6.3\%$ & $12.0\%$ & $6.2\%$ & $6.3\%$ & $11.7\%$ \\
  & Long-lived & $-$ & $-$ & $-$ & $-$ & $-$ & $-$\\
\end{tabularx}
  \caption{Selection efficiencies when requiring two isolated leptons with $p_T>10$~GeV, $|\eta|<2.4$, and either $|d_0|>0.1$~mm (long-lived) or $|d_0|<0.1$~mm (prompt). We consider signals emerging from $pp\to h \Delta \to b\bar b N N$ (upper), $pp\to Z \Delta \to b\bar b N N$ (middle) and $pp\to W \Delta \to jj NN$ (lower), for a scenario where $M_{W_R}= 6$~TeV and the heavy $N$ decays democratically into electrons and muons. Leptons are required to be isolated from any jet by $\Delta R>0.4$.}\label{tab:eff_6}
\end{table}
\begin{table}\small
\renewcommand{\arraystretch}{1.15}
  \centering
  \begin{tabularx}{.9\textwidth}{c c|*{6}{c}}
   \multicolumn{2}{c|}{$pp\to h \Delta \to b\bar b \, N N$}  & $e^+ e^-$ & $\mu^+ \mu^-$ & $e^\pm \mu^\mp$ & $e^\pm e^\pm$ & $\mu^\pm \mu^\pm$ & $e^\pm \mu^\pm$\\ \hline
  \multirow{2}{*}{$m_\Delta = 50$~GeV}  & Prompt & $0.1\%$ & $0.2\%$ & $0.3\%$ & $0.1\%$ & $0.1\%$ & $0.1\%$ \\
  & Long-lived & 4.9\% & 5.3\%& 10.1\% & 5.0\% & 4.8\%  & 9.8\%\\ \hdashline
  \multirow{2}{*}{$m_\Delta = 100$~GeV} & Prompt & 0.3\%& 0.2\% & 0.3\% & 0.1\% & 0.1\% & 0.1\% \\
  & Long-lived & 5.7\% & 5.9\%& 11.3\% & 5.4\% & 5.4\% & 10.1\%\\ \hdashline
  \multirow{2}{*}{$m_\Delta = 150$~GeV} & Prompt & 0.3\% & 0.2\%& 0.4\% & 0.1\%& 0.1\% & 0.4\% \\
  & Long-lived & 5.9\% & 6.3\%& 12.7\% & 5.8\% & 6.6\%& 11.8\% \\ \hdashline
  \multirow{2}{*}{$m_\Delta = 200$~GeV} & Prompt & 0.6\% & 0.4\%& 0.7\%& 0.3\%& 0.3\% & 0.6\%\\
  & Long-lived & 5.8\% & 6.2\%& 12.5\% & 6.0\% &6.4\%& 11.8\%\\ \hdashline
  \multirow{2}{*}{$m_\Delta = 250$~GeV} & Prompt & 0.7\% & 0.6\% &1.5\% & 0.6\% &0.6\% & 1.3\% \\
  & Long-lived & 5.0\% & 4.9\% & 10.0\%  & 4.8\% & 5.5\% & 9.7\%\\
\end{tabularx}\vspace{.3cm}
  \begin{tabularx}{.9\textwidth}{c c|*{6}{c}}
   \multicolumn{2}{c|}{$pp\to Z \Delta \to b\bar b \, N N$} & $e^+ e^-$ & $\mu^+ \mu^-$ & $e^\pm \mu^\mp$ & $e^\pm e^\pm$ & $\mu^\pm \mu^\pm$ & $e^\pm \mu^\pm$\\ \hline
  \multirow{2}{*}{$m_\Delta = 50$~GeV}  & Prompt & $0.1\%$ & $-$ & $0.1\%$ & $-$ & $-$ & $-$ \\
  & Long-lived & $1.3\%$ & $1.5\%$ & $2.5\%$ & 1.1\% & 1.3\%  & 2.5\%\\ \hdashline
  \multirow{2}{*}{$m_\Delta = 100$~GeV} & Prompt & 0.1\%& $-$ & $-$ & $-$ & $-$ & $-$ \\
  & Long-lived & 3.5\% & 3.3\%& 7.0\% & 3.5\% & 3.5\% & 6.8\%\\ \hdashline
  \multirow{2}{*}{$m_\Delta = 150$~GeV} & Prompt & 0.1\% & 0.1\%& 0.2\% & 0.1\%& 0.1\% & 0.1\% \\
  & Long-lived & 5.1\% & 5.1\%& 9.4\% & 5.3\% & 4.9\%& 9.7\% \\ \hdashline
  \multirow{2}{*}{$m_\Delta = 200$~GeV} & Prompt & 0.3\% & 0.2\%& 0.6\%& 0.3\%& 0.2\% & 0.4\%\\
  & Long-lived & 5.3\% & 5.2\%& 11.4\% & 5.1\% &5.4\%& 10.9\%\\ \hdashline
  \multirow{2}{*}{$m_\Delta = 250$~GeV} & Prompt & 0.7\% & 0.6\% &1.2\% & 0.5\% &0.4\% & 1.2\% \\
  & Long-lived & 4.5\% & 4.8\% &9.1\% & 4.1\% & 4.4\% & 8.6\%\\
\end{tabularx}\vspace{.3cm}
\begin{tabularx}{.9\textwidth}{c c|*{6}{c}}
\multicolumn{2}{c|}{$pp\to W \Delta \to jj \, N N$} & $e^+ e^-$ & $\mu^+ \mu^-$ & $e^\pm \mu^\mp$ & $e^\pm e^\pm$ & $\mu^\pm \mu^\pm$ & $e^\pm \mu^\pm$\\ \hline
  \multirow{2}{*}{$m_\Delta = 50$~GeV}  & Prompt & $-$ & $-$ & $-$ & $-$ & $-$ & $-$ \\
  & Long-lived & $1.1\%$ & $1.1\%$ & $2.4\%$ & $0.8\%$ & $1.1\%$ & $2.2\%$  \\ \hdashline
  \multirow{2}{*}{$m_\Delta = 100$~GeV}  & Prompt & $-$ & $-$ & $-$ & $-$ & $-$ & $-$ \\
  & Long-lived & $3.4\%$ & $3.5\%$ & $6.8\%$ & $3.3\%$ & $3.3\%$ & $6.5\%$  \\ \hdashline
  \multirow{2}{*}{$m_\Delta = 150$~GeV}  & Prompt & $0.1\%$ & $-$ & $0.1\%$ & $-$ & $-$ & $-$ \\
  & Long-lived & $4.5\%$ & $5.0\%$ & $10.6\%$ & $5.1\%$ & $4.4\%$ & $9.6\%$  \\ \hdashline
  \multirow{2}{*}{$m_\Delta = 200$~GeV}  & Prompt & $0.2\%$ & $0.1\%$ & $0.2\%$ & $-$ & $0.1\%$ & $0.1\%$ \\
  & Long-lived & $5.2\%$ & $5.7\%$ & $11.5\%$ & $5.2\%$ & $5.7\%$ & $11.0\%$  \\ \hdashline
  \multirow{2}{*}{$m_\Delta = 250$~GeV}  & Prompt & $0.3\%$ & $0.3\%$ & $0.5\%$ & $0.2\%$ & $0.2\%$ & $0.5\%$ \\
  & Long-lived & $4.6\%$ & $4.5\%$ & $9.0\%$ & $4.5\%$ & $4.5\%$ & $8.7\%$  \\
\end{tabularx}
\caption{Selection efficiencies when requiring two isolated leptons with $p_T>10$~GeV, $|\eta|<2.4$, and either $|d_0|>0.1$~mm (long-lived) or $|d_0|<0.1$~mm (prompt). We consider signals emerging from $pp\to h \Delta \to b\bar b N N$ (upper), $pp\to Z \Delta \to b\bar b N N$ (middle) and $pp\to W \Delta \to jj NN$ (lower), for a scenario where $M_{W_R}\!=\!30\,\mathrm{TeV}$ and the heavy $N$ decays democratically into electrons and muons. Leptons are required to be isolated from any jet by $\Delta R>0.4$.}
\label{tab:eff_30}
\end{table}

\begin{table}\small
\renewcommand{\arraystretch}{1.15}
  \centering
\begin{tabularx}{.9\textwidth}{c c|*{6}{c}}
\multicolumn{2}{c|}{$pp\to \Delta \Delta \to bb \, N N$} & $e^+ e^-$ & $\mu^+ \mu^-$ & $e^\pm \mu^\mp$ & $e^\pm e^\pm$ & $\mu^\pm \mu^\pm$ & $e^\pm \mu^\pm$\\ \hline
  \multirow{2}{*}{$m_\Delta = 50$~GeV}  & Prompt & $0.1\%$ & $0.1\%$ & $-$ & $0.1\%$ & $-$ & $-$ \\
  & Long-lived & $1.1\%$ & $1.0\%$ & $2.0\%$ & $1.0\%$ & $1.0\%$ & $2.0\%$  \\ \hdashline
  \multirow{2}{*}{$m_\Delta = 100$~GeV}  & Prompt & $1.9\%$ & $2.0\%$ & $3.9\%$ & $2.0\%$ & $1.9\%$ & $3.9\%$ \\
  & Long-lived & $2.8\%$ & $2.4\%$ & $5.7\%$ & $2.1\%$ & $2.3\%$ & $4.9\%$  \\ \hdashline
  \multirow{2}{*}{$m_\Delta = 150$~GeV}  & Prompt & $5.8\%$ & $5.3\%$ & $11.7\%$ & $5.4\%$ & $5.5\%$ & $11.2\%$ \\
  & Long-lived & $0.5\%$ & $0.5\%$ & $1.1\%$ & $0.2\%$ & $0.4\%$ & $0.8\%$  \\ \hdashline
  \multirow{2}{*}{$m_\Delta = 200$~GeV}  & Prompt & $7.4\%$ & $7.3\%$ & $13.9\%$ & $6.8\%$ & $6.8\%$ & $13.2\%$ \\
  & Long-lived & $0.4\%$ & $0.4\%$ & $0.7\%$ & $0.1\%$ & $0.1\%$ & $0.3\%$  \\ \hdashline
  \multirow{2}{*}{$m_\Delta = 250$~GeV}  & Prompt & $6.8\%$ & $6.6\%$ & $13.3\%$ & $6.2\%$ & $6.6\%$ & $13.1\%$ \\
  & Long-lived & $0.5\%$ & $0.5\%$ & $0.9\%$ & $0.2\%$ & $0.1\%$ & $0.3\%$  \\
\end{tabularx}\vspace{.3cm}
\begin{tabularx}{.9\textwidth}{c c|*{6}{c}}
\multicolumn{2}{c|}{$pp\to \Delta \Delta \to b\bar{b} \, N N$} & $e^+ e^-$ & $\mu^+ \mu^-$ & $e^\pm \mu^\mp$ & $e^\pm e^\pm$ & $\mu^\pm \mu^\pm$ & $e^\pm \mu^\pm$\\ \hline
  \multirow{2}{*}{$m_\Delta = 50$~GeV}  & Prompt & $-$ & $-$ & $-$ & $-$ & $-$ & $-$ \\
  & Long-lived & $1.1\%$ & $1.2\%$ & $2.5\%$ & $1.1\%$ & $1.2\%$ & $2.2\%$  \\ \hdashline
  \multirow{2}{*}{$m_\Delta = 100$~GeV}  & Prompt & $0.3\%$ & $0.2\%$ & $0.4\%$ & $0.1\%$ & $-$ & $0.1\%$ \\
  & Long-lived & $5.7\%$ & $5.8\%$ & $11.1\%$ & $5.5\%$ & $5.4\%$ & $11.1\%$  \\ \hdashline
  \multirow{2}{*}{$m_\Delta = 150$~GeV}  & Prompt & $0.4\%$ & $0.2\%$ & $0.5\%$ & $0.1\%$ & $0.1\%$ & $0.3\%$ \\
  & Long-lived & $6.2\%$ & $5.9\%$ & $12.1\%$ & $5.9\%$ & $6.3\%$ & $10.8\%$  \\ \hdashline
  \multirow{2}{*}{$m_\Delta = 200$~GeV}  & Prompt & $0.5\%$ & $0.3\%$ & $0.7\%$ & $0.3\%$ & $0.3\%$ & $0.6\%$ \\
  & Long-lived & $6.7\%$ & $6.7\%$ & $12.4\%$ & $5.9\%$ & $5.7\%$ & $12.3\%$  \\ \hdashline
  \multirow{2}{*}{$m_\Delta = 250$~GeV}  & Prompt & $0.9\%$ & $0.7\%$ & $1.6\%$ & $0.8\%$ & $0.6\%$ & $1.3\%$ \\
  & Long-lived & $5.1\%$ & $5.2\%$ & $10.0\%$ & $4.8\%$ & $4.8\%$ & $9.9\%$  \\
\end{tabularx}

\caption{Same as table~\ref{tab:eff_30}, but for $pp\to \Delta \Delta \to b\bar{b} NN$ , and for a scenario where $M_{W_R}= 6\,\mathrm{TeV}$ (upper) and  $M_{W_R}= 30\,\mathrm{TeV}$ (lower).}
\label{tab:eff_dds}
\end{table}

\clearpage
\bibliographystyle{JHEP}
\bibliography{bbNN}

\providecommand{\href}[2]{#2}\begingroup\raggedright\begin{thebibliography}{10}

\bibitem{Minkowski:1977sc}
P.~Minkowski, \emph{{$\mu \to e\gamma$ at a Rate of One Out of $10^{9}$ Muon Decays?}}, \href{https://doi.org/10.1016/0370-2693(77)90435-X}{\emph{Phys. Lett.} {\bfseries 67B} (1977) 421}.

\bibitem{Gell-Mann:1979vob}
M.~Gell-Mann, P.~Ramond and R.~Slansky, \emph{{Complex Spinors and Unified Theories}}, {\emph{Conf. Proc. C} {\bfseries 790927} (1979) 315} [\href{https://arxiv.org/abs/1306.4669}{{\ttfamily 1306.4669}}].

\bibitem{Glashow:1979nm}
S.L.~Glashow, \emph{{The Future of Elementary Particle Physics}}, \href{https://doi.org/10.1007/978-1-4684-7197-7_15}{\emph{NATO Sci. Ser. B} {\bfseries 61} (1980) 687}.

\bibitem{Mohapatra:1979ia}
R.N.~Mohapatra and G.~Senjanovi\'c, \emph{{Neutrino Mass and Spontaneous Parity Nonconservation}}, \href{https://doi.org/10.1103/PhysRevLett.44.912}{\emph{Phys. Rev. Lett.} {\bfseries 44} (1980) 912}.

\bibitem{Sawada:1979dis}
T.~Yanagida, \emph{{Proceedings: Workshop on the Unified Theories and the Baryon Number in the Universe}: {Tsukuba, Japan, February 13-14, 1979}}, .

\bibitem{Pati:1974yy}
J.C.~Pati and A.~Salam, \emph{{Lepton Number as the Fourth Color}}, \href{https://doi.org/10.1103/PhysRevD.10.275}{\emph{Phys. Rev.} {\bfseries D10} (1974) 275}.

\bibitem{Mohapatra:1974gc}
R.N.~Mohapatra and J.C.~Pati, \emph{{A Natural Left-Right Symmetry}}, \href{https://doi.org/10.1103/PhysRevD.11.2558}{\emph{Phys. Rev.} {\bfseries D11} (1975) 2558}.

\bibitem{Senjanovic:1978ev}
G.~Senjanovi\'c, \emph{{Spontaneous Breakdown of Parity in a Class of Gauge Theories}}, \href{https://doi.org/10.1016/0550-3213(79)90604-7}{\emph{Nucl. Phys.} {\bfseries B153} (1979) 334}.

\bibitem{Maiezza:2010ic}
A.~Maiezza, M.~Nemev\v{s}ek, F.~Nesti and G.~Senjanovi\'c, \emph{{Left-Right Symmetry at LHC}}, \href{https://doi.org/10.1103/PhysRevD.82.055022}{\emph{Phys. Rev.} {\bfseries D82} (2010) 055022} [\href{https://arxiv.org/abs/1005.5160}{{\ttfamily 1005.5160}}].

\bibitem{Beall:1981ze}
G.~Beall, M.~Bander and A.~Soni, \emph{{Constraint on the Mass Scale of a Left-Right Symmetric Electroweak Theory from the K(L) K(S) Mass Difference}}, \href{https://doi.org/10.1103/PhysRevLett.48.848}{\emph{Phys. Rev. Lett.} {\bfseries 48} (1982) 848}.

\bibitem{Senjanovic:1979cta}
G.~Senjanovi\'c and P.~Senjanovi\'c, \emph{{Suppression of Higgs Strangeness Changing Neutral Currents in a Class of Gauge Theories}}, \href{https://doi.org/10.1103/PhysRevD.21.3253}{\emph{Phys. Rev.} {\bfseries D21} (1980) 3253}.

\bibitem{Zhang:2007da}
Y.~Zhang, H.~An, X.~Ji and R.N.~Mohapatra, \emph{{General CP Violation in Minimal Left-Right Symmetric Model and Constraints on the Right-Handed Scale}}, \href{https://doi.org/10.1016/j.nuclphysb.2008.05.019}{\emph{Nucl. Phys.} {\bfseries B802} (2008) 247} [\href{https://arxiv.org/abs/0712.4218}{{\ttfamily 0712.4218}}].

\bibitem{Bertolini:2014sua}
S.~Bertolini, A.~Maiezza and F.~Nesti, \emph{{Present and Future K and B Meson Mixing Constraints on TeV Scale Left-Right Symmetry}}, \href{https://doi.org/10.1103/PhysRevD.89.095028}{\emph{Phys. Rev.} {\bfseries D89} (2014) 095028} [\href{https://arxiv.org/abs/1403.7112}{{\ttfamily 1403.7112}}].

\bibitem{Ecker:1985vv}
G.~Ecker and W.~Grimus, \emph{{CP violation and left-right symmetry}}, \href{https://doi.org/10.1016/0550-3213(85)90616-9}{\emph{Nucl. Phys. B} {\bfseries 258} (1985) 328}.

\bibitem{Kiers:2002cz}
K.~Kiers, J.~Kolb, J.~Lee, A.~Soni and G.-H.~Wu, \emph{{Ubiquitous CP violation in a top inspired left-right model}}, \href{https://doi.org/10.1103/PhysRevD.66.095002}{\emph{Phys. Rev. D} {\bfseries 66} (2002) 095002} [\href{https://arxiv.org/abs/hep-ph/0205082}{{\ttfamily hep-ph/0205082}}].

\bibitem{Maiezza:2014ala}
A.~Maiezza and M.~Nemev\v{s}ek, \emph{{Strong P invariance, neutron electric dipole moment, and minimal left-right parity at LHC}}, \href{https://doi.org/10.1103/PhysRevD.90.095002}{\emph{Phys. Rev.} {\bfseries D90} (2014) 095002} [\href{https://arxiv.org/abs/1407.3678}{{\ttfamily 1407.3678}}].

\bibitem{Bertolini:2019out}
S.~Bertolini, A.~Maiezza and F.~Nesti, \emph{{Kaon CP violation and neutron EDM in the minimal left-right symmetric model}}, \href{https://doi.org/10.1103/PhysRevD.101.035036}{\emph{Phys. Rev. D} {\bfseries 101} (2020) 035036} [\href{https://arxiv.org/abs/1911.09472}{{\ttfamily 1911.09472}}].

\bibitem{Kuchimanchi:2014ota}
R.~Kuchimanchi, \emph{{Leptonic CP problem in left-right symmetric model}}, \href{https://doi.org/10.1103/PhysRevD.91.071901}{\emph{Phys. Rev. D} {\bfseries 91} (2015) 071901} [\href{https://arxiv.org/abs/1408.6382}{{\ttfamily 1408.6382}}].

\bibitem{Senjanovic:2020int}
G.~Senjanovic and V.~Tello, \emph{{Strong CP violation: Problem or blessing?}}, \href{https://doi.org/10.1142/S0217751X23500677}{\emph{Int. J. Mod. Phys. A} {\bfseries 38} (2023) 2350067} [\href{https://arxiv.org/abs/2004.04036}{{\ttfamily 2004.04036}}].

\bibitem{Bezrukov:2009th}
F.~Bezrukov, H.~Hettmansperger and M.~Lindner, \emph{{keV sterile neutrino Dark Matter in gauge extensions of the Standard Model}}, \href{https://doi.org/10.1103/PhysRevD.81.085032}{\emph{Phys. Rev. D} {\bfseries 81} (2010) 085032} [\href{https://arxiv.org/abs/0912.4415}{{\ttfamily 0912.4415}}].

\bibitem{Nemevsek:2012cd}
M.~Nemev\v{s}ek, G.~Senjanovi\'c and Y.~Zhang, \emph{{Warm Dark Matter in Low Scale Left-Right Theory}}, \href{https://doi.org/10.1088/1475-7516/2012/07/006}{\emph{JCAP} {\bfseries 2012} (2012) 006} [\href{https://arxiv.org/abs/1205.0844}{{\ttfamily 1205.0844}}].

\bibitem{Nemevsek:2022anh}
M.~Nemev\v{s}ek and Y.~Zhang, \emph{{Dark Matter Dilution Mechanism through the Lens of Large-Scale Structure}}, \href{https://doi.org/10.1103/PhysRevLett.130.121002}{\emph{Phys. Rev. Lett.} {\bfseries 130} (2023) 121002} [\href{https://arxiv.org/abs/2206.11293}{{\ttfamily 2206.11293}}].

\bibitem{Nemevsek:2023yjl}
M.~Nemev\v{s}ek and Y.~Zhang, \emph{{Anatomy of diluted dark matter in the minimal left-right symmetric model}}, \href{https://doi.org/10.1103/PhysRevD.109.056021}{\emph{Phys. Rev. D} {\bfseries 109} (2024) 056021} [\href{https://arxiv.org/abs/2312.00129}{{\ttfamily 2312.00129}}].

\bibitem{Dev:2025fcv}
P.S.B.~Dev, J.~Heeck and A.~Thapa, \emph{{Decaying scalar dark matter in the minimal left-right symmetric model}},  \href{https://arxiv.org/abs/2501.14669}{{\ttfamily 2501.14669}}.

\bibitem{Mohapatra:1980yp}
R.N.~Mohapatra and G.~Senjanovi\'c, \emph{{Neutrino Masses and Mixings in Gauge Models with Spontaneous Parity Violation}}, \href{https://doi.org/10.1103/PhysRevD.23.165}{\emph{Phys. Rev.} {\bfseries D23} (1981) 165}.

\bibitem{Nemevsek:2012iq}
M.~Nemev\v{s}ek, G.~Senjanovi\'c and V.~Tello, \emph{{Connecting Dirac and Majorana Neutrino Mass Matrices in the Minimal Left-Right Symmetric Model}}, \href{https://doi.org/10.1103/PhysRevLett.110.151802}{\emph{Phys. Rev. Lett.} {\bfseries 110} (2013) 151802} [\href{https://arxiv.org/abs/1211.2837}{{\ttfamily 1211.2837}}].

\bibitem{Kriewald:2024cgr}
J.~Kriewald, M.~Nemev\v{s}ek and F.~Nesti, \emph{{Enabling precise predictions for left-right symmetry at colliders}}, \href{https://doi.org/10.1140/epjc/s10052-024-13614-8}{\emph{Eur. Phys. J. C} {\bfseries 84} (2024) 1306} [\href{https://arxiv.org/abs/2403.07756}{{\ttfamily 2403.07756}}].

\bibitem{Weinberg:1967tq}
S.~Weinberg, \emph{{A Model of Leptons}}, \href{https://doi.org/10.1103/PhysRevLett.19.1264}{\emph{Phys. Rev. Lett.} {\bfseries 19} (1967) 1264}.

\bibitem{ATLAS:2022vkf}
{\scshape ATLAS} collaboration, \emph{{A detailed map of Higgs boson interactions by the ATLAS experiment ten years after the discovery}}, \href{https://doi.org/10.1038/s41586-022-04893-w}{\emph{Nature} {\bfseries 607} (2022) 52} [\href{https://arxiv.org/abs/2207.00092}{{\ttfamily 2207.00092}}].

\bibitem{CMS:2022dwd}
{\scshape CMS} collaboration, \emph{{A portrait of the Higgs boson by the CMS experiment ten years after the discovery.}}, \href{https://doi.org/10.1038/s41586-022-04892-x}{\emph{Nature} {\bfseries 607} (2022) 60} [\href{https://arxiv.org/abs/2207.00043}{{\ttfamily 2207.00043}}].

\bibitem{Maiezza:2015lza}
A.~Maiezza, M.~Nemev\v{s}ek and F.~Nesti, \emph{{Lepton Number Violation in Higgs Decay at LHC}}, \href{https://doi.org/10.1103/PhysRevLett.115.081802}{\emph{Phys. Rev. Lett.} {\bfseries 115} (2015) 081802} [\href{https://arxiv.org/abs/1503.06834}{{\ttfamily 1503.06834}}].

\bibitem{Nemevsek:2016enw}
M.~Nemev\v{s}ek, F.~Nesti and J.C.~Vasquez, \emph{{Majorana Higgses at colliders}}, \href{https://doi.org/10.1007/JHEP04(2017)114}{\emph{JHEP} {\bfseries 04} (2017) 114} [\href{https://arxiv.org/abs/1612.06840}{{\ttfamily 1612.06840}}].

\bibitem{Senjanovic:1975rk}
G.~Senjanovi\'c and R.N.~Mohapatra, \emph{{Exact Left-Right Symmetry and Spontaneous Violation of Parity}}, \href{https://doi.org/10.1103/PhysRevD.12.1502}{\emph{Phys. Rev.} {\bfseries D12} (1975) 1502}.

\bibitem{Olness:1985bg}
F.I.~Olness and M.E.~Ebel, \emph{{Constraints on the Higgs Boson Masses in Left-right Electroweak Gauge Theories}}, \href{https://doi.org/10.1103/PhysRevD.32.1769}{\emph{Phys. Rev. D} {\bfseries 32} (1985) 1769}.

\bibitem{Gunion:1989in}
J.F.~Gunion, J.~Grifols, A.~Mendez, B.~Kayser and F.I.~Olness, \emph{{Higgs Bosons in Left-Right Symmetric Models}}, \href{https://doi.org/10.1103/PhysRevD.40.1546}{\emph{Phys. Rev. D} {\bfseries 40} (1989) 1546}.

\bibitem{Deshpande:1990ip}
N.G.~Deshpande, J.F.~Gunion, B.~Kayser and F.I.~Olness, \emph{{Left-right symmetric electroweak models with triplet Higgs}}, \href{https://doi.org/10.1103/PhysRevD.44.837}{\emph{Phys. Rev. D} {\bfseries 44} (1991) 837}.

\bibitem{Kiers:2005gh}
K.~Kiers, M.~Assis and A.A.~Petrov, \emph{{Higgs sector of the left-right model with explicit CP violation}}, \href{https://doi.org/10.1103/PhysRevD.71.115015}{\emph{Phys. Rev. D} {\bfseries 71} (2005) 115015} [\href{https://arxiv.org/abs/hep-ph/0503115}{{\ttfamily hep-ph/0503115}}].

\bibitem{Maiezza:2016ybz}
A.~Maiezza, G.~Senjanovi\'c and J.C.~Vasquez, \emph{{Higgs sector of the minimal left-right symmetric theory}}, \href{https://doi.org/10.1103/PhysRevD.95.095004}{\emph{Phys. Rev. D} {\bfseries 95} (2017) 095004} [\href{https://arxiv.org/abs/1612.09146}{{\ttfamily 1612.09146}}].

\bibitem{Maiezza:2016bzp}
A.~Maiezza, M.~Nemev\v{s}ek and F.~Nesti, \emph{{Perturbativity and mass scales in the minimal left-right symmetric model}}, \href{https://doi.org/10.1103/PhysRevD.94.035008}{\emph{Phys. Rev. D} {\bfseries 94} (2016) 035008} [\href{https://arxiv.org/abs/1603.00360}{{\ttfamily 1603.00360}}].

\bibitem{Chauhan:2018uuy}
G.~Chauhan, P.S.B.~Dev, R.N.~Mohapatra and Y.~Zhang, \emph{{Perturbativity constraints on $U(1)_{B-L}$ and left-right models and implications for heavy gauge boson searches}}, \href{https://doi.org/10.1007/JHEP01(2019)208}{\emph{JHEP} {\bfseries 01} (2019) 208} [\href{https://arxiv.org/abs/1811.08789}{{\ttfamily 1811.08789}}].

\bibitem{Mohapatra:2019qid}
R.N.~Mohapatra, G.~Yan and Y.~Zhang, \emph{{Ameliorating Higgs induced flavor constraints on TeV scale $W_R$}}, \href{https://doi.org/10.1016/j.nuclphysb.2019.114764}{\emph{Nucl. Phys. B} {\bfseries 948} (2019) 114764} [\href{https://arxiv.org/abs/1902.08601}{{\ttfamily 1902.08601}}].

\bibitem{Christensen:2008py}
N.D.~Christensen and C.~Duhr, \emph{{FeynRules - Feynman rules made easy}}, \href{https://doi.org/10.1016/j.cpc.2009.02.018}{\emph{Comput. Phys. Commun.} {\bfseries 180} (2009) 1614} [\href{https://arxiv.org/abs/0806.4194}{{\ttfamily 0806.4194}}].

\bibitem{Alloul:2013bka}
A.~Alloul, N.D.~Christensen, C.~Degrande, C.~Duhr and B.~Fuks, \emph{{FeynRules 2.0 - A complete toolbox for tree-level phenomenology}}, \href{https://doi.org/10.1016/j.cpc.2014.04.012}{\emph{Comput. Phys. Commun.} {\bfseries 185} (2014) 2250} [\href{https://arxiv.org/abs/1310.1921}{{\ttfamily 1310.1921}}].

\bibitem{Degrande:2011ua}
C.~Degrande, C.~Duhr, B.~Fuks, D.~Grellscheid, O.~Mattelaer and T.~Reiter, \emph{{UFO - The Universal FeynRules Output}}, \href{https://doi.org/10.1016/j.cpc.2012.01.022}{\emph{Comput. Phys. Commun.} {\bfseries 183} (2012) 1201} [\href{https://arxiv.org/abs/1108.2040}{{\ttfamily 1108.2040}}].

\bibitem{Darme:2023jdn}
L.~Darm\'e et~al., \emph{{UFO 2.0: the \textquoteleft{}Universal Feynman Output\textquoteright{} format}}, \href{https://doi.org/10.1140/epjc/s10052-023-11780-9}{\emph{Eur. Phys. J. C} {\bfseries 83} (2023) 631} [\href{https://arxiv.org/abs/2304.09883}{{\ttfamily 2304.09883}}].

\bibitem{ATLAS:2019fgd}
{\scshape ATLAS} collaboration, \emph{{Search for new resonances in mass distributions of jet pairs using 139 fb$^{-1}$ of $pp$ collisions at $\sqrt{s}=13$ TeV with the ATLAS detector}}, \href{https://doi.org/10.1007/JHEP03(2020)145}{\emph{JHEP} {\bfseries 03} (2020) 145} [\href{https://arxiv.org/abs/1910.08447}{{\ttfamily 1910.08447}}].

\bibitem{CMS:2019gwf}
{\scshape CMS} collaboration, \emph{{Search for high mass dijet resonances with a new background prediction method in proton-proton collisions at $\sqrt{s} =$ 13 TeV}}, \href{https://doi.org/10.1007/JHEP05(2020)033}{\emph{JHEP} {\bfseries 05} (2020) 033} [\href{https://arxiv.org/abs/1911.03947}{{\ttfamily 1911.03947}}].

\bibitem{CMS:2023gte}
{\scshape CMS} collaboration, \emph{{Search for W' bosons decaying to a top and a bottom quark in leptonic final states in proton-proton collisions at $ \sqrt{s} $ = 13 TeV}}, \href{https://doi.org/10.1007/JHEP05(2024)046}{\emph{JHEP} {\bfseries 05} (2024) 046} [\href{https://arxiv.org/abs/2310.19893}{{\ttfamily 2310.19893}}].

\bibitem{ATLAS:2023ibb}
{\scshape ATLAS} collaboration, \emph{{Search for vector-boson resonances decaying into a top quark and a bottom quark using pp collisions at $ \sqrt{s} $ = 13 TeV with the ATLAS detector}}, \href{https://doi.org/10.1007/JHEP12(2023)073}{\emph{JHEP} {\bfseries 12} (2023) 073} [\href{https://arxiv.org/abs/2308.08521}{{\ttfamily 2308.08521}}].

\bibitem{Senjanovic:2014pva}
G.~Senjanovi\'c and V.~Tello, \emph{{Right Handed Quark Mixing in Left-Right Symmetric Theory}}, \href{https://doi.org/10.1103/PhysRevLett.114.071801}{\emph{Phys. Rev. Lett.} {\bfseries 114} (2015) 071801} [\href{https://arxiv.org/abs/1408.3835}{{\ttfamily 1408.3835}}].

\bibitem{Senjanovic:2015yea}
G.~Senjanovi\'c and V.~Tello, \emph{{Restoration of Parity and the Right-Handed Analog of the CKM Matrix}}, \href{https://doi.org/10.1103/PhysRevD.94.095023}{\emph{Phys. Rev. D} {\bfseries 94} (2016) 095023} [\href{https://arxiv.org/abs/1502.05704}{{\ttfamily 1502.05704}}].

\bibitem{Keung:1983uu}
W.-Y.~Keung and G.~Senjanovic, \emph{{Majorana Neutrinos and the Production of the Right-handed Charged Gauge Boson}}, \href{https://doi.org/10.1103/PhysRevLett.50.1427}{\emph{Phys. Rev. Lett.} {\bfseries 50} (1983) 1427}.

\bibitem{Frank:2023epx}
M.~Frank, B.~Fuks, A.~Jueid, S.~Moretti and O.~Ozdal, \emph{{A novel search strategy for right-handed charged gauge bosons at the Large Hadron Collider}}, \href{https://doi.org/10.1007/JHEP02(2024)150}{\emph{JHEP} {\bfseries 02} (2024) 150} [\href{https://arxiv.org/abs/2312.08521}{{\ttfamily 2312.08521}}].

\bibitem{Nemevsek:2011hz}
M.~Nemevsek, F.~Nesti, G.~Senjanovic and Y.~Zhang, \emph{{First Limits on Left-Right Symmetry Scale from LHC Data}}, \href{https://doi.org/10.1103/PhysRevD.83.115014}{\emph{Phys. Rev. D} {\bfseries 83} (2011) 115014} [\href{https://arxiv.org/abs/1103.1627}{{\ttfamily 1103.1627}}].

\bibitem{Mitra:2016kov}
M.~Mitra, R.~Ruiz, D.J.~Scott and M.~Spannowsky, \emph{{Neutrino Jets from High-Mass $W_R$ Gauge Bosons in TeV-Scale Left-Right Symmetric Models}}, \href{https://doi.org/10.1103/PhysRevD.94.095016}{\emph{Phys. Rev. D} {\bfseries 94} (2016) 095016} [\href{https://arxiv.org/abs/1607.03504}{{\ttfamily 1607.03504}}].

\bibitem{Nemevsek:2018bbt}
M.~Nemev\v{s}ek, F.~Nesti and G.~Popara, \emph{{Keung-Senjanovi\'c process at the LHC: From lepton number violation to displaced vertices to invisible decays}}, \href{https://doi.org/10.1103/PhysRevD.97.115018}{\emph{Phys. Rev. D} {\bfseries 97} (2018) 115018} [\href{https://arxiv.org/abs/1801.05813}{{\ttfamily 1801.05813}}].

\bibitem{CMS:2022krd}
{\scshape CMS} collaboration, \emph{{Search for new physics in the lepton plus missing transverse momentum final state in proton-proton collisions at $\sqrt{s} =$ 13 TeV}}, \href{https://doi.org/10.1007/JHEP07(2022)067}{\emph{JHEP} {\bfseries 07} (2022) 067} [\href{https://arxiv.org/abs/2202.06075}{{\ttfamily 2202.06075}}].

\bibitem{CMS:2022ncp}
{\scshape CMS} collaboration, \emph{{Search for new physics in the \ensuremath{\tau} lepton plus missing transverse momentum final state in proton-proton collisions at $ \sqrt{s} $ = 13 TeV}}, \href{https://doi.org/10.1007/JHEP09(2023)051}{\emph{JHEP} {\bfseries 09} (2023) 051} [\href{https://arxiv.org/abs/2212.12604}{{\ttfamily 2212.12604}}].

\bibitem{ATLAS:2019lsy}
{\scshape ATLAS} collaboration, \emph{{Search for a heavy charged boson in events with a charged lepton and missing transverse momentum from $pp$ collisions at $\sqrt{s} = 13$ TeV with the ATLAS detector}}, \href{https://doi.org/10.1103/PhysRevD.100.052013}{\emph{Phys. Rev. D} {\bfseries 100} (2019) 052013} [\href{https://arxiv.org/abs/1906.05609}{{\ttfamily 1906.05609}}].

\bibitem{Ruiz:2017nip}
R.~Ruiz, \emph{{Lepton Number Violation at Colliders from Kinematically Inaccessible Gauge Bosons}}, \href{https://doi.org/10.1140/epjc/s10052-017-4950-2}{\emph{Eur. Phys. J. C} {\bfseries 77} (2017) 375} [\href{https://arxiv.org/abs/1703.04669}{{\ttfamily 1703.04669}}].

\bibitem{Nemevsek:2023hwx}
M.~Nemev\v{s}ek and F.~Nesti, \emph{{Left-right symmetry at an FCC-hh}}, \href{https://doi.org/10.1103/PhysRevD.108.015030}{\emph{Phys. Rev. D} {\bfseries 108} (2023) 015030} [\href{https://arxiv.org/abs/2306.12104}{{\ttfamily 2306.12104}}].

\bibitem{Dev:2015kca}
P.S.B.~Dev, D.~Kim and R.N.~Mohapatra, \emph{{Disambiguating Seesaw Models using Invariant Mass Variables at Hadron Colliders}}, \href{https://doi.org/10.1007/JHEP01(2016)118}{\emph{JHEP} {\bfseries 01} (2016) 118} [\href{https://arxiv.org/abs/1510.04328}{{\ttfamily 1510.04328}}].

\bibitem{Gluza:2016qqv}
J.~Gluza, T.~Jelinski and R.~Szafron, \emph{{Lepton number violation and \textquoteleft{}Diracness\textquoteright{} of massive neutrinos composed of Majorana states}}, \href{https://doi.org/10.1103/PhysRevD.93.113017}{\emph{Phys. Rev. D} {\bfseries 93} (2016) 113017} [\href{https://arxiv.org/abs/1604.01388}{{\ttfamily 1604.01388}}].

\bibitem{Das:2016akd}
A.~Das, N.~Nagata and N.~Okada, \emph{{Testing the 2-TeV Resonance with Trileptons}}, \href{https://doi.org/10.1007/JHEP03(2016)049}{\emph{JHEP} {\bfseries 03} (2016) 049} [\href{https://arxiv.org/abs/1601.05079}{{\ttfamily 1601.05079}}].

\bibitem{Das:2017hmg}
A.~Das, P.S.B.~Dev and R.N.~Mohapatra, \emph{{Same Sign versus Opposite Sign Dileptons as a Probe of Low Scale Seesaw Mechanisms}}, \href{https://doi.org/10.1103/PhysRevD.97.015018}{\emph{Phys. Rev. D} {\bfseries 97} (2018) 015018} [\href{https://arxiv.org/abs/1709.06553}{{\ttfamily 1709.06553}}].

\bibitem{Helo:2018rll}
J.C.~Helo, H.~Li, N.A.~Neill, M.~Ramsey-Musolf and J.C.~Vasquez, \emph{{Probing neutrino Dirac mass in left-right symmetric models at the LHC and next generation colliders}}, \href{https://doi.org/10.1103/PhysRevD.99.055042}{\emph{Phys. Rev. D} {\bfseries 99} (2019) 055042} [\href{https://arxiv.org/abs/1812.01630}{{\ttfamily 1812.01630}}].

\bibitem{Dawson:2021jcl}
S.~Dawson, P.P.~Giardino and S.~Homiller, \emph{{Uncovering the High Scale Higgs Singlet Model}}, \href{https://doi.org/10.1103/PhysRevD.103.075016}{\emph{Phys. Rev. D} {\bfseries 103} (2021) 075016} [\href{https://arxiv.org/abs/2102.02823}{{\ttfamily 2102.02823}}].

\bibitem{Senjanovic:2016vxw}
G.~Senjanovi\'c and V.~Tello, \emph{{Probing Seesaw with Parity Restoration}}, \href{https://doi.org/10.1103/PhysRevLett.119.201803}{\emph{Phys. Rev. Lett.} {\bfseries 119} (2017) 201803} [\href{https://arxiv.org/abs/1612.05503}{{\ttfamily 1612.05503}}].

\bibitem{Senjanovic:2018xtu}
G.~Senjanovic and V.~Tello, \emph{{Disentangling the seesaw mechanism in the minimal left-right symmetric model}}, \href{https://doi.org/10.1103/PhysRevD.100.115031}{\emph{Phys. Rev. D} {\bfseries 100} (2019) 115031} [\href{https://arxiv.org/abs/1812.03790}{{\ttfamily 1812.03790}}].

\bibitem{Kiers:2022cyc}
J.~Kiers, K.~Kiers, A.~Szynkman and T.~Tarutina, \emph{{Disentangling the seesaw mechanism in the left-right model: An algorithm for the general case}}, \href{https://doi.org/10.1103/PhysRevD.107.075001}{\emph{Phys. Rev. D} {\bfseries 107} (2023) 075001} [\href{https://arxiv.org/abs/2212.14837}{{\ttfamily 2212.14837}}].

\bibitem{Alwall:2014hca}
J.~Alwall, R.~Frederix, S.~Frixione, V.~Hirschi, F.~Maltoni, O.~Mattelaer et~al., \emph{{The automated computation of tree-level and next-to-leading order differential cross sections, and their matching to parton shower simulations}}, \href{https://doi.org/10.1007/JHEP07(2014)079}{\emph{JHEP} {\bfseries 07} (2014) 079} [\href{https://arxiv.org/abs/1405.0301}{{\ttfamily 1405.0301}}].

\bibitem{Patel:2015tea}
H.H.~Patel, \emph{{Package-X: A Mathematica package for the analytic calculation of one-loop integrals}}, \href{https://doi.org/10.1016/j.cpc.2015.08.017}{\emph{Comput. Phys. Commun.} {\bfseries 197} (2015) 276} [\href{https://arxiv.org/abs/1503.01469}{{\ttfamily 1503.01469}}].

\bibitem{Patel:2016fam}
H.H.~Patel, \emph{{Package-X 2.0: A Mathematica package for the analytic calculation of one-loop integrals}}, \href{https://doi.org/10.1016/j.cpc.2017.04.015}{\emph{Comput. Phys. Commun.} {\bfseries 218} (2017) 66} [\href{https://arxiv.org/abs/1612.00009}{{\ttfamily 1612.00009}}].

\bibitem{Hahn:1998yk}
T.~Hahn and M.~Perez-Victoria, \emph{{Automatized one loop calculations in four-dimensions and D-dimensions}}, \href{https://doi.org/10.1016/S0010-4655(98)00173-8}{\emph{Comput. Phys. Commun.} {\bfseries 118} (1999) 153} [\href{https://arxiv.org/abs/hep-ph/9807565}{{\ttfamily hep-ph/9807565}}].

\bibitem{NNPDF:2021njg}
{\scshape NNPDF} collaboration, \emph{{The path to proton structure at 1\% accuracy}}, \href{https://doi.org/10.1140/epjc/s10052-022-10328-7}{\emph{Eur. Phys. J. C} {\bfseries 82} (2022) 428} [\href{https://arxiv.org/abs/2109.02653}{{\ttfamily 2109.02653}}].

\bibitem{Buckley:2014ana}
A.~Buckley, J.~Ferrando, S.~Lloyd, K.~Nordstr\"om, B.~Page, M.~R\"ufenacht et~al., \emph{{LHAPDF6: parton density access in the LHC precision era}}, \href{https://doi.org/10.1140/epjc/s10052-015-3318-8}{\emph{Eur. Phys. J. C} {\bfseries 75} (2015) 132} [\href{https://arxiv.org/abs/1412.7420}{{\ttfamily 1412.7420}}].

\bibitem{Artoisenet:2012st}
P.~Artoisenet, R.~Frederix, O.~Mattelaer and R.~Rietkerk, \emph{{Automatic spin-entangled decays of heavy resonances in Monte Carlo simulations}}, \href{https://doi.org/10.1007/JHEP03(2013)015}{\emph{JHEP} {\bfseries 03} (2013) 015} [\href{https://arxiv.org/abs/1212.3460}{{\ttfamily 1212.3460}}].

\bibitem{Alwall:2014bza}
J.~Alwall, C.~Duhr, B.~Fuks, O.~Mattelaer, D.G.~\"Ozt\"urk and C.-H.~Shen, \emph{{Computing decay rates for new physics theories with FeynRules and MadGraph 5\_aMC@NLO}}, \href{https://doi.org/10.1016/j.cpc.2015.08.031}{\emph{Comput. Phys. Commun.} {\bfseries 197} (2015) 312} [\href{https://arxiv.org/abs/1402.1178}{{\ttfamily 1402.1178}}].

\bibitem{Bierlich:2022pfr}
C.~Bierlich et~al., \emph{{A comprehensive guide to the physics and usage of PYTHIA 8.3}}, \href{https://doi.org/10.21468/SciPostPhysCodeb.8}{\emph{SciPost Phys. Codeb.} {\bfseries 2022} (2022) 8} [\href{https://arxiv.org/abs/2203.11601}{{\ttfamily 2203.11601}}].

\bibitem{Conte:2012fm}
E.~Conte, B.~Fuks and G.~Serret, \emph{{MadAnalysis 5, A User-Friendly Framework for Collider Phenomenology}}, \href{https://doi.org/10.1016/j.cpc.2012.09.009}{\emph{Comput. Phys. Commun.} {\bfseries 184} (2013) 222} [\href{https://arxiv.org/abs/1206.1599}{{\ttfamily 1206.1599}}].

\bibitem{Conte:2014zja}
E.~Conte, B.~Dumont, B.~Fuks and C.~Wymant, \emph{{Designing and recasting LHC analyses with MadAnalysis 5}}, \href{https://doi.org/10.1140/epjc/s10052-014-3103-0}{\emph{Eur. Phys. J. C} {\bfseries 74} (2014) 3103} [\href{https://arxiv.org/abs/1405.3982}{{\ttfamily 1405.3982}}].

\bibitem{Conte:2018vmg}
E.~Conte and B.~Fuks, \emph{{Confronting new physics theories to LHC data with MADANALYSIS 5}}, \href{https://doi.org/10.1142/S0217751X18300272}{\emph{Int. J. Mod. Phys. A} {\bfseries 33} (2018) 1830027} [\href{https://arxiv.org/abs/1808.00480}{{\ttfamily 1808.00480}}].

\bibitem{Araz:2021akd}
J.Y.~Araz, B.~Fuks, M.D.~Goodsell and M.~Utsch, \emph{{Recasting LHC searches for long-lived particles with MadAnalysis~5}}, \href{https://doi.org/10.1140/epjc/s10052-022-10511-w}{\emph{Eur. Phys. J. C} {\bfseries 82} (2022) 597} [\href{https://arxiv.org/abs/2112.05163}{{\ttfamily 2112.05163}}].

\bibitem{CMS:2021dzb}
{\scshape CMS} collaboration, \emph{{Search for a right-handed W boson and a heavy neutrino in proton-proton collisions at $ \sqrt{s} $ = 13 TeV}}, \href{https://doi.org/10.1007/JHEP04(2022)047}{\emph{JHEP} {\bfseries 04} (2022) 047} [\href{https://arxiv.org/abs/2112.03949}{{\ttfamily 2112.03949}}].

\bibitem{ATLAS:2023cjo}
{\scshape ATLAS} collaboration, \emph{{Search for heavy Majorana or Dirac neutrinos and right-handed W gauge bosons in final states with charged leptons and jets in pp collisions at $\sqrt{s}=13$~TeV with the ATLAS detector}}, \href{https://doi.org/10.1140/epjc/s10052-023-12021-9}{\emph{Eur. Phys. J. C} {\bfseries 83} (2023) 1164} [\href{https://arxiv.org/abs/2304.09553}{{\ttfamily 2304.09553}}].

\bibitem{deFavereau:2013fsa}
{\scshape DELPHES 3} collaboration, \emph{{DELPHES 3, A modular framework for fast simulation of a generic collider experiment}}, \href{https://doi.org/10.1007/JHEP02(2014)057}{\emph{JHEP} {\bfseries 02} (2014) 057} [\href{https://arxiv.org/abs/1307.6346}{{\ttfamily 1307.6346}}].

\bibitem{Cacciari:2008gp}
M.~Cacciari, G.P.~Salam and G.~Soyez, \emph{{The anti-$k_t$ jet clustering algorithm}}, \href{https://doi.org/10.1088/1126-6708/2008/04/063}{\emph{JHEP} {\bfseries 04} (2008) 063} [\href{https://arxiv.org/abs/0802.1189}{{\ttfamily 0802.1189}}].

\bibitem{Cacciari:2011ma}
M.~Cacciari, G.P.~Salam and G.~Soyez, \emph{{FastJet User Manual}}, \href{https://doi.org/10.1140/epjc/s10052-012-1896-2}{\emph{Eur. Phys. J.} {\bfseries C72} (2012) 1896} [\href{https://arxiv.org/abs/1111.6097}{{\ttfamily 1111.6097}}].

\bibitem{ATLAS:2023dxj}
{\scshape ATLAS} collaboration, \emph{{Electron and photon efficiencies in LHC Run 2 with the ATLAS experiment}}, \href{https://doi.org/10.1007/JHEP05(2024)162}{\emph{JHEP} {\bfseries 05} (2024) 162} [\href{https://arxiv.org/abs/2308.13362}{{\ttfamily 2308.13362}}].

\bibitem{Araz:2020lnp}
J.Y.~Araz, B.~Fuks and G.~Polykratis, \emph{{Simplified fast detector simulation in MADANALYSIS 5}}, \href{https://doi.org/10.1140/epjc/s10052-021-09052-5}{\emph{Eur. Phys. J. C} {\bfseries 81} (2021) 329} [\href{https://arxiv.org/abs/2006.09387}{{\ttfamily 2006.09387}}].

\bibitem{Araz:2023axv}
J.Y.~Araz, A.~Buckley and B.~Fuks, \emph{{Searches for new physics with boosted top quarks in the MadAnalysis 5 and Rivet frameworks}}, \href{https://doi.org/10.1140/epjc/s10052-023-11779-2}{\emph{Eur. Phys. J. C} {\bfseries 83} (2023) 664} [\href{https://arxiv.org/abs/2303.03427}{{\ttfamily 2303.03427}}].

\bibitem{ATLAS:2023nze}
{\scshape ATLAS} collaboration, \emph{{Performance of the reconstruction of large impact parameter tracks in the inner detector of ATLAS}}, \href{https://doi.org/10.1140/epjc/s10052-023-12024-6}{\emph{Eur. Phys. J. C} {\bfseries 83} (2023) 1081} [\href{https://arxiv.org/abs/2304.12867}{{\ttfamily 2304.12867}}].

\bibitem{ATLAS:2019fwx}
{\scshape ATLAS} collaboration, \emph{{Search for displaced vertices of oppositely charged leptons from decays of long-lived particles in $pp$ collisions at $\sqrt {s}$ =13 TeV with the ATLAS detector}}, \href{https://doi.org/10.1016/j.physletb.2019.135114}{\emph{Phys. Lett. B} {\bfseries 801} (2020) 135114} [\href{https://arxiv.org/abs/1907.10037}{{\ttfamily 1907.10037}}].

\bibitem{Heeck:2013rpa}
J.~Heeck and W.~Rodejohann, \emph{{Neutrinoless Quadruple Beta Decay}}, \href{https://doi.org/10.1209/0295-5075/103/32001}{\emph{EPL} {\bfseries 103} (2013) 32001} [\href{https://arxiv.org/abs/1306.0580}{{\ttfamily 1306.0580}}].

\bibitem{Barabash:2019enn}
A.S.~Barabash, P.~Hubert, A.~Nachab and V.I.~Umatov, \emph{{Search for triple and quadruple $\beta$ decay of $^{150}$Nd}}, \href{https://doi.org/10.1103/PhysRevC.100.045502}{\emph{Phys. Rev. C} {\bfseries 100} (2019) 045502} [\href{https://arxiv.org/abs/1906.07180}{{\ttfamily 1906.07180}}].

\bibitem{Lepage:1977sw}
G.P.~Lepage, \emph{{A New Algorithm for Adaptive Multidimensional Integration}}, \href{https://doi.org/10.1016/0021-9991(78)90004-9}{\emph{J. Comput. Phys.} {\bfseries 27} (1978) 192}.

\bibitem{Lepage:2020tgj}
G.P.~Lepage, \emph{{Adaptive multidimensional integration: VEGAS enhanced}}, \href{https://doi.org/10.1016/j.jcp.2021.110386}{\emph{J. Comput. Phys.} {\bfseries 439} (2021) 110386} [\href{https://arxiv.org/abs/2009.05112}{{\ttfamily 2009.05112}}].

\bibitem{ParticleDataGroup:2024cfk}
{\scshape Particle Data Group} collaboration, \emph{{Review of particle physics}}, \href{https://doi.org/10.1103/PhysRevD.110.030001}{\emph{Phys. Rev. D} {\bfseries 110} (2024) 030001}.

\end{thebibliography}\endgroup

\end{document}